\documentclass{aa}  
\usepackage{longtable}
\usepackage{caption}

\usepackage{graphicx}
\usepackage{txfonts}
\usepackage[fleqn]{amsmath}	% Advanced maths commands
\usepackage{amssymb}	% Extra maths symbols
\usepackage{mathtools}
\usepackage[breaklinks=true]{hyperref}
\hypersetup{
  colorlinks=true,   %% links colored instead of frames
  citecolor=blue,    %% links' color
  urlcolor=blue,     %% external hyperlinks
  linkcolor=blue,     %% internal latex links (eg Fig)
}
\bibpunct{(}{)}{;}{a}{}{,} %% natbib format for A&A and ApJ
\usepackage{lscape}

\newcommand{\lxluv}{$L_{\rm X}-L_{\rm UV}$\xspace}

\DeclareRobustCommand{\ion}[2]{%
\relax\ifmmode
\ifx\testbx\f@series
{\mathbf{#1\,\mathsc{#2}}}\else
{\mathrm{#1\,\mathsc{#2}}}\fi
\else\textup{#1\,{\mdseries\textsc{#2}}}%
\fi}

\newcommand{\xmm}{XMM--\emph{Newton}\xspace}

\newcommand{\xspec}{\textsc{xspec}\xspace}

\newcommand{\pc}{\phantom{0}}

\begin{document}

    \title{Quasars as standard candles VI: spectroscopic validation of the cosmological sample}
    
      %\author{E. Lusso\inst{1,2}\thanks{\email{elisabeta.lusso@unifi.it}}, S. Bisogni\inst{2}, G. Risaliti\inst{1,2}, E. Nardini\inst{2} \and others. 
   \author{Bartolomeo Trefoloni\inst{1,2}
   \thanks{\email{bartolomeo.trefoloni@unifi.it}},
    Elisabeta~Lusso\inst{1,2},
    Emanuele~Nardini\inst{2}, 
    Guido~Risaliti\inst{1,2},
    Alessandro~Marconi\inst{1,2},
    Giada~Bargiacchi\inst{3,4},
    Andrea~Sacchi\inst{5},
	Matilde~Signorini\inst{1,2}
	      }
          
% List of institutions
\institute{
$^{1}$Dipartimento di Fisica e Astronomia, Universit\`a di Firenze, via G. Sansone 1, 50019 Sesto Fiorentino, Firenze, Italy\\
$^{2}$INAF -- Osservatorio Astrofisico di Arcetri, Largo Enrico Fermi 5, I-50125 Firenze, Italy\\
$^{3}$Scuola Superiore Meridionale, Largo S. Marcellino 10, I-80138, Napoli\\
$^{4}$Istituto Nazionale di Fisica Nucleare (INFN), Sez. di Napoli, Complesso Univ. Monte S. Angelo, Via Cinthia 9, I-80126, Napoli, Italy\\
$^{5}$Center for Astrophysics | Harvard \& Smithsonian, 60 Garden Street, Cambridge, MA 02138, USA\\
}

\titlerunning{Quasars as standard candles VI: spectroscopic validation of the cosmological sample}
\authorrunning{B. Trefoloni et al.}
%\date{Received XX; accepted XX}
\date{\today}

   \abstract{
   \textit{Context:} A sample of quasars has been recently assembled to investigate the non-linear relation between their monochromatic luminosities at 2500\,\AA\, and 2 keV and to exploit quasars as a new class of \textit{standardized candles}. The use of this technique for cosmological purposes relies on the non-evolution with redshift of the UV--optical spectral properties of quasars, as well as on the absence of possible contaminants such as dust extinction and host-galaxy contribution.
   \newline
   \textit{Aims:} We search for possible clues of significant evolution with redshift, presence of dust extinction and host-galaxy contamination of the spectral properties of our cosmological quasar sample. 
   \newline
   \textit{Methods:} We produced composite spectra in different bins of redshift and accretion parameters (black hole mass, bolometric luminosity), to search for any possible evolution of the spectral properties of the continuum of the composites with these parameters.\newline
   \textit{Results:} We found a remarkable similarity amongst the various stacked spectra.  Apart from the well known evolution of the emission lines with luminosity (i.e. the Baldwin effect) and black hole mass (i.e. the virial relation), the overall shape of the continuum, produced by the accretion disc, does not show any statistically significant trend with black-hole mass ($M_{\rm BH}$), bolometric luminosity ($L_{\rm bol}$), or redshift ($z$). The composite spectrum of our quasar sample is consistent with negligible levels of both intrinsic reddening (with a colour excess $E(B-V)\lesssim 0.01$) and host-galaxy emission (less than 10\%) in the optical. We tested whether unaccounted dust extinction could explain the discrepancy between our cosmographic fit of the Hubble--Lema\^{\i}tre diagram and the concordance $\rm{\Lambda}$CDM model. We found that the average colour excess required to solve the tension should increase with redshift up to unphysically high values $(E(B-V)\simeq0.1$ at $z>3$). This would also imply that the intrinsic emission of quasars is much bluer and more luminous than ever reported in observed spectra.
   \newline
   \textit{Conclusions:} The similarity of quasar spectra across the parameter space excludes a significant evolution of the average continuum properties with any of the explored parameters, confirming the reliability of our sample for cosmological applications. Lastly, dust reddening cannot account for the observed tension between the Hubble--Lema\^{\i}tre diagram of quasars and the $\rm{\Lambda}$CDM model.}

\abstract{A sample of quasars has been recently assembled to investigate the non-linear relation between their monochromatic luminosities at 2500\,\AA\, and 2 keV and to exploit quasars as a new class of \textit{standardized candles}. The use of this technique for cosmological purposes relies on the non-evolution with redshift of the UV--optical spectral properties of quasars, as well as on the absence of possible contaminants such as dust extinction and host-galaxy contribution. We address these possible issues by analysing the spectral properties of our cosmological quasar sample. We produced composite spectra in different bins of redshift and accretion parameters (black hole mass, bolometric luminosity), to investigate any possible evolution of the spectral properties of the continuum of the composites with these parameters. We found a remarkable similarity amongst the various stacked spectra. Apart from the well known evolution of the emission lines with luminosity (i.e. the Baldwin effect) and black hole mass (i.e. the virial relation), the overall shape of the continuum, produced by the accretion disc, does not show any statistically significant trend with black-hole mass ($M_{\rm BH}$), bolometric luminosity ($L_{\rm bol}$), or redshift ($z$). The composite spectrum of our quasar sample is consistent with negligible levels of both intrinsic reddening (with a colour excess $E(B-V)\lesssim 0.01$) and host-galaxy emission (less than 10\%) in the optical. We tested whether unaccounted dust extinction could explain the discrepancy between our cosmographic fit of the Hubble--Lema\^{\i}tre diagram and the concordance $\rm{\Lambda}$CDM model. The average colour excess required to solve the tension should increase with redshift up to unphysically high values ($E(B-V)\simeq0.1$ at $z>3$) that  would imply that the intrinsic emission of quasars is much bluer and more luminous than ever reported in observed spectra. The similarity of quasar spectra across the parameter space excludes a significant evolution of the average continuum properties with any of the explored parameters, confirming the reliability of our sample for cosmological applications. Lastly, dust reddening cannot account for the observed tension between the Hubble--Lema\^{\i}tre diagram of quasars and the $\rm{\Lambda}$CDM model.}


   \maketitle
%
%-------------------------------------------------------------------

\section{Introduction}

Quasars are the most luminous persistent sources in the Universe and, as such, they allow us to investigate its evolution up to redshifts $z>6$  (e.g. \citealt{mortlock2011luminous,banados2018,yang2020poniua, wang2021luminous, zappacosta2023hyperluminous}). Thanks to their ubiquity and high luminosity, these sources were early recognized as powerful cosmological probes. Since the '70s, correlations between quasars' observables have been employed to determine their distances. The discovery of the Baldwin effect (the emission-line equivalent widths anti-correlate with the quasar luminosity; \citealt{baldwin1977}) envisaged that quasars could be utilised as cosmological probes. Techniques involving similar relations have been subsequently developed (see Section 6 in \citealp{czerny2018astronomical} for a detailed review). Yet, the large dispersion as well as the scarcity of suitable sources, have in most cases limited their applicability for measuring %to measure 
the expansion parameters (e.g., $H_0$). Other attempts to extract cosmological information from quasars are based on the broad line region (BLR, \citealt{elviskarovska2002}) or dust \citep{yoshii2014new} reverberation mapping, also in combination with spectro-astrometric measurements (SARM; \citealt{wang2020parallax}), the Eigenvector 1 main sequence (\citealt{marzianisulentic2014}), or the X-ray excess variance (\citealt{lafranca2014}).

In the last decade, the efforts of our group have been mostly devoted to develop and refine a novel technique based on the non-linear relation between the X-ray and UV luminosity (\lxluv) in quasars (or, equivalently, the $\alpha_{\rm{OX}}-L_{\rm UV}$ relation, e.g. \citealp{vignali03, lusso2010}), which has long been known (\citealp{avnitananbaum79}). The \lxluv relation, according to our current knowledge, reflects the interplay between the UV photons coming from the disc (\citealt{shakura1973black, novikov1973astrophysics}) and the hot ``corona'' (\citealt{haardt1991two, haardt1993x}) that boosts their energy producing the X-ray emission. Despite the theoretical efforts (e.g. \citealt{nicastro2000, merloni2003, lusso2017quasars, arcodia2019}), a comprehensive description of the physical mechanism underlying the \lxluv relation is still missing. Even the choice of the monochromatic luminosities at 2 keV and 2500\,\AA\ can be regarded as rather arbitrary, being mainly due to historical and empirical reasons. Indeed, we have argued that better proxies, such as the \ion{Mg}{ii} emission line (see \citealt{signorini2023quasars} for further details), of the physical mechanisms responsible for the observed relation can be exploited.

In addition to important implications for accretion physics, the non-linearity of the \lxluv relation, combined with the non-evolution with redshift of its parameters (\citealp{risaliti2019cosmological}), makes it a useful tool to determine the luminosity distance of quasars and so transform them into a new class of \textit{standardizable candles}. The \lxluv relation has indeed proved to be a fruitful cosmological tool
after it was demonstrated that most of the observed scatter is not intrinsic but rather due to observational issues, as the dispersion can be narrowed down to 0.09 dex in case of samples with high quality data (\citealt{sacchi2022quasars}). 
This new category of standard candles comes particularly convenient when it comes to populate the intermediate redshifts between the farthest supernovae at $z\sim2$ \citep{scolnic2018} and the cosmic microwave background (CMB). Despite a good agreement with the $\rm{\Lambda}$CDM model in the low redshift regime of the Hubble-Lema\^{\i}tre diagram, a 3--4$\,\sigma$ tension with the model, driven by the quasar data at earlier cosmic epochs ($z>1.5$), has been reported (\citealt{risaliti2019cosmological, bargiacchi2021cosmography, sacchi2022quasars, bargiacchi2022quasar,lenart2023}).

In \citet[][henceforth L20]{lusso2020} we discussed how we produced our latest compilation of 2,421 quasars for cosmological purposes, describing how the measurements of X-ray and UV luminosities were performed. However, in order to validate the selection criteria adopted to build the sample, we still need to address two main issues that could be present to some extent: i) an evolution with redshift of the slope of the optical and UV continuum, ii) the contamination of host-galaxy in the optical and flux attenuation due to dust and gas. While we discuss the validity of the cosmological techniques adopted to analyse the sample in a companion paper (Risaliti et al. 2024, in preparation), in this work we mainly focus on a spectral confirmation of the robustness of the selection criteria. In particular, we aim at demonstrating, by means of composite spectra, that the continuum of our sample (in which the UV proxy $L_{2500\,\AA}$ is evaluated) does not show any appreciable evolution with redshift. At the same time, the spectral composites are a useful tool to reveal the presence of residual (i.e., still present despite the selection cuts) dust extinction and/or host-galaxy contribution.

To the end of investigating any possible spectral diversity among the objects constituting our cosmological sample, we employed stacked templates. The lack of a significant evolution of the continuum among the stacked spectra, representative of different regions of the parameter space, would ensure the universality of the accretion mechanism powering the optical--UV emission.

In this paper, we analyse the average spectral properties of the UV and optical data of the cosmological sample described in L20, with three main goals: i) directly verify how close the average spectral properties of our sample are to those of typical blue quasars, by means of composite spectra in different regions of the $M_{\rm BH}-L_{\rm bol}-z$ parameter space; ii) estimate the degree of host-galaxy contamination; iii) quantify the effect of intrinsic dust absorption in our sample of quasars. The last two effects would affect our estimates of $f_{2500\,\AA}$
 and so undermine the cosmological application of the \lxluv relation.
Employing spectral data to understand -- and, in case, account for -- such possible observational effects, which can alter both X-ray and UV measurements, would definitely corroborate the evidence that the observed discrepancy with the concordance cosmological model is not due to observational issues.

The paper is structured as follows. A brief summary about the selection of the quasar sample is provided in Section \ref{sec:dataset}, whilst the analysis performed on the sample spectra is described in Section \ref{sec:methods}. Results are discussed in Section \ref{results} and conclusions are drawn in Section \ref{sec:discussion}.

\section{The data set}
\label{sec:dataset}
The quasar sample analysed here was published by L20 and is made of 2,421 sources within the redshift range $0.13 < z < 5.42$. The data set was produced by merging seven different subsamples available from the literature:
\begin{itemize}
    \item the \textit{Sloan Digital Sky Survey} data release 14 (SDSS-DR14, \citealt{paris2018sloan}) was cross-matched with the \textit{XMM-Newton}  4XMM-DR9 (\citealt{webb2020}) catalogue of serendipitous sources, %SDSS-4XMM, 
    the second release of the \textit{Chandra Source Catalogue} (CSC2.0, \citealt{evans2010}) and the AGN sample described in \citet{menzel2016}, belonging to the \textit{XMM–Newton} XXL survey (XMM–XXL, PI: Pierre), producing respectively the SDSS-4XMM, the SDSS-Chandra, and the SDSS-XXL samples.
    \item local quasars ($0.009<z<0.1$) with available UV data from the \textit{International Ultraviolet Explorer} (IUE) in the Minkulski Archive for Space Telescopes (MAST) and X-ray data were included in the sample to anchor the normalisation of the quasar Hubble diagram with Type Ia supernovae.
    \item To improve the coverage at high redshifts, three samples of quasars with pointed observations were added. We added the sample of 29 luminous quasars at redshift $3.0<z<3.3$ with dedicated \textit{XMM-Newton} X-ray observations described in \citet{nardini2019}, the 53 high redshift ($z>4.01$) quasars from \citet{salvestrini2019}, and the 25 quasars at $z>6$ from \citet{vito2019}.
\end{itemize}

In order to produce such a compilation, several selection criteria were adopted. We thus redirect the interested reader to L20 (in particular, their Section 5) for a more comprehensive description, while here we briefly highlight the filters adopted to define the sample. All the selected sources are radio quiet\footnote{We applied the usual criteria based on the ratio between the flux density at 6 cm and 2500\,\AA, i.e. AGN with $R = F(6\,{\rm cm})/F(2500\,\AA)<10$ are considered radio quiet \citep{kellermann89}.} and not flagged as broad absorption line (BAL) quasars according to the SDSS-DR14 quasar catalogue (\citealt{paris2018sloan}), which was the latest release at the time of the L20 analysis. A broadband photometric spectral energy distribution (SED) was build for each source, and only quasars with UV--optical colours close to the ones of typical blue quasars (\citealt{richards2006spectral}) were retained. In addition, steep X-ray spectra ($1.7 <\Gamma< 2.8$) and deep enough X-ray observations (to avoid the Eddington bias), were required. Such properties make the L20 sample a truly unique sample of blue quasars, sharing a high degree of homogeneity in terms of broadband properties and observational data. The present work is focused on the spectral properties of the sources with available SDSS data, which constitute the bulk ($\sim$93\%) of the sample.

\section{Methods}
\label{sec:methods}

\begin{figure*}[h!]
\centering
\includegraphics[scale=0.4]{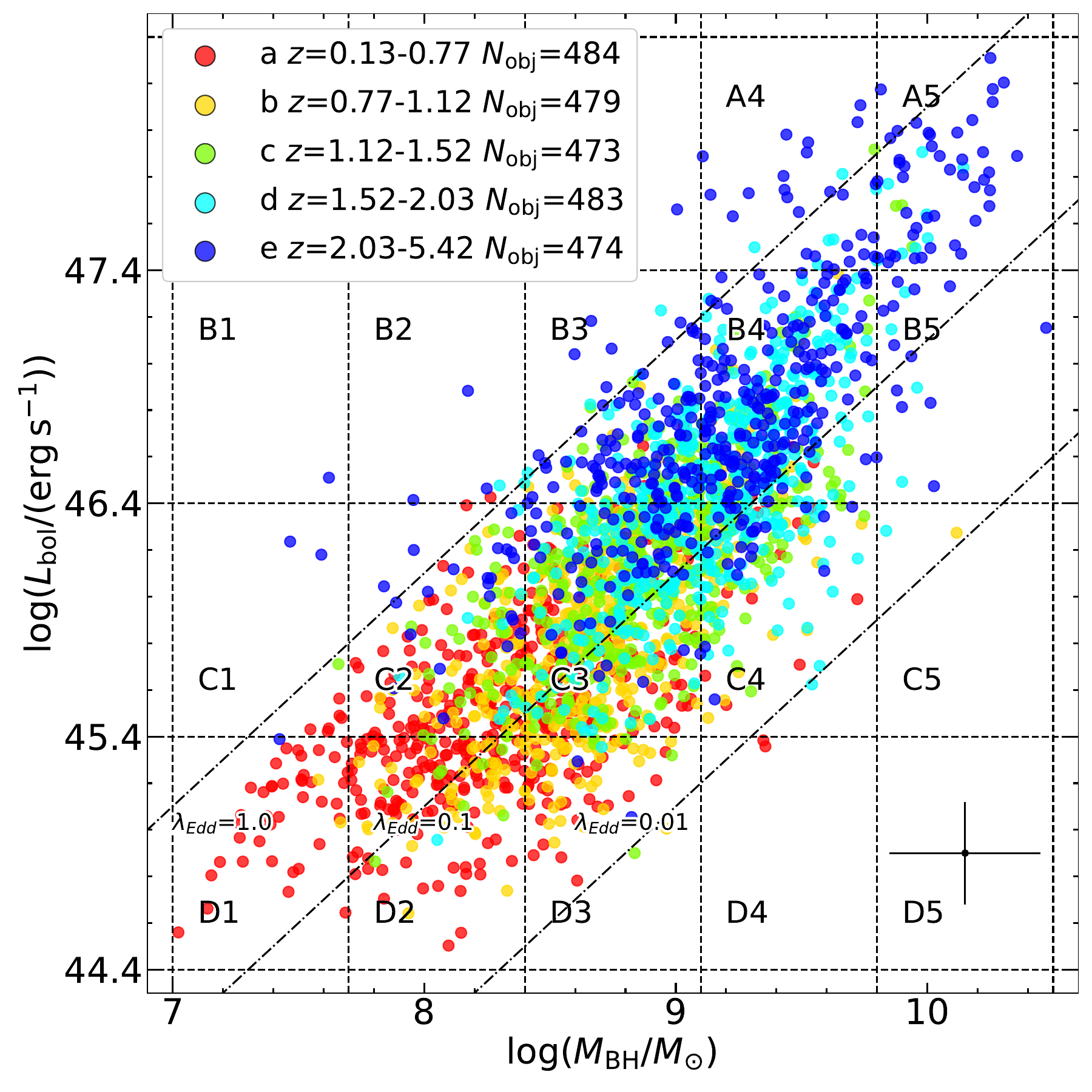}
\caption{$\log M_{\rm BH}$\,--\,$\log L_{\rm bol}$ parameter space for the quasar sample published by \cite{lusso2020}, with colour-coded redshifts. Dot-dashed lines represent constant $\lambda_{\rm Edd}$ values. Typical uncertainties on $\log(M_{\rm BH})$ and $\log(L_{\rm bol})$ are represented in the bottom right corner.}
\label{fig:space}
\end{figure*}

\subsection{The parameter space}
\label{sec:paramspace}
Our aim is to study whether the continuum properties of our sample evolve with redshift or any of the governing parameters of the quasar spectrum, namely the bolometric luminosity ($L_{\rm bol}$), black hole mass ($M_{\rm BH}$), and accretion rate (parametrised through the Eddington ratio $\lambda_{\rm Edd}=L_{\rm bol}/L_{\rm Edd}$, where $L_{\rm Edd}$ is the luminosity at the Eddington limit). %and redshift. 
To this end, we considered the latest quasar catalogue by 
\citet[][henceforth WS22]{wu2022catalog}, who compiled continuum and emission-line properties for 750,414 broad-line quasars from the SDSS DR16, including physical quantities such as single-epoch virial $M_{\rm BH}$, $L_{\rm bol}$, and $\lambda_{\rm Edd}$. We also gathered the latest spectral data available for each source in our sample, hence the spectra of $\sim$4\% of the sources belong to the DR17 release, where calibration files have been updated \citep{abdurro2022seventeenth}.

Fig.~\ref{fig:space} presents the $\log L_{\rm bol}$\,--\,$\log M_{\rm BH}$ plane for our sample. 
The partition of the plane was defined by seeking the balance between a large number of bins to improve the sensitivity on the parameters ($L_{\rm bol}$, $M_{\rm BH}$) that control the average spectrum, and a statistically robust number of objects in each region.
We further divided the sample into five redshift intervals, $z$\,=\,[0.13, 0.77, 1.12, 1.52, 2.03, 5.42], each one containing a similar number of objects (roughly 480 objects per bin).
The resulting grid is shown in Fig. \ref{fig:space}. The width of the $\log M_{\rm BH}$ ($\log L_{\rm bol}$) bins is 0.7 dex (1 dex), large enough to encompass the large systematic uncertainties associated to the estimate of this quantity, typically 0.3 dex (but up to 0.4 dex in case of \ion{C}{iv}-based $M_{\rm BH}$, e.g., \citealt{shen2011, kratzer2015mean, rakshit2020spectral}). The points of the grid are $\log M_{\rm BH}$\,=\,[7.0, 7.7, 8.4, 9.1, 9.8, 10.5] and $\log L_{\rm bol}$\,=\,[44.4, 45.4, 46.4, 47.4, 48.4]. This procedure defines a 3D grid made of 5\,$\times$\,5\,$\times$\,4\,=\,100 bins.
The individual cells are labelled using the following nomenclature: increasing capital letters (A, B, C, D) denote decreasing bolometric luminosity across the bins, while increasing numbers (1, 2, 3, 4, 5) mark increasing average black hole masses, and increasing lower-case letters (a, b, c, d, e) represent increasing redshift intervals, which are also colour-coded according to the values in the legend.

Clearly, not all the bins are densely, nor uniformly populated. Lower redshift quasars preferentially occupy the lower luminosity and black hole mass region, whereas higher redshift sources are observed at higher values of both $L_{\rm bol}$ and $M_{\rm BH}$. Moreover, there are cells in the $L_{\rm bol}$\,--\,$M_{\rm BH}$ plane where the barycentre of the sources is close to the edge of a given interval (e.g., B2, B5, C1). In these cases, low sample statistics would produce not only low signal-to-noise (SNR) composite spectra, but also a stack not fully representative of its nominal locus in the parameter space. 

To select the regions in the $\log M_{\rm BH}$\,--\,$\log L_{\rm bol}$ plane populated enough to be included in the stacking analysis, we quantified the probability that a statistical fluctuation in terms of $\log M_{\rm BH}$ and $\log L_{\rm bol}$ of objects in adjacent bins pushed them into a given region.
We simulated 1000 mock samples starting from the actual distribution of $\log M_{\rm BH}$ and $\log L_{\rm bol}$ in each bin within the grid. The values of $\log M_{\rm BH}$ and $\log L_{\rm bol}$ are then scattered by a value drawn from a Gaussian distribution whose amplitude was set by the total uncertainty on each given data point. In doing so, we conservatively assumed the largest possible systematic uncertainty on both parameters and summed it in quadrature to the statistical uncertainty reported in WS22. For the bolometric corrections, we assumed a systematic uncertainty of $\sim$50\% on $\log L_{\rm bol}$ for each data point in the plane, suitable in case the bolometric correction is performed on the rest-frame 1350 \AA\ luminosity (see \citealt{shen2011}), which translates into $\lesssim$\,0.22 dex. The systematic uncertainty on $\log M_{\rm BH}$ was assumed to be $0.40$ dex, a value based on the \ion{C}{iv} calibration (i.e. the least reliable one). As a consequence of the random scatter, the points of the mock sample distribution could escape the bin where the original data were instead contained. We then computed the average $\log M_{\rm BH}$ and $\log L_{\rm bol}$ of the resulting scattered points and retained only the regions within the parameter space where in at least 90\% of the mock samples the average $\log M_{\rm BH}$ and $\log L_{\rm bol}$ remained within the bin. We refer to bins retained by this procedure as bins meeting the ``representativeness criterion''.
In general, 44 bins out of 100 in the parameter space do not contain any object, while 25 bins are excluded by the aforementioned criterion.
The whole selection procedure ultimately excluded 69 scarcely populated bins (with less than 9 objects), 
and it ensures that the composite spectrum is truly representative of its associated region in the parameter space. At the end of this procedure, 2,316 objects were retained in the parameter space.

\subsection{Building the stacks}
\label{sec:stacking}
The spectra were collected according to their PLATE, MJD and FIBERID reported in WS22.\footnote{The spectrum of SDSS J124406.96+113524.2 could not be found. However, this quasar belongs to a region where there are $\sim$60 objects, thus we expect that its inclusion would not play any significant difference in the resulting stack.}
We produced the stacked spectra and adopted a bootstrap recipe for the evaluation of the uncertainty on the stacks (more quantitative assessments on how different stacking prescriptions affect the results are explored in Appendix \ref{app:stack_opt}), following the steps described below in each of the selected regions:

\begin{itemize}
    \item Each spectrum was corrected for Galactic absorption assuming the value for the colour excess $E(B-V)$ available in the WS22 catalogue (\citealt{schlafly2011measuring}) and a \citet{fitzpatrick1999correcting} extinction curve. Then, the de-reddened spectra were shifted to the rest frame.
    \item All the spectra in the same redshift bin were resampled, by means of linear interpolation, onto a fixed wavelength grid. The grid was defined in the interval $\lambda_1-\lambda_2$ with a fixed dispersion $\Delta \lambda$ so as to cover the shortest and the longest rest-frame wavelengths in each redshift interval. The values for $\lambda_1, \lambda_2, \Delta \lambda$ are listed in Table \ref{tbl:tbl_1}. The value for $\Delta \lambda$ was chosen in order to preserve the SDSS observed frame resolution also in the rest frame spectra, similarly to what described in Section 3.2 of \citet[][VB01]{vandenberk2001}, assuming an observed frame spectral resolution of $\sim$2000 at 6500\,\AA.
    \item We also applied a few quality cuts. Objects with more than 10\% of the pixels flagged as bad ({\sc and\_mask}\,>\,0) in their SDSS spectra were excluded. Also objects where the normalisation wavelength $\bar{\lambda}$ (the wavelength chosen to scale all the spectra in the cell) was not covered in the rest-frame spectrum were excluded.
    
    \item The average flux in each spectral channel and the relative uncertainty were evaluated adopting a bootstrap technique. In brief, a number of spectra equal to the total number of objects in the bin is drawn, allowing for replacement. These spectra are normalised by their flux value at the reference wavelength $\bar{\lambda}$, and an average stack is produced. This procedure is repeated with new draws until the desired number of stacks is reached. Subsequently, in each spectral channel the distribution of the normalised fluxes is computed, a 3$\sigma$ clipping is applied, and the median value (less affected by possible strong outliers) is chosen. The uncertainty on the median value was evaluated as the semi-inter-percentile range between 1$^{\rm st}$ and 99$^{\rm th}$ percentile. %As a consistency check, we verified that the  uncertainty derived from bootstrap agreed (within a factor of 10) with the uncertainty determined by measuring the standard deviation of the flux in relatively featureless sections of the combined spectrum.

    %\item In order to further clean the stacks from defective pixels in the original spectra, the composite spectrum is run through a filter so designed: for each spectral channel (with the only exception of the first and last four) a box 8 channels wide is selected and the median absolute dispersion ($MAD = (\sum_{i=1}^{i=8} |\Tilde{f} - f_{i}|)/8$, being $\Tilde{f}$ the median flux in the box) is evaluated. If the flux in such channel deviated by more than 3 times the $MAD$ from the median value in the box, the flux of such pixel is excluded from the stack.

\end{itemize}

The information about the relevant quantities adopted in each redshift interval is shown in Table \ref{tbl:tbl_1}.
We explored different possible ways of combining the spectra, averaging the flux in each spectral channel and estimating the uncertainty. How these assumptions affect the resulting composites and the subsequent results, is presented in Appendices \ref{app:stack_opt} and \ref{app:test_bin}. All the composite spectra are presented in Appendix \ref{app:synopt}.

\begin{table}[h!]
\centering
\begin{tabular}{c c c c c c}
 \hline \noalign{\smallskip}
 Redshift interval & $N_{\rm obj}$ & $\lambda_1$ & $\lambda_2$ & $\Delta \lambda$ & $\bar{\lambda}$  \\
 (1) & (2) & (3) & (4)  & (5) & (6) \\
 \hline \noalign{\smallskip}
 0.13--0.77 & 484 & 2062 & 9200 & 2.2  & 5100 \\ 
 
 0.77--1.12 & 479 & 1722 & 5876 & 1.7 & 4200 \\
 
 1.12--1.52 & 473 & 1448 & 4906 & 1.4 & 3100 \\
 
 1.52--2.03 & 483 & 1205 & 4127 & 1.2 & 2200 \\
 
 2.03--5.42 & 474 & 569 & 3432 & 0.7 & 1450 \\ %[0.5ex] 
 \hline
 \end{tabular}
\caption{Stacking parameters: redshift interval (1), number of sources within the redshift interval (2), lower (3) and upper (4) wavelengths of the wavelength grid, fixed dispersion (5), and reference wavelength (6). All wavelengths are reported in units of \AA.}
\label{tbl:tbl_1}
\end{table}

% PLOT STACK
\begin{figure*}[h!]
\centering
\includegraphics[width=\linewidth,clip]{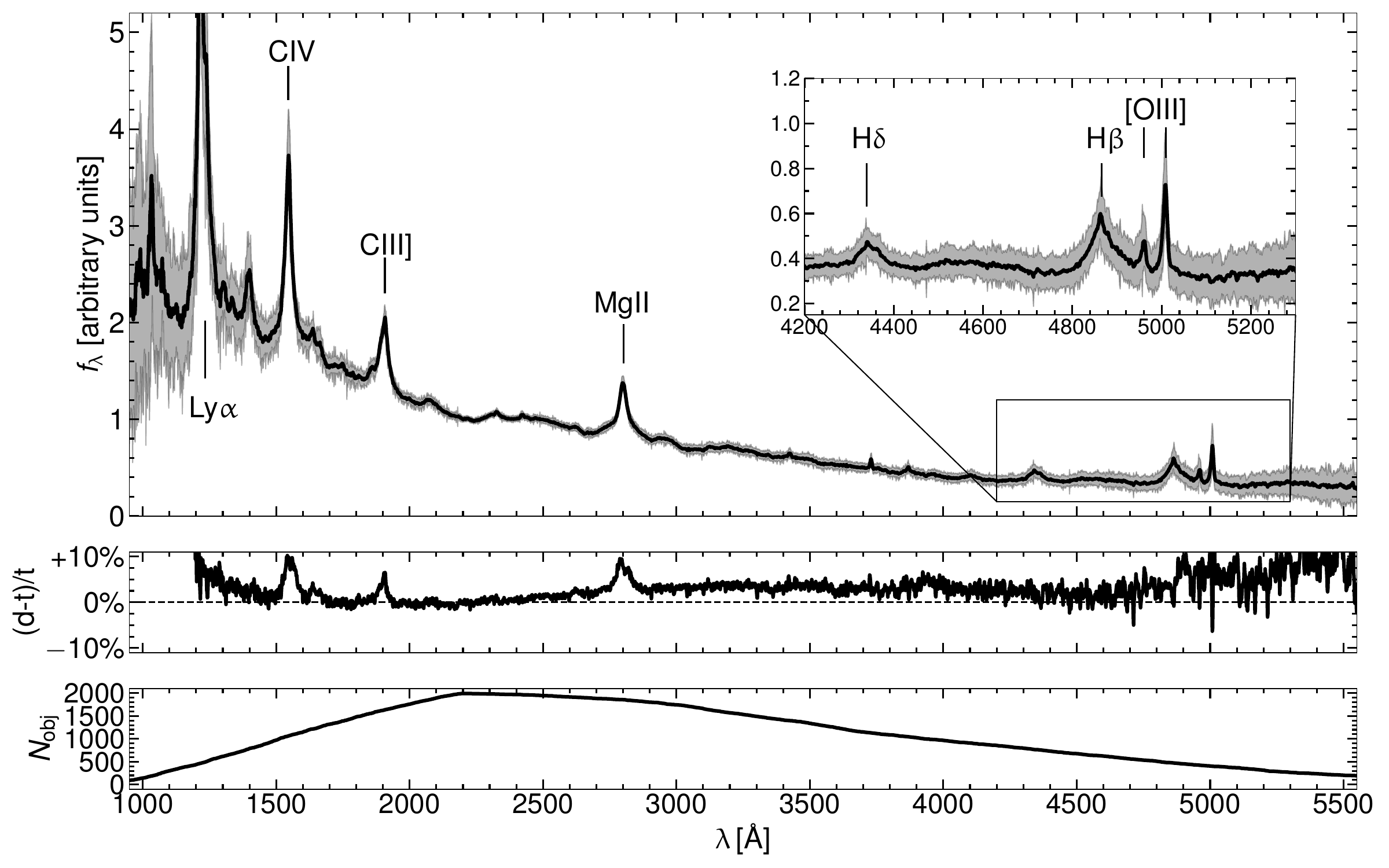}
\caption{Spectral composite of the entire L20 sample (\textit{top panel}). The Ly$\alpha$ emission line is cut for visualization purposes. The inset zooms in on the H$\beta$--[\ion{O}{iii}] region. The main emission lines are labelled accordingly. \textit{Middle panel:} relative residuals with respect to the VB01 template. \textit{Bottom panel:} number of objects contributing to each spectral channel. The composite spectrum of the L20 sample is available in its entirety in a machine-readable form in the online journal.}
\label{fig:full_stack}
\end{figure*}

\subsection{Spectral fit of the composites}
\label{sec:fits}
To estimate the characteristic spectral quantities of the continuum (chiefly the slope $\alpha_{\lambda}$ of the power law $f_{\lambda}\propto \lambda^{-\alpha_{\lambda}}$) and the emission lines, in particular the rest-frame equivalent width (EW), the full width at half-maximum (FWHM), and the peak offset velocity ($v_{\rm off}$), we performed the spectral fits of all of our composite spectra. To this end we adopted a custom-made code, based on the IDL MPFIT package \citep{Markwardt2009}, which takes advantage of the Levenberg-Marquardt technique \citep{more1978levenberg} to solve the least-squares problem. Our main goal is a faithful reproduction of the continuum, in order to investigate whether the continuum slope and consequently the 2500\,\AA\ flux exhibit any significant evolution with the accretion parameters.

\begin{figure*}[h!]
     \centering
         \centering
         \includegraphics[width=0.49\textwidth]{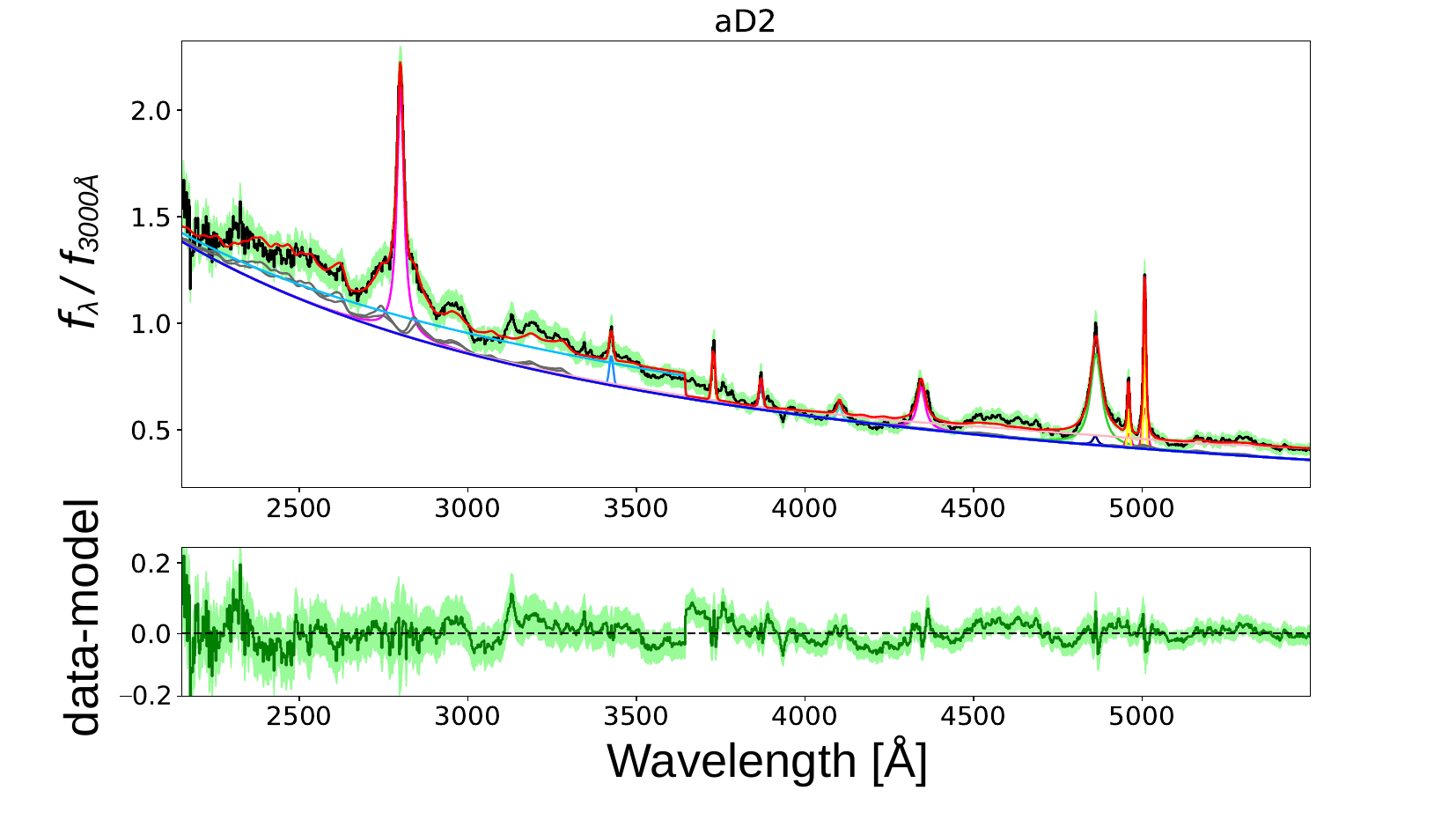}
     \hfill
         \centering
         \includegraphics[width=0.49\textwidth]{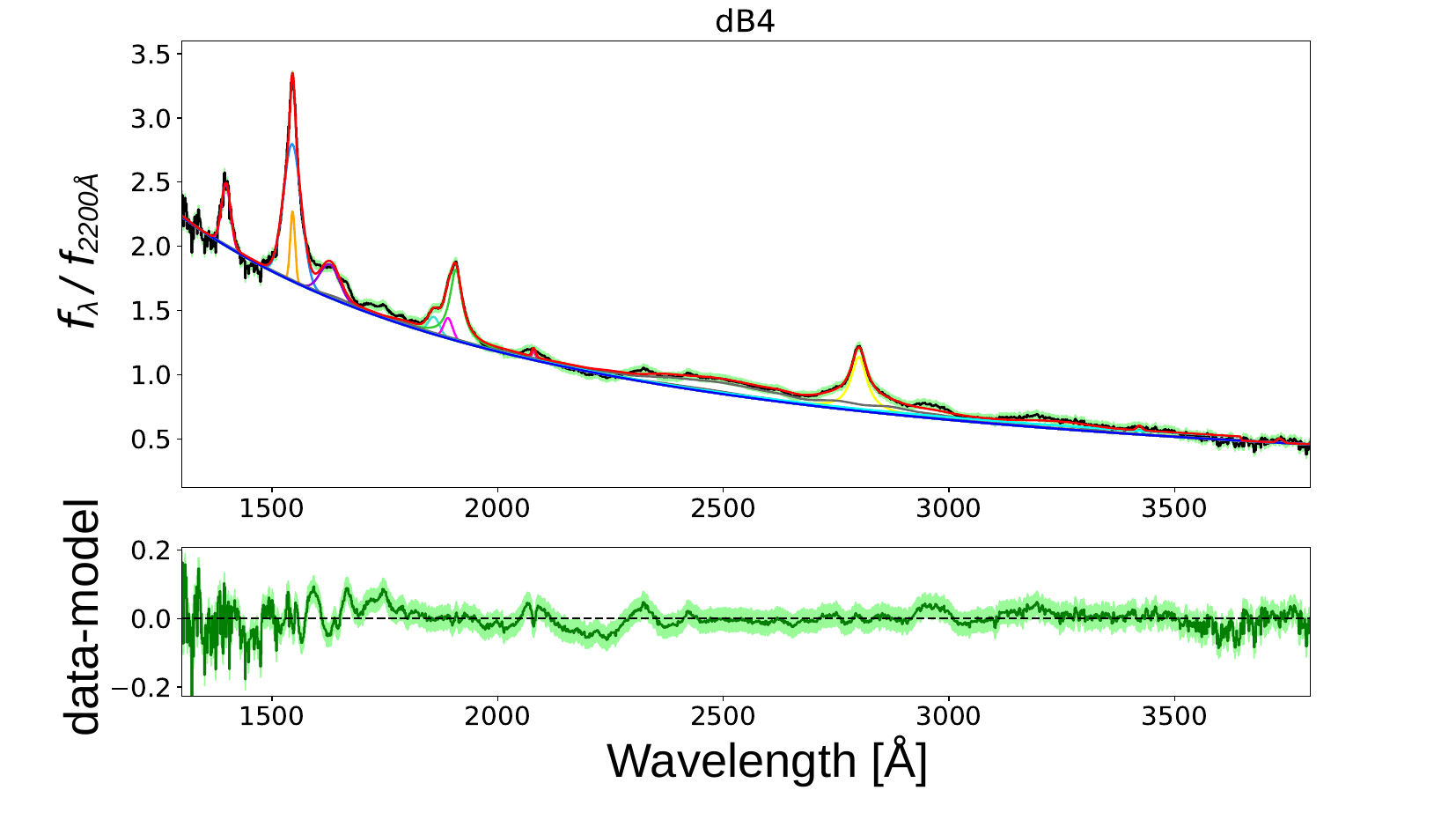}
\caption{Examples of a low-redshift bin (left) and a high-redshift bin (right) fit. The continuum power law is marked in blue, \ion{Fe}{ii} templates in grey, emission lines in different colours (orange, purple, green, cyan). The total model is depicted in red.}
\label{fig:fit_ex}
\end{figure*}

The baseline model is given by a continuum power law pivoting at different emission-free wavelengths depending on the redshift interval (3000\,\AA\ in the three bins at $z<$\,1.52, 2200\,\AA\ in the two bins at $z\geq$\,1.52), a set of emission lines, the Balmer continuum, and a set of iron templates convolved with a Gaussian line-of-sight velocity distribution whose weights are free parameters in the fit. Each line required a different decomposition: the \ion{C}{iv} line was generally fitted with two Lorentzian profiles (one broad, one narrow).
A single broad Lorentzian profile combined with the iron templates is able to adequately reproduce the \ion{Mg}{ii} profile. The H$\beta$ was modelled by considering a core narrow Gaussian component whose FWHM could not exceed $\sim$1000 km s$^{-1}$ and a broad component whose analytical shape (Gaussian or Lorentzian) was chosen each time to best reproduce the data. The [\ion{O}{iii}] doublet was modelled with Gaussian or Lorentzian profiles with a ratio of the [\ion{O}{iii}]$\lambda$4959\AA\, to $\lambda$5007\AA\, components fixed at one to three. In addition, other weaker forbidden ([\ion{Ne}{v}]\,$\lambda$3426, [\ion{O}{ii}]\,$\lambda$3727, [\ion{Ne}{iii}\,$\lambda$3869]) and permitted (H$\gamma$, H$\delta$) emission lines were included to improve the residuals. Above $\log(L_{\rm bol}/({\rm erg\,s^{-1}}))\gtrsim 46$ the host galaxy contribution at optical wavelengths is expected to be negligible with respect to the quasar emission (\citealt{shen2011, jalan2023empirical}). Therefore, we only added an elliptical galaxy template from \cite{assef2010low} to the model, to account for possible galactic emission, only in the two lowest redshift bins, where the bulk of the sample has $L_{\rm bol}$ close or below above value and the rest-frame wavelengths sampled in these redshift ranges could be imprinted by galactic emission.
The fits were performed adopting the same wavelength interval within the same redshift bin to avoid mismatches between the adopted continuum windows. The fitting wavelengths are listed in Table \ref{tbl:tbl_2} and examples of a low- and a high-redshift fits are shown in the left and right panels of Fig. \ref{fig:fit_ex}, respectively.

\begin{table}[h!]
\centering
%\hspace{8mm}Fit parameters
\begin{tabular}{c c c c} 
 \hline \noalign{\smallskip}
 Redshift interval & $\lambda_{1}$ & $\lambda_{2}$ & $\bar{\lambda}$ \\[0.5ex] 
 (1) & (2) & (3) & (4) \\
 \hline \noalign{\smallskip}
 0.13--0.77 & 2150 & 5500 & 3000 \\ 
 0.77--1.12 & 2150 & 5500 & 3000 \\
 1.12--1.52 & 1700 & 4500 & 3000 \\
 1.52--2.03 & 1300 & 3800 & 2200 \\
 2.03--5.42 & 1300 & 3200 & 2200 \\ %[1ex] 
  \hline
\end{tabular}
\caption{Fit parameters: redshift interval (1), lower (2) and upper (3) wavelengths, and normalisation wavelength (4) of the stacks. All wavelengths are reported in units of \AA.}
\label{tbl:tbl_2}
\end{table}

The uncertainties on the relevant parameters were estimated by means of mock samples. For each composite we produced 100 mock samples by adding a random value to the actual flux, extracted from a Gaussian distribution whose amplitude was set by the uncertainty value in that spectral channel. We then fitted the mock samples and evaluated the 3$\sigma$ clipped distribution of the relevant parameters. The final uncertainty on the reported parameters was estimated as the standard deviation of such a distribution.

\section{Results}
\label{results}

\subsection{The full sample and literature composite spectra}
\label{sec:comparison}

As a reference for the whole sample we produced a stacked spectrum collecting all the spectra of the L20 sample matching our quality criteria. The result is shown in Fig. \ref{fig:full_stack}. The procedure adopted to produce our template is the same as described in \ref{sec:stacking}. The normalisation wavelength $\bar{\lambda}$ was set to 2200\,\AA, as this is the only relatively emission-line-free continuum window available for most of the objects in our sample. 
To place our sample in a broader context, we compared our new template with other literature average spectra, namely \citet{francis1991high}, \citet{brotherton2001composite}, VB01, \citet{selsing2016x}, and \citet{harris2016composite}.

The composite spectrum by \citet{francis1991high} was produced using 718 objects from the Large Bright Quasar Survey (LBQS) with absolute magnitude in the $B$-band $M_B$\,$\leq$\,$-20.5$ and redshifts in the range 0.05\,<\,$z$\,<\,3.36; this sample is strongly biased towards luminous objects at high redshift. 
The FIRST Bright Quasar Survey (FBQS, \citealp[]{brotherton2001composite}) took advantage of FIRST radio observations of known point-like sources with colour indices\footnote{More details about the POSS-I photometric system can be found in \citet{evans1989photometric}.} $O-E<2.0$ and $E<17.8$ from the Palomar Observatory Sky Survey I (POSS-I). This match yielded 636 sources, here we show only the composite spectrum made of the radio-quiet sources as a comparison.

The first SDSS quasar template was produced by VB01, who selected 2204 sources between $0.004\leq z \leq 4.789$ based on the SDSS colour difference \citep{richards01}, a non-stellar continuum, and at least one broad (FWHM\,$>$\,500 km s$^{-1}$) permitted emission line.
The sample adopted by \citet{harris2016composite} was instead based on the Baryon Oscillation Spectroscopic Survey (BOSS; \citealt{dawson2012baryon}), as part of SDSS-III (\citealt{eisenstein2011sdss}), aimed at characterizing the effects of the baryon acoustic oscillations (BAO) on the distribution of luminous red galaxies and studying the Ly$\alpha$ forest in high-redshift quasars. These purposes resulted in a sample of 102,150 objects between 2.1\,$\leq$\,$z$\,$\leq$\,3.5 (see \citealt{harris2016composite} for more details about the sample selection). Given the redshift range covered by such sources, the comparison with this composite spectrum was possible only in the UV up to $\sim$3000\,\AA. 
Lastly, \citet{selsing2016x} observed with XSHOOTER/{\it VLT} a sample of 7 bright ($r\lesssim 17$) quasars between 1.0\,$<$\,$z$\,$<$\,2.1, for which galactic contribution should be regarded as negligible.

\begin{figure}[h!]
\centering
\includegraphics[scale=0.22]{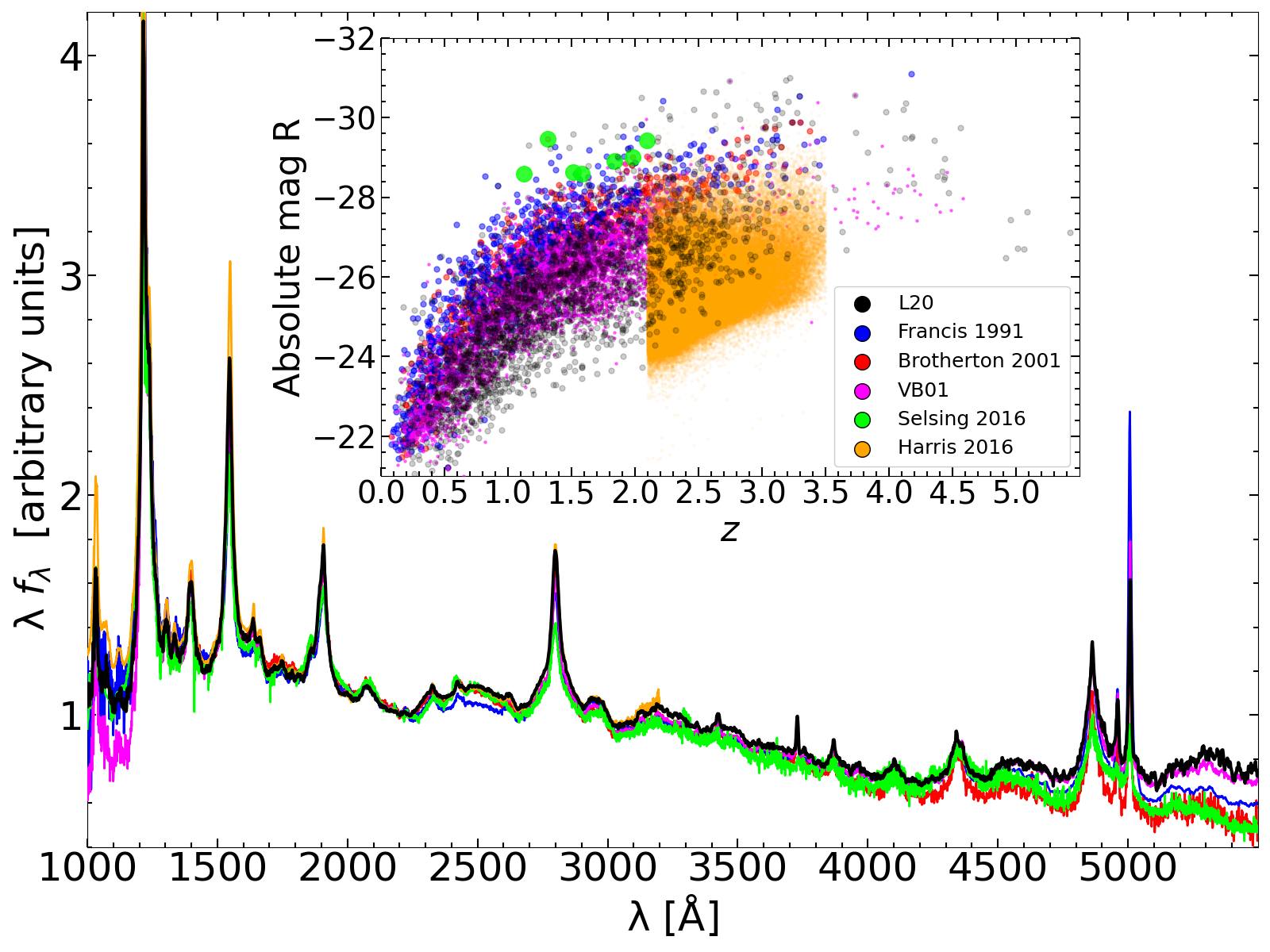}
\caption{Comparison between our quasar template and others from the literature. All the templates are scaled by their emission at 2200\,\AA. The agreement on the UV side, apart from minor differences in the \ion{Fe}{ii} pseudo-continuum, specifically the 2250-2650\,\AA\, and the 3090-3400\,\AA\, bumps, is apparent. The Ly$\alpha$ emission line is cut for visualization purposes. The inset shows the distribution of the absolute $R$ magnitude against the redshift for the different samples reported.}
\label{fig:tpl_comparison}
\end{figure}

The comparison between our template and the literature ones is shown in Fig.\,\ref{fig:tpl_comparison}. Note that, the $y$-axis represents $\lambda f_{\lambda}$, rather than the usual flux density. The agreement is generally good, especially in the region  between 1100--4000\,\AA. At shorter wavelengths, the intergalactic medium (IGM) can affect the composites differently, according to the distribution in redshift of the individual spectra upon which the stacks are based (e.g. \citealt{madau1995radiative, faucher08, prochaska2009direct}). On the red side, instead, some intrinsic differences start to become visible. Our composite is almost indistinguishable from the VB01 composite, but these two seem flatter than the others.
There are several explanations for this slight difference in the long-wavelength end of the composites. First, the colour selections performed to select the quasar candidates is not homogeneous among the different samples, thus the SDSS objects that constitute the VB01 sample as well as ours could present slightly redder indices in the optical range below $z=0.8$. Secondly, it is possible that the host galaxy, whose contribution is negligible in the UV (see Sec. \ref{sec:hgc}) plays a minor role in flattening the optical side of the spectra. VB01 estimated the galaxy contribution to be roughly 7--15\% at the \ion{Ca}{ii}\,$\lambda3934$ wavelengths, and a similar fraction can be estimated in our global composite. On the other hand, the other surveys and the \citet{selsing2016x} sample are made, on average, of more luminous objects for the same redshift (see inset in Fig. \ref{fig:tpl_comparison}). It is therefore likely that the galactic contribution in those templates is even lower than in ours.

Throughout this work, we will refer to the full sample template as a reference to compare to the composite spectra of the individual bins in the $M_{\rm BH}$\,--\,$L_{\rm bol}$\,--\,$z$ space, unless differently stated.

% PLOT TEMPLATES

\subsection{Non-evolution with redshift}
A key observational feature bridging local and high-redshift quasars is the apparent similarity of their optical--UV continuum up to primordial times (e.g. \citealt{banados2018, yang2020poniua,wang2021luminous}). However, if the quasar spectra showed even a subtle evolution with redshift, the accretion physics underlying the X-ray to UV relation could also be evolving, making quasars unsuitable candidates as standard candles. Given the wide range in terms of redshift covered by our sample, we expect to be capable of detecting any such possible effect. This would appear as a systematic difference in the slope of the continuum produced by the accretion process at different cosmic times.
%, once we consider, as we show later on in the text, that both the effects of $M_{\rm BH}$ and $L_{\rm bol}$ do not seem to drive significant differences in the continuum of our composite spectra.}

\begin{figure*}[h!]
\centering
\includegraphics[width=\linewidth,clip]{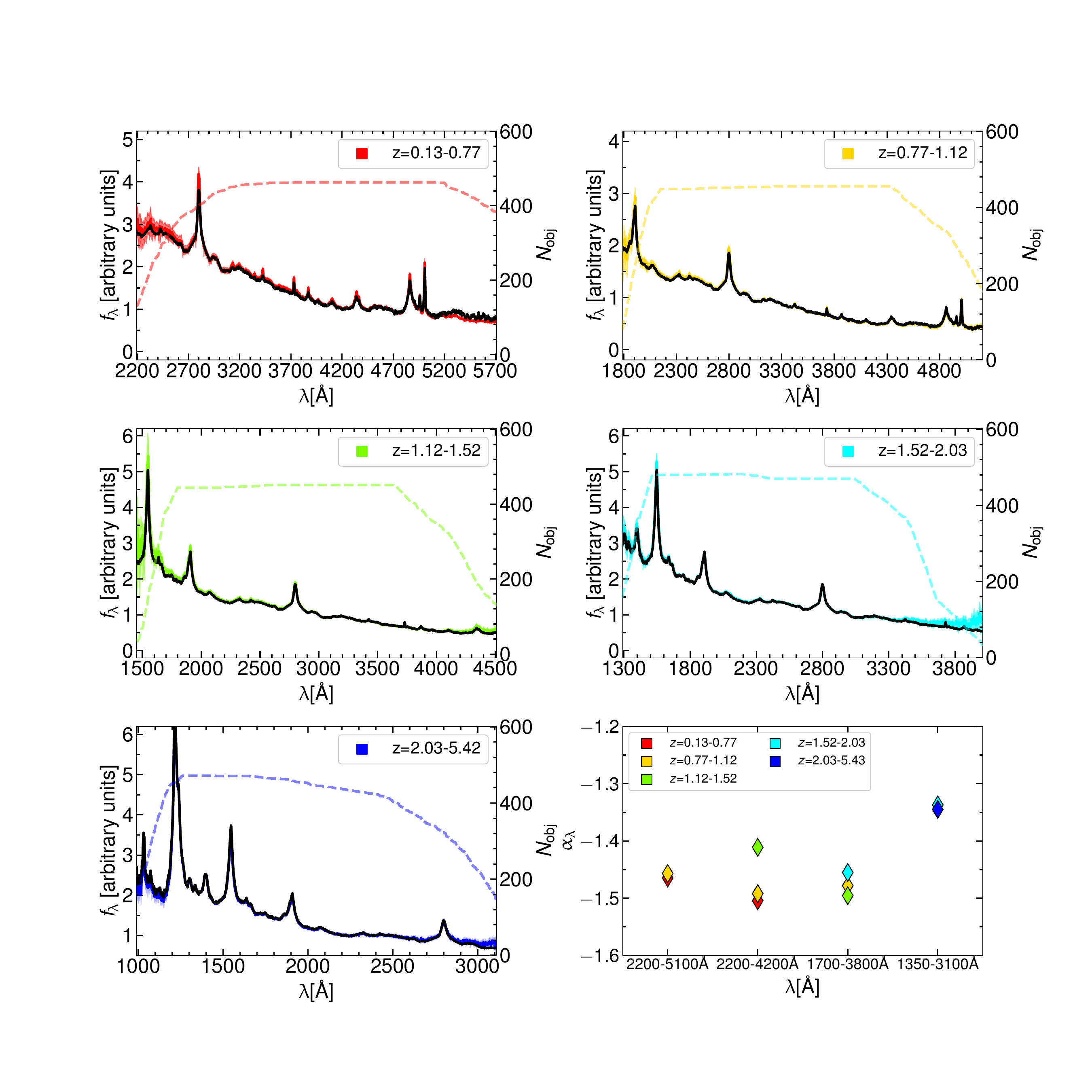}
\caption{Composite spectra divided in redshift bins superimposed to the full sample composite. The dashed line denotes the number of objects contributing to the stack in each spectral channel. The composite spectra in each bin are scaled to match the template respectively at 4200\,\AA\, for the low redshift bin, 3000\,\AA\, for the $z=[0.77-1.12)$, $z=[1.12-1.52)$, $z=[1.52-2.03)$ bins, and 2200\,\AA\, for the highest redshift bin. In the bottom right panel we show the slopes of the continuum evaluated in different pairs of continuum windows, which are generally consistent.}
\label{fig:zbin_stacks}
\end{figure*}

As a very basic step to check for any evolution with redshift of the spectral properties within our sample, in Fig. \ref{fig:zbin_stacks} we compare the average spectra in the redshift bins with the global composite (described in Sec. \ref{sec:comparison}), obtained by stacking the spectra of all the sources in the sample. We find a very good agreement with the global composite at all redshifts, without any clear evidence of evolution, apart from the known trends of the emission lines, whose FWHM varies with $M_{\rm BH}$ (i.e. the virial relation; e.g. \citealt{peterson04}) and whose EW anti-correlates with $L_{\rm bol}$ (i.e. the Baldwin effect; \citealt{baldwin1977}). 

The visual agreement is confirmed by a more quantitative assessment presented in the bottom right panel in Fig.\,\ref{fig:zbin_stacks}. There we show the slope of the continuum evaluated in continuum intervals sampled at different redshifts by the spectral composites. The slopes are generally consistent within $1\sigma$, and no systematic trends are observed. The optimal agreement between the composite spectrum in every redshift bin and the global composite, as well as among the different redshift composites, demonstrates the non-evolution of the average continuum properties of the sample across cosmic time.

In addition, as discussed in Section \ref{sec:comparison}, the global sample composite exhibits a remarkable agreement with the average SDSS spectrum described in VB01, meaning that the stacks in the different redshift bins are also similar to the spectrum of a typical blue quasar. Since even this direct comparison reveals that the average continuum properties of our sample do not suggest any redshift evolution, we now investigate and characterise in more detail how our sample behaves across the $\log M_{\rm BH}$\,--\,$\log L_{\rm bol}$ parameter space.

% PLOT STACK

% PLOT UV SLOPE
\begin{figure*}[h!]
\centering
\includegraphics[width=\linewidth,clip]{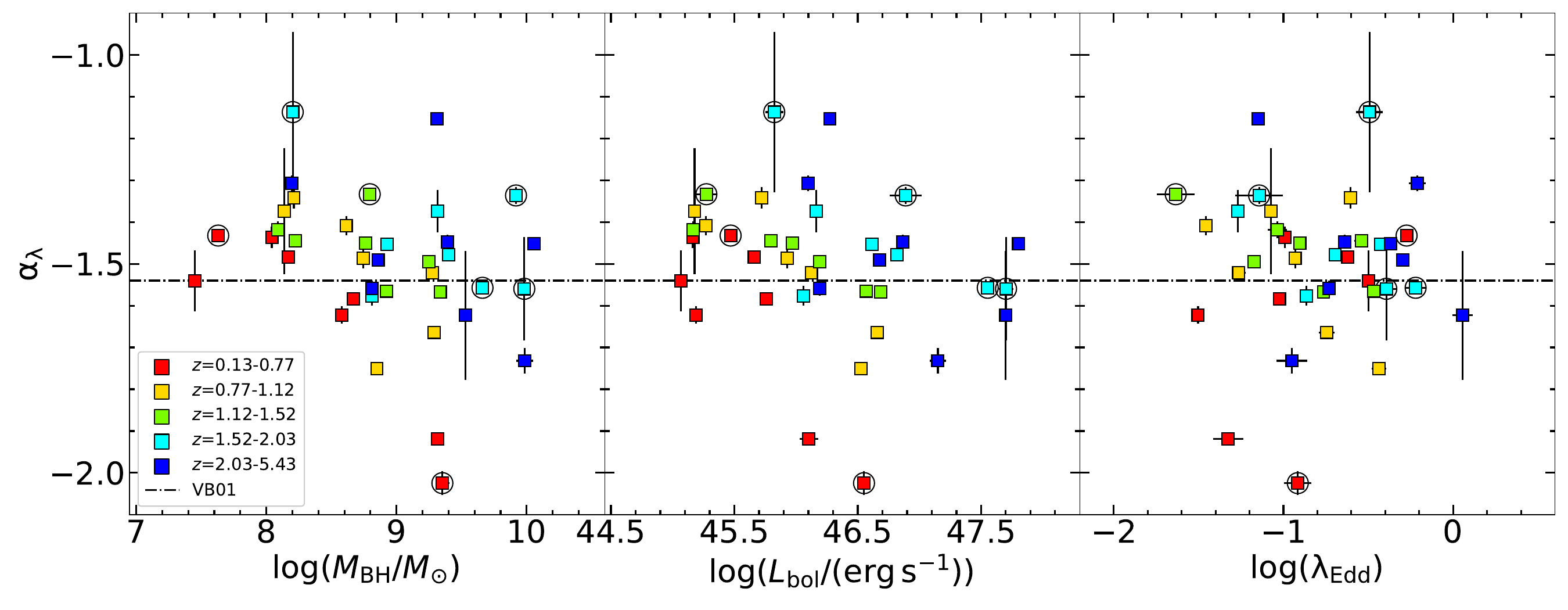}
\caption{Values of the slope of the continuum produced by our fits. Bins not meeting the representativeness criterion are marked by a circle. The slope of the VB01 template (dot-dashed line) is shown as an external reference. Different pairs of line-free windows were adopted to define the continuum in each bin.}
\label{fig:slope_pars}
\end{figure*}

\subsection{Analysis of the evolution of the continuum slope of the stacks}
\label{sec:cont_slope}

The fits of the stacked spectra returned the slope of the continuum in each bin of the parameter space. We thus explored whether any possible evolution of the continuum slope exists throughout our sample. With the term slope ($\alpha_{\lambda}$), we denote the slope of the continuum power law underlying the emission lines roughly between 1300--5500\,\AA. This wavelength range is sampled differently in the different redshift bins: for instance in the lowest redshift bin only the interval 2200--5500\,\AA\ is available, while in the highest redshift bin the composite spectra were fitted over the interval 1300--3200\,\AA. We also underline that, throughout this work, we never assign a physical meaning to the slope of the continuum, adopting it exclusively as a comparative parameter. The physical meaning of such a parameter would require the assumption of a physical continuum model, whose implications go beyond the scope of this work.

In Fig. \ref{fig:slope_pars}, we show the values of the slope in the redshift bins as a function of $\log M_{\rm BH}$, $\log L_{\rm bol}$, and $\log\lambda_{\rm Edd}$. 
To avoid dust reddening or galaxy contamination, a selection in terms of photometric colours has been performed when constructing the cosmological sample (see L20, Section 5.1). In brief, for each object, the slope of the power law in the $\log\nu$\,--\,$\log(\nu f_{\nu})$ space was computed in the rest-frame ranges 3000--10000\,\AA\  ($\Gamma_1$) and 1450--3000\,\AA\ ($\Gamma_2$). Objects whose photometric colours were compatible with those of typical blue quasars, $\Gamma_1$\,=\,0.82 and $\Gamma_2$\,=\,0.40 \citep[][]{richards2006spectral}, only tolerating mild reddening, i.e. $E(B-V)$\,<\,0.1, were selected. Converting the photometric indices $\Gamma$ in terms of spectral slopes as $\alpha_{\lambda}=-\Gamma - 1$, the expected slopes in the 3000--10000\,\AA\ and the 1450--3000\,\AA\ ranges, respectively covered by the low and high redshift bins, are $\alpha_{\lambda,1}=-1.82$ and $\alpha_{\lambda,2}=-1.40$. Although this process operates a sort of pre-selection on the possible continuum slopes, we note that only roughly 20\% of the starting sample is excluded according to this colour selection.

In Fig. \ref{fig:slope_pars} we show the slopes of the continuum evaluated in the bins of the parameter space. We mark with a hollow circle the values of the bins with a low $N_{\rm obj}$ which would be excluded as a result of the representativeness criterion described in Sec.\,\ref{sec:paramspace}. The values cluster around $\alpha_{\lambda}=-1.50$ with a standard deviation of $\sigma_{\alpha}=0.14$, showing little or no evolution from low to high redshift. Considering also the excluded bins, the results are not significantly altered as the mean value negligibly shifts to $\alpha_{\lambda}=-1.51$ and the standard deviation to $\sigma_{\alpha}=0.17$. The slope average value is close to those expected from the \citet{richards2006spectral} typical photometric index and from the templates of blue quasars described in Sec. \ref{sec:comparison}, whose continuum slopes, evaluated using our fitting routine, are reported in the inset panel of Fig. \ref{fig:hd_sl_distr}. We recall that the slopes are evaluated using different continuum windows at different redshifts. For a fixed value of $M_{\rm BH}$ (or equivalently $L_{\rm bol}$), the distribution in the observed slopes is therefore partly due to the spread along the other axis of the parameter space, which is anyway modest, and partly driven by the different continuum windows.

We assessed the degree of correlation between the slope and the axes of the parameter space by means of the Spearman rank correlation index $S$ (correlation increases in modulus from 0 to 1). We do not find any significant trend with any of the accretion parameters, (i.e. $|S|\leq0.3$ and associated $p-{\rm values} \, >0.05$). The inclusion of the bins excluded as a result of the representativeness criterion, does not change significantly the correlation index, as we show in Appendix \ref{app:correlations}.

\begin{figure}[h!]
\centering
\includegraphics[width=\linewidth,clip]{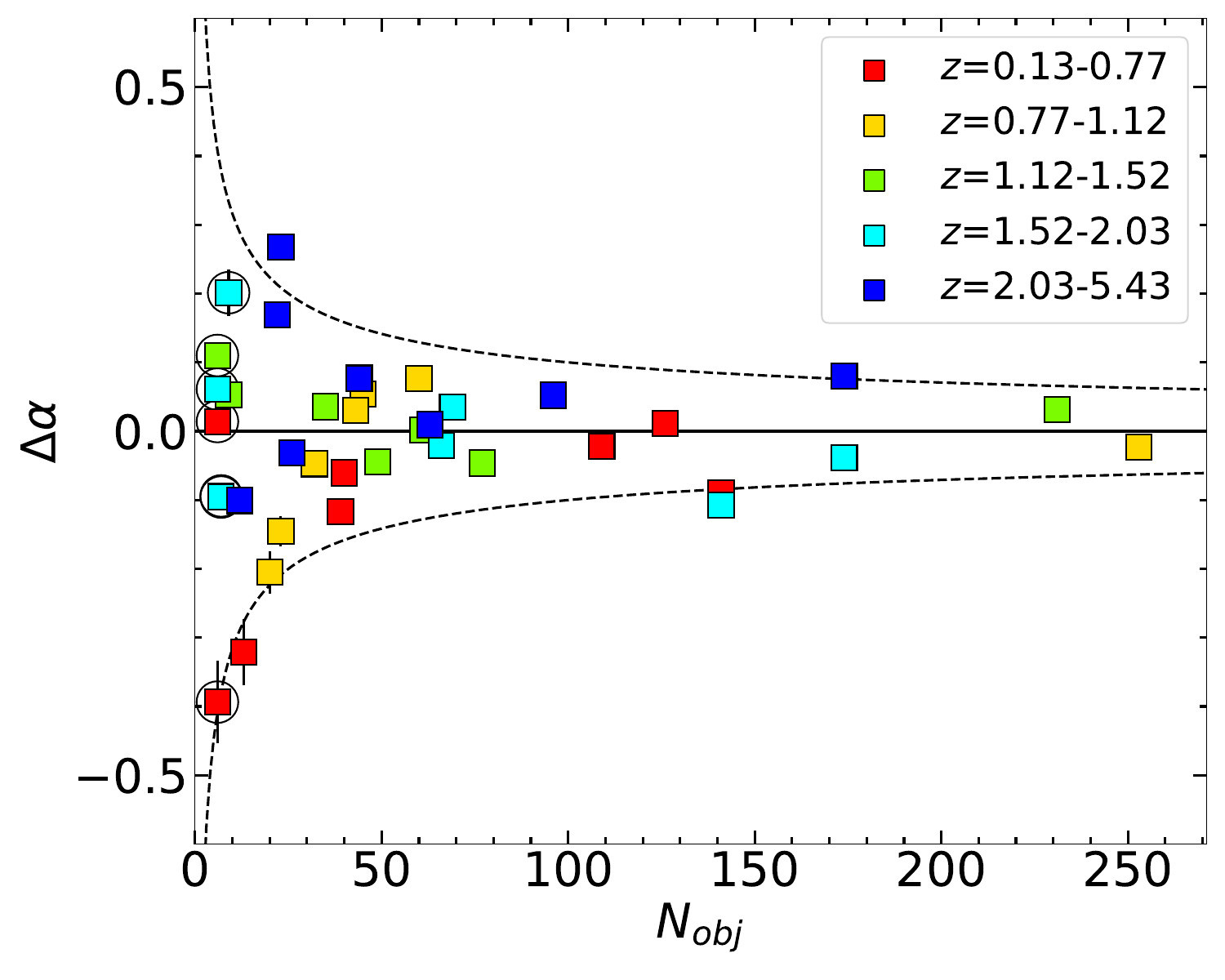}
\caption{Values of $\Delta \alpha$ ($\Delta \alpha=1-\alpha_{\rm bin}/\alpha_{FS}$) in the bins of our parameter space as a function of the objects in each bin. Bins not meeting the representativeness criterion are marked by a circle. The dashed lines defines the $\sim1/\sqrt{N_{\rm obj}}$ envelope characterizing the standard error on the average slope assuming a standard deviation $\sigma_{\alpha}=1$.}
\label{fig:delta_sl}
\end{figure}

In addition to exploring the evolution of the continuum within our parameter space, we also explored how the number of sources in each bin affects the continuum properties with respect to the global template. In particular, in Fig. \ref{fig:delta_sl} we present the relative difference between the continuum slope of the continuum for each bin ($\alpha_{\rm bin}$) and the full sample composite ($\alpha_{\rm FS}$) evaluated in the same wavelength range $\Delta \alpha=1-\alpha_{\rm bin}/\alpha_{FS}$ against the number of objects in the bin. Regardless of the redshift, the difference decreases in modulus for increasingly populated bins, as if the samples were drawn from a single population. 
In this picture, the spread in the observed slopes results from sample variance, rather than from an actual evolution with the accretion parameters explored.

% PLOT UV SLOPE

% PLOT EXAMPLE
\begin{figure*}[h!]
\centering
\includegraphics[width=\linewidth,clip]{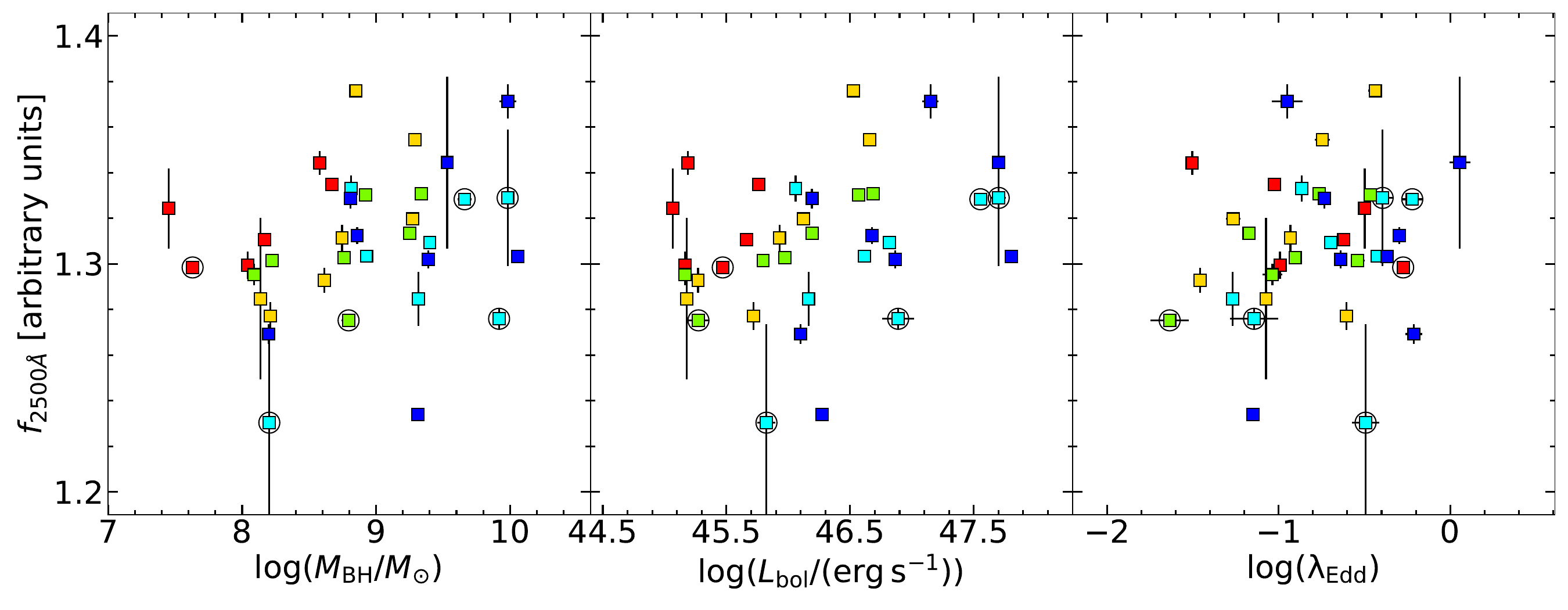}
\caption{Values of the normalised 2500\,\AA\ fluxes along the axes of the parameter space, assuming a common normalisation of the continuum at 3000\,\AA\, in all the bins. Bins not meeting the representativeness criterion are marked by a circle.}
\label{fig:f2500_pars}
\end{figure*}

\subsection{Comparison of the 2500\,\AA\, monochromatic fluxes}

We produced the normalised monochromatic 2500\,\AA\, fluxes ($f$\textsubscript{2500\,\AA}) by interpolating the continuum resulting from the best fit slope and assuming a unitary monochromatic flux at 3000\,\AA. 
In Fig. \ref{fig:f2500_pars} we present $f$\textsubscript{2500\,\AA} against $\log M_{\rm BH}$, $\log L_{\rm bol}$, and $\log\lambda_{\rm Edd}$ in order to study any possible evolution. 
In principle, a systematic change of the $f$\textsubscript{2500\,\AA} with the accretion parameters would require a correction to \textit{standardise} the UV proxy. However there is evidence against this possibility.

In \citet{signorini2023quasars} we showed that the monochromatic 2500\,\AA\, flux is a good UV proxy for the \lxluv relation, although other wavelengths in the range 1350--5100\,\AA\, can perform equally well (see also \citealt{jin2024wavelength}). Therefore, even assuming different slopes, the UV term of the \lxluv relation would hardly be affected.

Here we aimed at quantifying the effect on the monochromatic 2500\,\AA\, fluxes of the differences in the spectroscopic slopes within our sample. The slopes are comprised between roughly [$-$2.0, $-$1.1], with the extremal values found in scarcely populated bins (respectively 9 and 6 objects), which would be excluded as a result of the procedure described in Sec. \ref{sec:paramspace}. Since our goal is to corroborate our sample for its cosmological application, we can estimate how much the spread in the observed slopes affects, for instance, the distance modulus (DM) estimates. Here we show that, even assuming a systematic evolution in the continuum slope with one of the explored parameters (an occurrence which is not supported by our analysis), the variation induced in the $f$\textsubscript{2500\AA} would propagate into a negligible difference in the DM for our scopes. We can easily check this by expressing DM in terms of $f$\textsubscript{2500\AA} adopting the expression (see \citealt{risaliti2015hubble, risaliti2019cosmological, lusso2020}):

\begin{equation}
\begin{split}
    {\rm DM} & = 5 \left[\frac{\log f_{\rm X} - \beta -\gamma\,(\log f_{2500\,\AA}+27.5)}{2(\gamma -1)} - \frac{1}{2} \log (4\pi) +28.5 \right] \\
    & - 5 \, \log (10 \, {\rm pc}),
\end{split}
\label{eq:dm}
\end{equation}
where $f_{\rm X}$ and $f_{2500\,\AA}$ are, respectively, the X-ray and UV flux densities in units of erg s$^{-1}$ cm$^{-2}$ Hz$^{-1}$. The parameter $\log f_{2500\,\AA}$ is normalised to the value of 27.5, whereas the logarithm of the  luminosity distance is normalised to 28.5. The $\gamma$\,=\,0.600\,$\pm$\,0.015 term is the slope of the \lxluv relation reported in L20, and $\beta$ is the intercept.
Here we can express the 2500\,\AA\, monochromatic flux in a power-law form pivoting at 5100\,\AA\, (i.e. the farthest among the explored characteristic wavelengths). By differentiating Eq.\,\ref{eq:dm} with respect to the slope of the continuum $\alpha_{\lambda}$, we get
\begin{equation}
    \Delta DM = \left | -\frac{5\gamma}{2(\gamma - 1)} \log \left(\frac{2500}{5100} \right) \right |  \Delta \alpha_{\lambda}.
\end{equation}
We can then substitute the standard deviation of the observed slopes ($\sigma_{\alpha}$=0.17) as a measure of the spread of the $\Delta \alpha_{\lambda}$ term. The DM variation $\Delta$DM is $\sim$0.2 in modulus, is not sufficient to explain the differences observed with the $\Lambda$CDM at high redshifts (see \ref{fig:hd_reddening}). If, as we believe, 2500\,\AA\, is a better approximation to the UV pivot point than 5100\,\AA\,, the average $\Delta$ DM is expected to be even smaller.

\subsection{Host-galaxy contamination}
\label{sec:hgc}
Although the broadband properties of the objects included in the sample are designed to minimise the additional contribution of the host galaxy in the quasar spectrum, here we perform an a posteriori test based on the spectral data to estimate the average degree of host-galaxy emission in our sample. In \cite{rakshit2020spectral} a host-galaxy--quasar decomposition was performed using a principal component analysis (PCA, \citealt{yip2004spectral}) in two windows close to 4200\,\AA\ and 5100\,\AA, obtaining reliable results for $\sim$12,700 objects below redshift $z$\,$\leq$\,0.8. We employed these measurements to assess the level of residual (i.e. not filtered out by our criteria) host-galaxy contamination within the quasars of our sample. With this aim, we built a purposefully designed sample from the latest WS22 catalogue by adopting the following criteria:
\begin{itemize}
    \item We selected all the objects below $z$\,$\leq$\,0.8.
    \item We considered only the objects for which the host-galaxy fraction indices at 4200\,\AA\ (HOST\_FR\_4200) and 5100\,\AA\ (HOST\_FR\_5100) were both evaluated, with values ranging continuously from 0 (quasar-dominated emission) to 1 (galaxy-dominated emission). If any of the two indices was flagged as not measured (i.e. $=-999$), the object was discarded.
    \item We defined the host galaxy fraction index (HGF) and created 10 bins of width $\Delta$HGF\,=\,0.1. A source was a given HGF index when both the 4200\,\AA\ and the 5100\,\AA\ fractions fell within the same HGF bin in order to define an unambiguous index. The sample so constructed has 5748 objects. 
    
\end{itemize}

\begin{figure}[h!]
\centering
\includegraphics[width=\linewidth,clip]{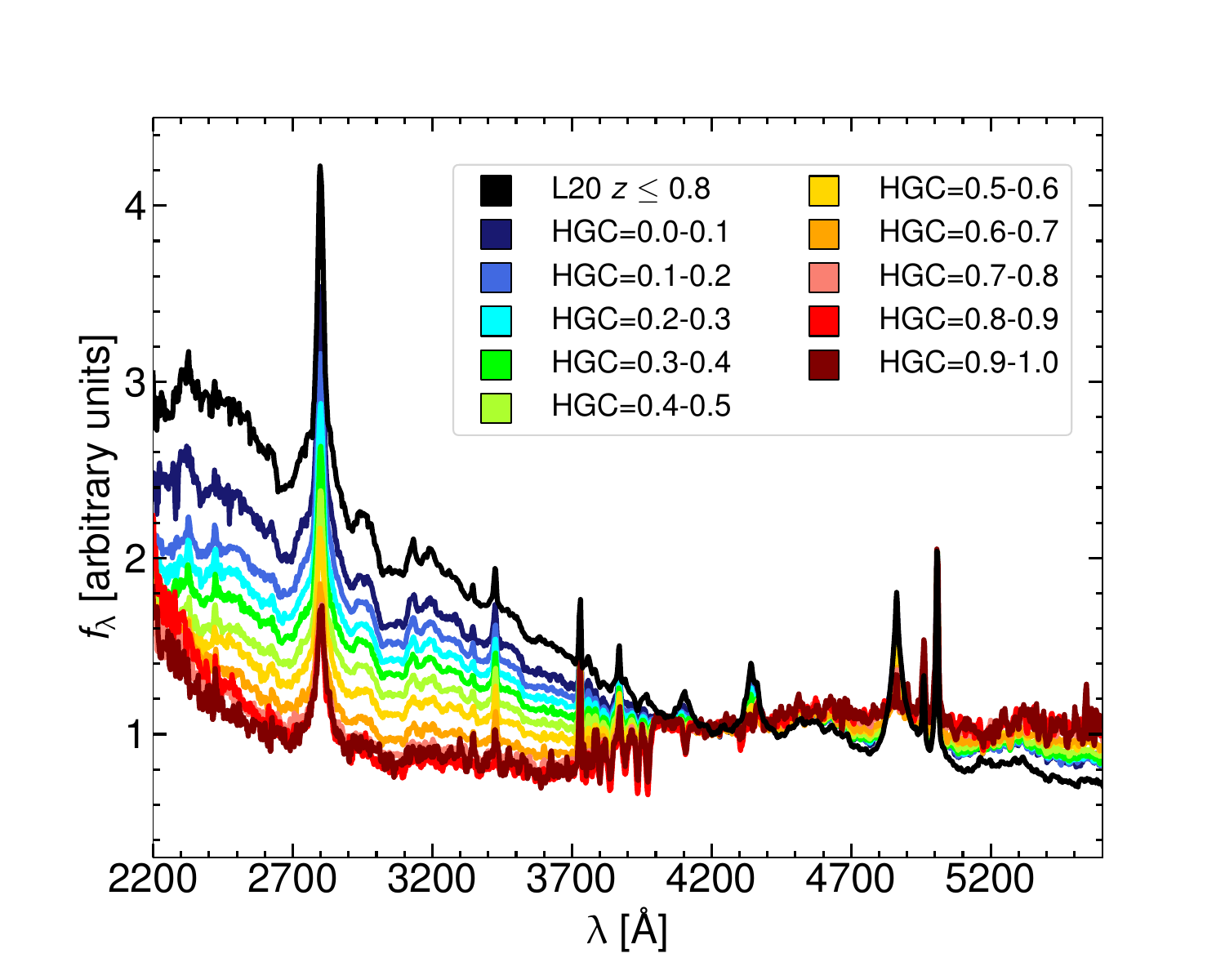}
\caption{Comparison between the stacks in bins of host-galaxy fraction (HGF) and our reference spectrum (black) for the $z$\,$\leq$\,0.8 interval, as colour-coded in the legend. All the spectra are scaled to their 4200\,\AA\ flux. Increasingly host-dominated composites exhibit flatter UV continua.}
\label{fig:hgc_stack}
\end{figure}

We then stacked the spectra of the objects in each bin according to the same procedure described in Sec.~\ref{sec:stacking}. In Fig. \ref{fig:hgc_stack} we show the stacked spectra as a function of their HGF index. To perform a fair comparison, we also produced a stack of all the 529 sources in our sample with $z$\,$\leq$\,0.8 (shown in black). On average, the quasars in the low redshift tail of our sample retain minimal levels ($\lesssim$\,10\%) of host-galaxy contamination below 5400\,\AA, in good agreement with VB01 who report a galaxy flux fraction between 7--15\% on the basis of the \ion{Ca}{ii}\,$\lambda$3934 absorption feature. At redder wavelengths, the host contribution is expected to become increasingly important. In addition, we also note that the difference is terms of continuum slope and line strength is more evident in the UV region of the spectra, where our template exhibits the best match with the other samples of blue quasars. This test confirms that our UV proxy for the \lxluv relation is not systematically affected by host galaxy contamination.

% PLOT HGC STACKS

\subsection{Intrinsic extinction}
\label{sec:extinction}
Our photometric selection is designed to exclude dust reddened quasars, by retaining only sources near the locus in the $\Gamma_1$\,--\,$\Gamma_2$ plane where blue unobscured quasars are expected to reside (\citealt{richards2006spectral}). We again checked, a posteriori, that the spectra composing our sample only accept minimal levels of extinction, and we bring three arguments in favour of this hypothesis. 

First, as a general consideration, we find a very good agreement between our stacks and the VB01 SDSS template, which is made of sources whose colour-selection had been designed to select blue quasars \citep{richards01}. The agreement is excellent -- especially on the bluer UV side -- also with the other average spectra shown in Sec.~\ref{sec:comparison}. This implies that, if the quasars used to build such templates are not redder than the average, the same should hold for our sample. We also recall that, by selection, the VB01 template was built using objects with at least one broad emission line, thus the presence of Type 2 obscured objects should be limited.

%PLOT EXT STACKS
\begin{figure}[h!]
\centering
\includegraphics[width=\linewidth,clip]{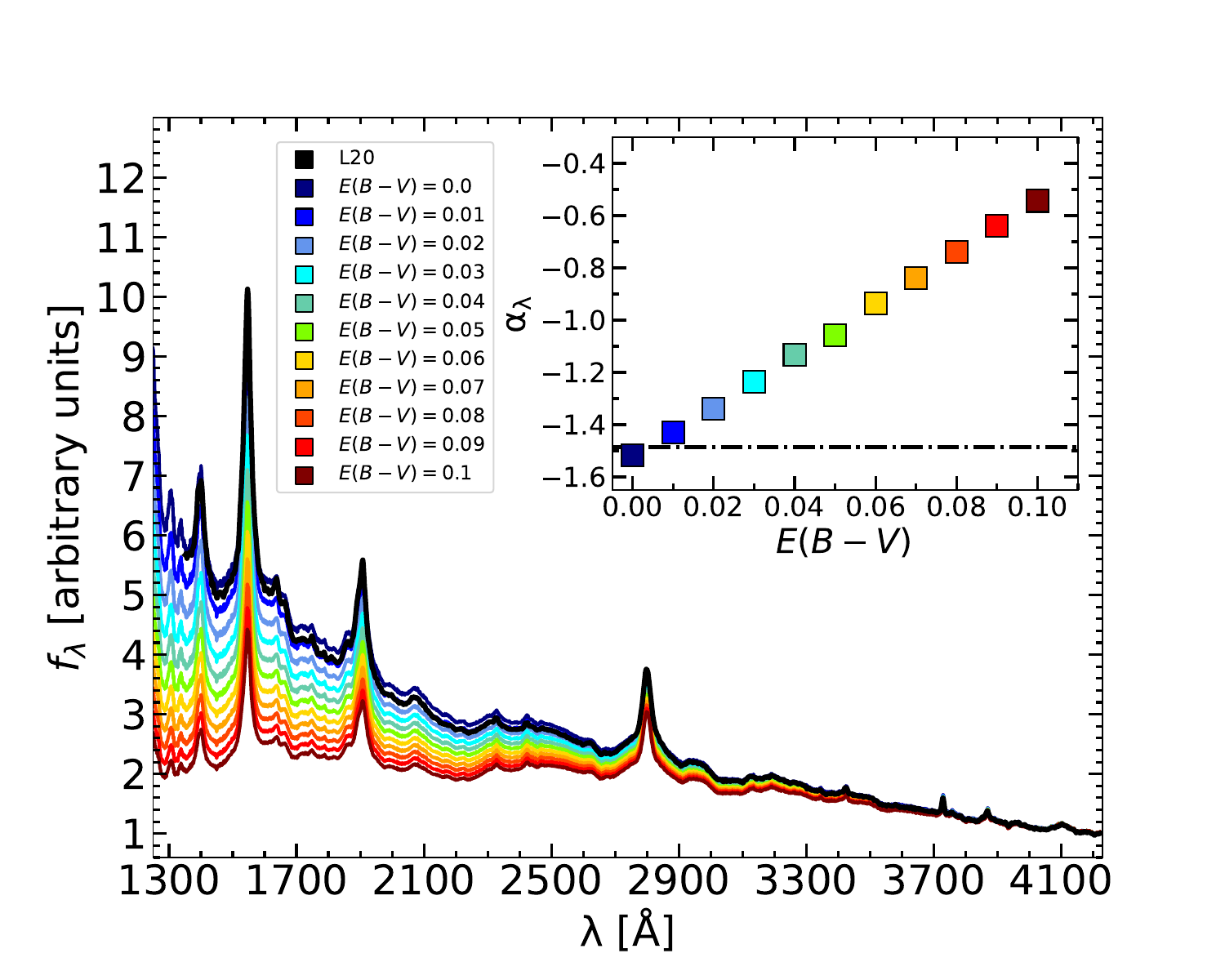}
    \caption[flux]{Extinction of the VB01 template 
    according to the colour coded values. All the spectra are scaled by their 4200-\AA\ flux, where  extinction becomes less relevant. The black spectrum is our full sample stack, in very good agreement with a low degree of extinction.
    The inset shows the best-fit slope in the range 1300--4100\,\AA\, of the VB01 template reddened as a function of the $E(B-V)$ value. The black dot-dashed line represents the slope of the continuum fitted to our stack. Minimal levels of extinction are allowed to match the reddened template with our composite. We note that the visual match between the L20 and the VB01 reddened template depends slightly on the scaling wavelength. The slope derived from the fits, however, does not depend on such a wavelength.}
\label{fig:ext_vdb}
\end{figure}
% PLOT Ha/Hb
\begin{figure}[h!]
\centering
\includegraphics[width=\linewidth,clip]{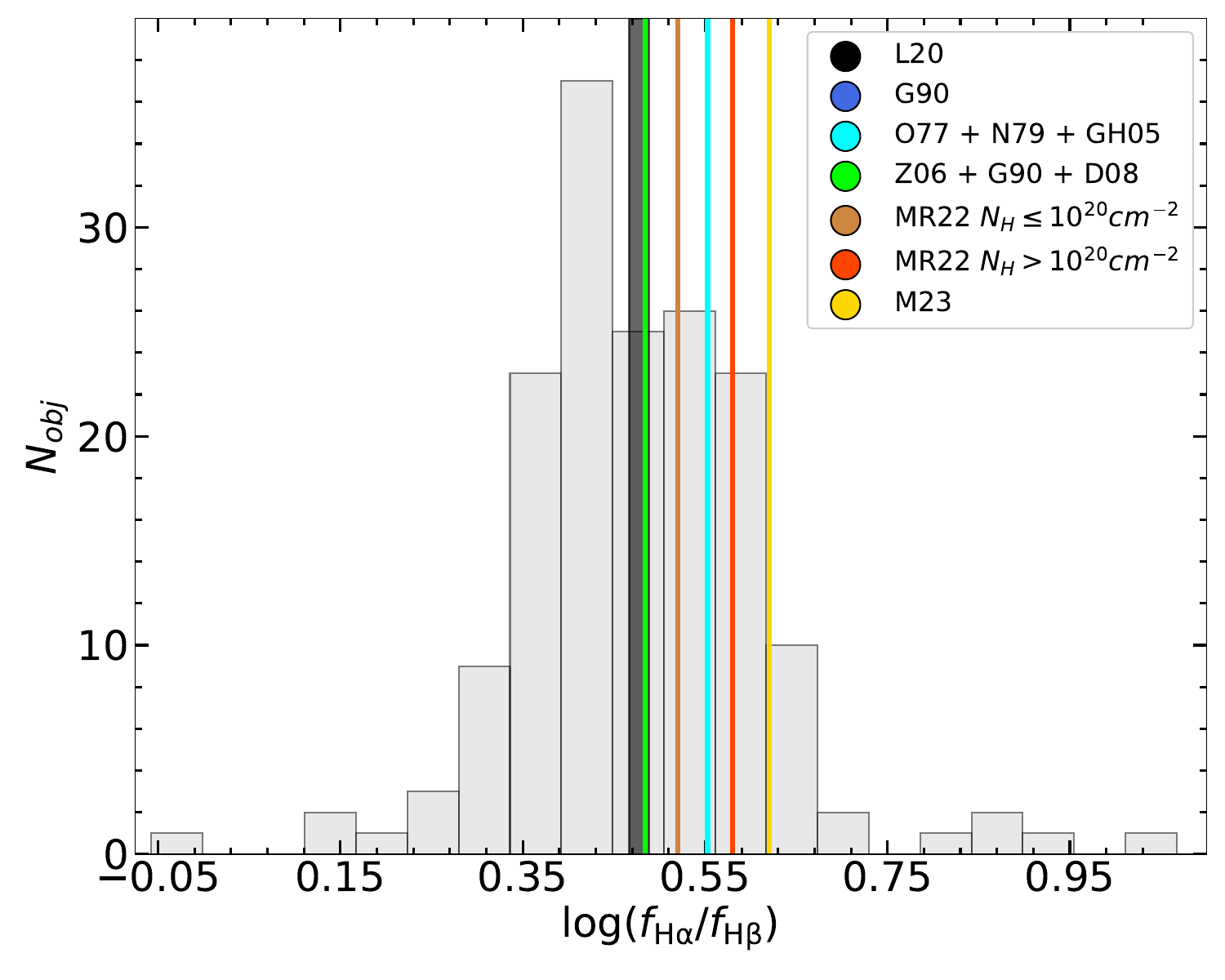}
\caption{Distribution of $\log$(H$\alpha$\textsubscript{br}/H$\beta$\textsubscript{br}) for our sources with available data. The mean values from other literature samples (\citealp[O77]{osterbrock1977spectrophotometry}, \citealp[N79]{neugebauer1979absolute}, \citealp[G90]{goodrich1990pa}, \citealp[GH05]{greene2005estimating}, \citealp[Z06]{zhou2006comprehensive}, \citealp[D07]{dong2007broad}, \citealp[MR22]{mejia2022bass}, \citealp[M23]{ma2023variability}) are also shown for reference according to the colour code. The mean of our values is shown in black. The BASS sample described in \citet{mejia2022bass} has been split according to the column density $N_{\rm H}$ inferred from the X-ray analysis.}
\label{fig:ha_hb_distr}
\end{figure}

As a second test, to get a more quantitative assessment of the the effect on the spectral slope of increasing reddening, we reddened the VB01 template under the reasonable hypothesis that this is representative of the intrinsic spectrum of a typical blue quasar. We reddened the template with increasing $E(B-V)$ from 0 to 0.1 with steps of 0.01, using the extinction curve described in \cite{gordon2016panchromatic} for an average Small Magellanic Cloud (SMC) extinction curve, which according to several authors is a good approximation of the AGN extinction curve (see e.g. \citealt{brotherton2001composite, hopkins2004dust}), and then checked the agreement with our full sample stack. A more detailed discussion on how a different extinction curve can affect this procedure is presented in Appendix \ref{app:extinction_curves}, where other prescriptions are explored. In brief, we found that the variation of the continuum slope for a given colour excess derived adopting the \citet{gordon2016panchromatic}, used to account for intrinsic reddening throughout this work, is the closest to the average (see Fig. \ref{fig:ext_test}). The result of the reddening procedure is shown in Fig. \ref{fig:ext_vdb}. The best match is obtained for minimal levels of intrinsic reddening, i.e. $E(B-V)$\,$\lesssim$\,0.01, an estimate similar to that provided by \citet{reichard2003continuum} on the basis of spectral data from the SDSS early data release. To quantify this effect, we performed a spectral fit of the reddened templates in the region between 2200--3400\,\AA, and compared the slope of the continuum with that obtained by the fit of our stack in the same region. This test confirmed the finding emerging from a crude comparison of the overall spectral shape, that is a continuum slope consistent with a low degree of intrinsic extinction. The slopes resulting from the spectral fit of the reddened templates are shown in the inset of Fig. \ref{fig:ext_vdb}. We note that, choosing a different template than VB01 would produce analogous results, given the similarity of all the explored templates in the near UV.

Finally, we estimated the Balmer decrement of the low-redshift sources with available broad H$\alpha$ and H$\beta$ emission lines to assess possible levels of intrinsic extinction of the broad line region. \cite{dong2007broad} computed the distribution of the broad-line Balmer decrement H$\alpha$\textsubscript{br}/H$\beta$\textsubscript{br}\,$\equiv$\,$f$(H$\alpha$\textsubscript{br})/$f$(H$\beta$\textsubscript{br}) for a sample of 446 low-redshift ($z$\,$\lesssim$\,0.35) blue AGN, finding that the distribution can be well approximated by a log-normal, with a mean value of $\langle$log(H$\alpha$\textsubscript{br}/H$\beta$\textsubscript{br})$\rangle$\,=\,0.486 and a standard deviation of 0.03. A value of log(3.1)=0.49 is generally adopted for the Balmer decrement in the narrow line region in AGN \citep{halpern1983low,gaskell1984theoretical}, while it is found to vary widely according to different BLR conditions (e.g., \citealp{dong2007broad} and references therein).

% PLOT HD REDDENING
\begin{figure*}[h!]
\centering
\includegraphics[width=\linewidth,clip]{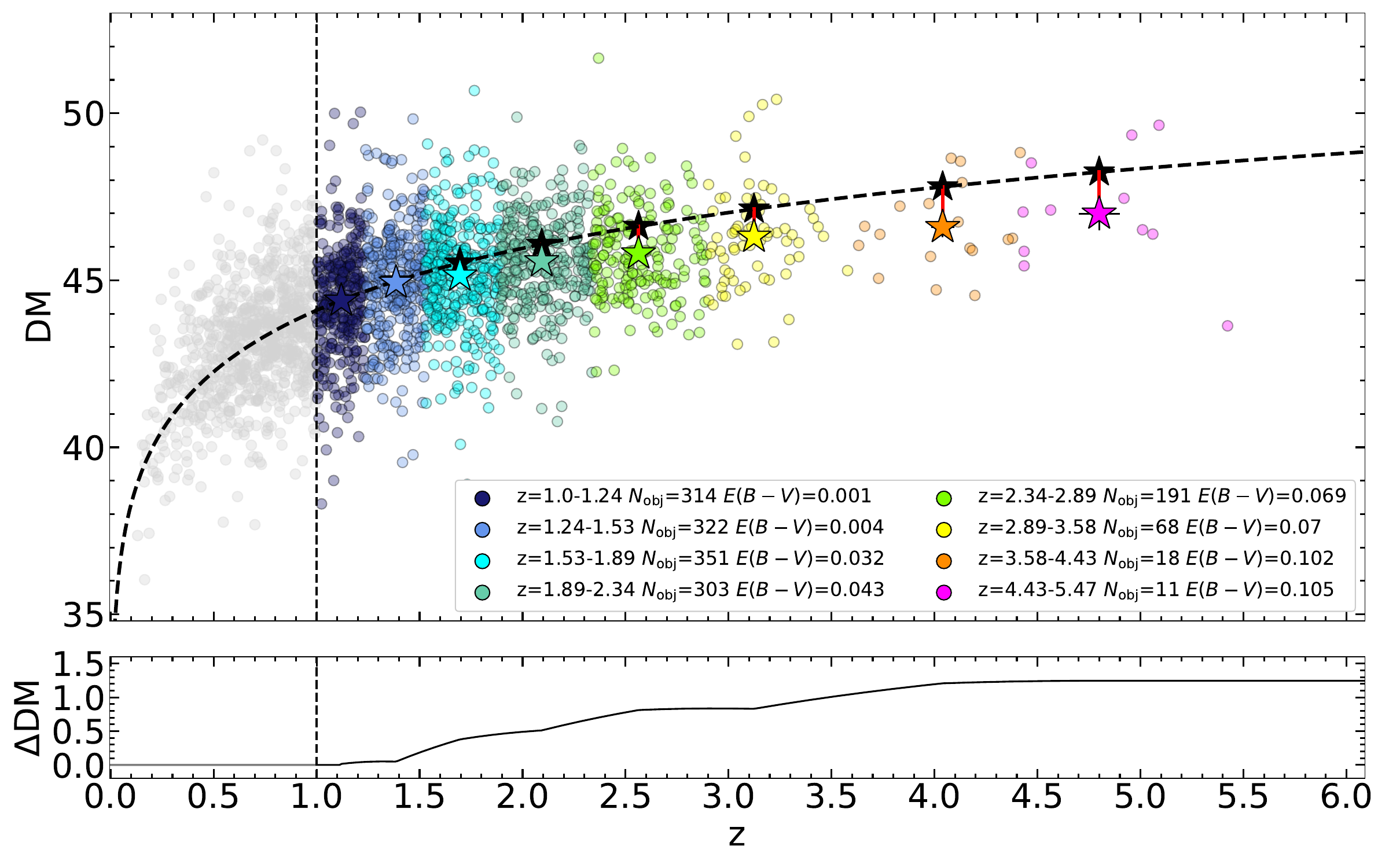}
\caption{Hubble diagram of the sources used to investigate the possible effects of local reddening (\textit{top panel}). The redshift bins used for this purpose are colour-coded in the legend, together with the number of objects per bin and the corresponding $E(B-V)$. Coloured stars represent the average values in each redshift bin, while black stars mark the $\mathrm{\Lambda}$CDM model expectations, with red segments joining these two values. Increasingly high values of $E(B-V)$ are needed in order to produce the observed tension with the concordance cosmological model. \textit{Bottom panel}: difference between the predicted and observed distance modulus ($\Delta$DM). Values are not extrapolated below $z=1.0$ and above $z=4.8$.}
\label{fig:hd_reddening}
\end{figure*}

In this context, large Balmer decrements are interpreted as due to local reddening, as also suggested by large ratios between infrared emission and broad lines \citep{dong2005partially} and significant X-ray absorption \citep{wang2009x}. In Fig. \ref{fig:ha_hb_distr}, we show the distribution of $\log$(H$\alpha$\textsubscript{br}/H$\beta$\textsubscript{br}) ratios assuming the values reported in the WS22 catalogue for the 167 objects in our sample with reliable H$\alpha$ and H$\beta$ measurements 
(see Sec. 4 of WS22 for the detailed criteria). The mean value of the distribution is  $\langle$log(H$\alpha$\textsubscript{br}/H$\beta$\textsubscript{br})$\rangle$\,=\,0.48, and the standard deviation is 0.14. Under the assumption that the sample of blue quasars described in \cite{dong2007broad} retains, on average, minimal levels of BLR extinction (see their Sec. 4.1), we find that the intrinsic extinction is negligible, at least for the low redshift sample.

Finally, the combination of the three tests proposed here strongly suggests that intrinsic reddening should be minimal accross the entire redshift range covered by our sample.

\subsection{Is the $\rm{\Lambda}$CDM tension driven by reddening?}
\label{sec:hd_reddening}
One of the main consequences of reddening would be the underestimate of the 2500\,\AA\ monochromatic fluxes. A relatively small amount of extinction, such as $E(B-V)$\,$\simeq$\,0.1, would diminish the 5100-\AA\ flux by $\sim$25\%, but the extincted radiation would be roughly double (depending on the assumed extinction curve) at 2500\,\AA, as a result of the differential cross section of dust. The net effect of extinction is thus the underestimation of the DM, enhancing the tension between our cosmographic best-fit model and the high-redshift ($z$\,$\gtrsim$\,2--3) extrapolation of the $\rm{\Lambda}$CDM model \citep{risaliti2019cosmological, bargiacchi2023tensions}. 
Here we focus on the possibility that the observed tension with the $\rm{\Lambda}$CDM model is caused by local dust extinction affecting the UV measurements. To this end, we performed the following exercise, neglecting any possible effects on the X-ray fluxes. We assumed that the difference between the DM predicted by $\mathrm{\Lambda}$CDM and that based on the \lxluv relation is entirely driven by local (i.e. at the source redshift) extinction on $f$\textsubscript{2500\,\AA}. Therefore, assuming a standard extinction curve, we can evaluate the average $E(B-V)$ by which $f$\textsubscript{2500\,\AA} should be de-extincted in order to match the predicted DM.
With this quantity in hand, we can de-redden the UV spectra and check how much their intrinsic shapes would differ from those observed. In the following, the term ``de-reddened'' will be used to indicate the extinction-corrected spectral fluxes that produce a DM complying with the $\rm{\Lambda}$CDM model.

Fig. \ref{fig:hd_reddening} shows the samples used for this purpose. We gathered spectral data of objects at $z$\,$\geq$\,1, and divided them into 8 logarithmically equi-spaced redshift bins. For each of them, we evaluated the average redshift and DM\textsubscript{obs} values.
The DM can be written in terms of $f_{2500\,\AA}$ by using Eq. \ref{eq:dm} as:
\begin{equation}
    {\rm DM} = -\frac{5}{2}\, \frac{\gamma}{\gamma - 1} \log f_{2500\,\AA} + C, 
\end{equation}
where $C$ includes the X-ray flux contribution, the normalisation $\beta$, and the other constants.  If we assume that the measured DM values are affected by reddening, we can evaluate the amount of extinction needed to place the intrinsic DM exactly onto the $\rm{\Lambda}$CDM expectations and de-redden $f_{2500\,\AA}$ accordingly. Such a correction can be expressed as (see Appendix \ref{app:dm_reddening} for the explicit derivation):

\begin{equation}\label{eq:dm_ext}
    E(B-V) = \Delta {\rm DM}\,\frac{1}{R_V}\,\frac{A_V}{A_{2500\,\AA}}\,\frac{1-\gamma}{\gamma},
\end{equation}

where $\Delta {\rm DM}$ is the difference between the average distance modulus in a given redshift bin according to the $\rm{\Lambda}$CDM model and that observed based on the \lxluv relation.
Assuming a selective to total extinction ratio $R_{V}$\,=\,3.1 and the SMC extinction curve from \citet{gordon2016panchromatic} used above, we retrieve the values listed in the legend of Fig. \ref{fig:hd_reddening}.

\begin{figure}[h!]
\centering
\includegraphics[width=\linewidth,clip]{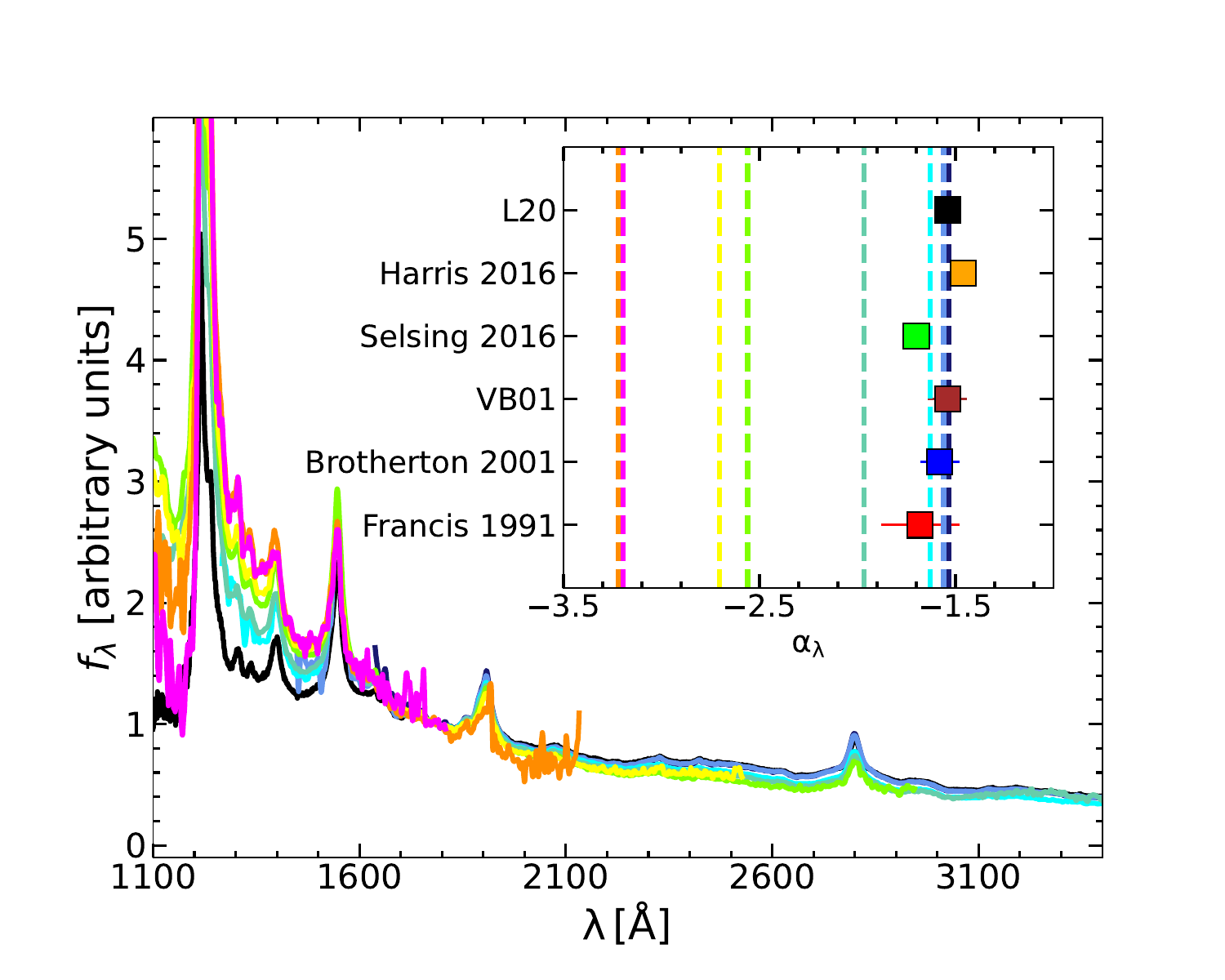}
\caption{De-reddened stacks, according to the same colour code as in Fig. \ref{fig:hd_reddening}, scaled by their mean emission between 1770 \AA\ and 1810 \AA. As a comparison also the VB01 template is shown in black as an external reference, rather than the usual L20 template. The rising continuum slopes with increasing redshift (and extinction) are clearly observed. Inset: UV continuum slopes of the literature samples described in Section \ref{sec:comparison} and plotted in Fig. \ref{fig:tpl_comparison} (squares), against the slopes of the the de-reddened spectra (dashed lines).}
\label{fig:hd_sl_distr}
\end{figure}

Next, we produced the composite spectrum in each bin, de-reddened it by applying the $E(B-V)$ values found in the previous step, and performed the spectral fits to each of them to evaluate the slope of the continuum.
We show the so derived spectra and the slopes of the continuum obtained by means of spectral fits in Fig. \ref{fig:hd_sl_distr}. The slopes characterising the continua of the de-reddened spectra are not compatible with the expectations based on literature samples, as shown in the inset of Fig. \ref{fig:hd_sl_distr}.
Indeed, moving towards higher redshift, under the hypothesis of local reddening, we would select intrinsically bluer objects with increasing amounts of extinction, whose UV slopes are anomalously steeper with respect to those typically found in blue quasars, while still delivering similar spectral shapes at all redshifts.

With the aim %end 
to explore the consequences of de-reddening the spectra by such amounts in terms of quasar physics, we built the semi-analytic (i.e. using our bin-averaged data and adopting analytic approximations from the literature in the other bands) intrinsic quasar SED in each redshift bin and computed the hard X-ray (2--10 keV) to bolometric corrections ($k_{\rm bol,X}$) to be compared with other literature samples. Anomalously high $k_{\rm bol,X}$ would further testify the selection of a very peculiar class of objects at high redshift, seemingly at odds with the typical spectral properties upon which the sample is built, and which we have verified so far.

In order to produce the average SEDs we built a semi-analytical parametrization, similar to that discussed in \citealt{marconi2004local} (M04). For wavelengths $\lambda$\,>\,1 $\mu$m we adopted a power-law spectrum $L_{\nu}$\,$\propto$\,$\nu^{\alpha_{\nu}}$, with slope $\alpha_{\nu,\rm IR}$\,=\,2 as in the Rayleigh-Jeans tail of the disc. Between 1 $\mu$m\,>\,$\lambda$\,>\,1200\,\AA\ we adopted the spectral index $\alpha_{\nu,\rm UV}$ found in the fits of the de-reddened spectra, under the assumption that at high redshift and high luminosity the continuum can be approximated as a single power law. In the range 1200\,\AA\,>\,$\lambda$\,>\,600\,\AA\ we chose the average far UV spectral slope $\alpha_{\nu,\rm FUV}$\,=\,$-1.70$ described in \citet{lusso2015}, who took into account a statistical correction on the neutral hydrogen column density of the inter-galactic medium. On the X-ray side, we used the \textit{pexrav} model (\citealt{magdziarz1995angle}) in the \xspec software (\citealt{arnaud1996}) to simulate a power-law spectrum with a photon index $\Gamma$ equal to the average photometric photon index derived from \xmm photometry (see Sec.~\ref{sec:xray}) with cutoff at 500 keV, and a reflection component with solid angle of 2$\pi$, inclination angle of $\cos\,i$\,=\,0.5, and solar metal abundances.\footnote{This is a simplistic description of the X-ray spectrum, as we do not take into account the so-called ``soft excess'' component, nor any additional spectral complexity. However, for the sake of simplicity and consistence with M04 we adhere to this description.} Each X-ray spectrum was then scaled to the average 2 keV luminosity in each bin.

Lastly, we connected the 600\,\AA\ and the 0.5 keV  luminosities with a power law. We defined 8 template SEDs according to the 8 values of $L_{2500\,\AA,\rm int}$ and $\alpha_{\nu,\rm UV}$, and used them to compute the hard X-ray (2--10 keV) bolometric corrections $k_{\rm bol,X}$. We estimated the uncertainty on $k_{\rm bol,X}$ by building 100 mock SEDs in each bin assuming Gaussian distributions for the adopted parameters with standard deviations $\sigma(\alpha_{\nu,\rm IR})$\,=\,0.1, $\sigma(\alpha_{\nu, \rm UV})$ equal to the uncertainty on the continuum slope evaluated from the fit of each de-reddened spectrum, $\sigma(L_{2500\,\AA,\rm int})$ equal to the standard deviation on the mean of the $L_{2500\,\AA,\rm int}$ values in each redshift bin, $\sigma(\alpha_{\nu,\rm FUV})$\,=\,0.6, $\sigma(L_{2\,\rm keV})$ and $\sigma(\Gamma)$ equal to the standard deviation on the mean of the $L_{2\,\rm keV}$ and $\Gamma$ distributions in each redshift bin.

\begin{figure}[h!]
\centering
\includegraphics[scale=0.4]{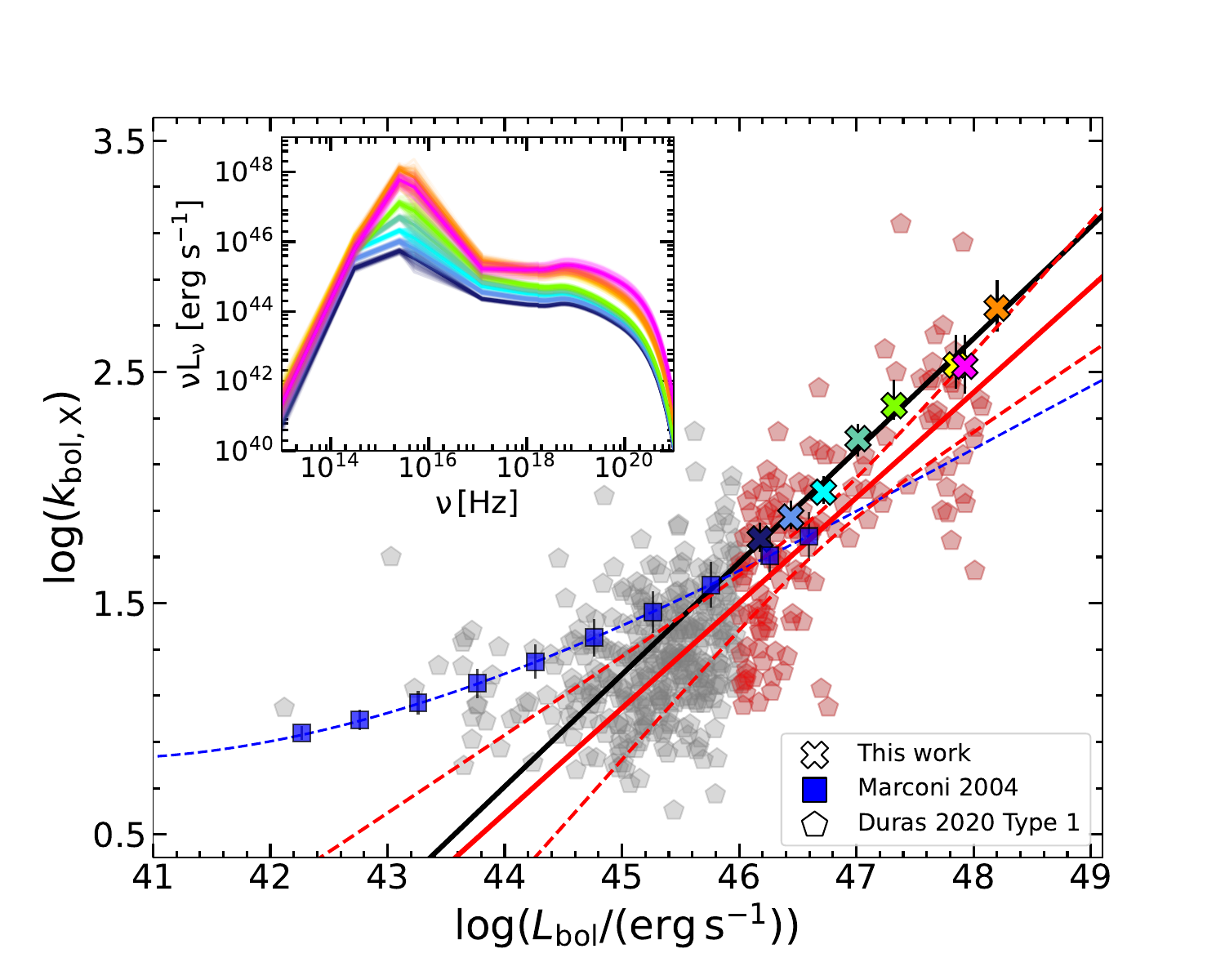}
\caption{Average hard X-ray bolometric corrections in each redshift bin. As a term of comparison, we report the $k_{\rm bol,X}$ for the semi-analytical SED templates from M04 (blue squares) and the empirical ones from D20 (pentagons). The blue dashed line represents the best-fit reported in M04. Black and red continuous lines show, respectively, the best fit of our sample and the D20 objects in the high $L_{\rm bol}$ ($\geq$\,10$^{46}$ erg s$^{-1}$) regime. The red dashed lines represent the 95\% confidence interval of the best-fit regression line for the selected high-luminosity objects from D20 (red pentagons). Our values of $k_{\rm bol,X}$ from the high-redshift, de-reddened SEDs are inconsistent with the extrapolation of the lower luminosity trend from M04, and steadily offset with respect to our best fit of the D20 high-luminosity type-1 quasars. Inset: average SEDs in each redshift bin, according to the same colour code as in Fig. \ref{fig:hd_reddening}. The shaded area is given by the individual mock SEDs in each bin.}
\label{fig:kbol_sed}
\end{figure}

The average SEDs resulting from this procedure are shown in Fig.~\ref{fig:kbol_sed}, together with other literature samples as considered in \citealt{duras2020universal} (D20). Our $k_{\rm bol,X}$ values from the de-reddened SEDs clearly detach from the high-luminosity extrapolation of the M04 best-fit relation. At the same time, they are all located on the upper envelope of the D20 best fit for luminous type-1 sources. This implies that, if such sources existed, we would be selecting an exotic population made of the most luminous, yet dust-extincted quasars. These sources would exhibit the highest known bolometric conversion factors, while still showing the UV--optical colours and spectral properties of typical blue quasars. Moreover, if our SED templates are consistent by construction with M04, which then represents a proper comparison, the same does not hold for the empirical SEDs in D20. Indeed, we did not include any physically motivated emission in the spectral region between the far UV and the soft X-rays, as we only linked the $L_{600\,\AA}$ and $L_{0.5\,{\rm keV}}$ with a straight power law, while in the latter work the authors could also rely on far-UV photometry from GALEX. 
In order to quantify the possible inconsistency of our average estimates with those from the other literature samples, we performed a basic fit of the $k_{\rm bol,X}$\,--\,$L_{\rm bol}$ relation in our de-reddened sources. We adopted a simple power-law functional form, which is a good approximation for the high luminosity ($L_{\rm bol}$\,$\geq$\,$10^{46}$ erg s$^{-1}$) end of the D20 distribution:
\begin{equation}%\label{eq:dm_ext}
    \log (k_{\rm bol,X}) = a + b \left[\log \left(\frac{L_{\rm bol}}{10^{47.5}\,{\rm erg \,  s}^{-1}}\right)\right].
\end{equation}
The best fit parameters and their respective uncertainties are listed in Table \ref{tbl:tbl_4}. In Fig. \ref{fig:kbol_sed} we show the $k_{\rm bol,X}$\,--\,$L_{\rm bol}$ relation for our sample and the D20 compilation of type-1 objects. The bolometric corrections of our higher redshift bins are located outside the 95\% confidence interval for the best fit of the high-luminosity tail of the D20 distribution (red pentagons). The offset $a$ of the best fit relation for our sources is significantly higher (3.7$\sigma$) than for those from D20. We checked that this result is not biased by the choice of the luminosity threshold for the objects included in the fit, by performing again the fit increasing the threshold by steps of $\Delta\log(L_{\rm bol}/{\rm erg\,s^{-1})}$\,=\,0.5 from 45.5 to 47. In each case, the high-redshift ($z$\,$\gtrsim$\,2.5) bolometric correction factors remain located above the 95\% confidence interval. In conclusion, the bolometric correction estimates derived from our average de-reddened SEDs are significantly offset with respect to the  expectations for high-luminosity type-1 quasars, hinting at a seemingly exotic -- and yet unknown -- population.

\begin{table}[h!]
\centering
%\begin{tabular}{||c | c | c | c ||}
\begin{tabular}{ccc}
\hline
                & $a$ & $b$  \\ 
 \hline     \noalign{\smallskip}
 This work  & 2.402$\pm$0.042 & 0.484$\pm$0.050  \\ 
 %\hline
 D20        & 2.184$\pm$0.041 & 0.455$\pm$0.040  \\ 
 \hline

\end{tabular}
\caption{Results of the fit of the $k_{\rm bol,X}$\,--\,$L_{\rm bol}$ relation for this work and the template SEDs of the most luminous sources in D20.}
\label{tbl:tbl_4}
\end{table}

%PLOT Kbol SEDs

As a result of all the tests performed in this Section, we conclude that a dust extinction scenario as an explanation for the tension with the $\Lambda$CDM model emerging in the quasar Hubble--Lema\^{\i}tre diagram seems disfavoured for several reasons:

\begin{itemize}
    \item Significant amounts of extinction at high redshift, in combination with steeper than the average UV spectra, are required in order to reconcile the tension with the $\Lambda$CDM model, unless we assume a cosmological evolution from high-redshift extremely blue sources to low-redshift typical quasars.
    \item The reddening should increase with redshift, in opposition to the current paradigm of increasing dust with time (see, e.g., \citealp{aoyama2018cosmological}, and references therein). 
    \item The dust content in the high-redshift quasars should increase by the exact amount needed to compensate for the steepening of the intrinsic spectrum, thus producing the typical spectrum of blue quasars at all redshifts.
    \item The hard X-ray bolometric correction factors derived from the average de-reddened SEDs are in tension with the high luminosity extrapolation from known sources.

\end{itemize}

All the above considerations would imply the selection of exotic objects at high redshift whose emission properties are intrinsically different from those at lower redshift and luminosity, while their apparent broadband properties are consistent with those of typical blue quasars.

\subsection{The X-ray stacks}
\label{sec:xray}
The availability of X-ray data, which underlies the creation of the L20 sample, allowed us to address the question on the evolution of the shape of X-ray emission with respect to the accretion parameters and redshift. However, at this stage, a full stacking of the X-ray spectra, taking into account the largely different redshift intervals sampled, as well as the different response functions is still unfeasible. For this reason, we adopted a slightly different approach and built ``photometric X-ray stacks''. The intrinsic X-ray spectrum of an unobscured quasar above 0.2 keV can be roughly described as a power law with spectral index $\alpha = 1 - \Gamma$, %with $\Gamma$ being 
where $\Gamma$ is the photon index. For our sample the best estimate for the $\Gamma$ parameter is the photometric photon index ($\Gamma_{\rm ph}$) derived following the procedure described in Section 4 of L20 (where also a more detailed description of photometric %$\Gamma$ 
spectral indices and of their limitations can be found). Although the photometric $\Gamma$ values are somewhat less reliable than those spectroscopically derived ($\Gamma_{\rm sp}$), the lack of a systematic offset in the $\Delta \Gamma$ distribution (defined as $\Delta \Gamma = \Gamma_{\rm sp} - \Gamma_{\rm ph}$) and its non-evolution with redshift allow us to use these estimates in a statistical sense. For each source we built a synthetic spectrum between 0.2 and 10 keV using $\Gamma_{\rm ph}$ and the 2 keV flux reported in L20. Then, the procedure to build the X-ray stacks is analogous to that explained in Sec.~\ref{sec:stacking}. 

\begin{figure*}[h!]
\centering
\includegraphics[scale=0.35]{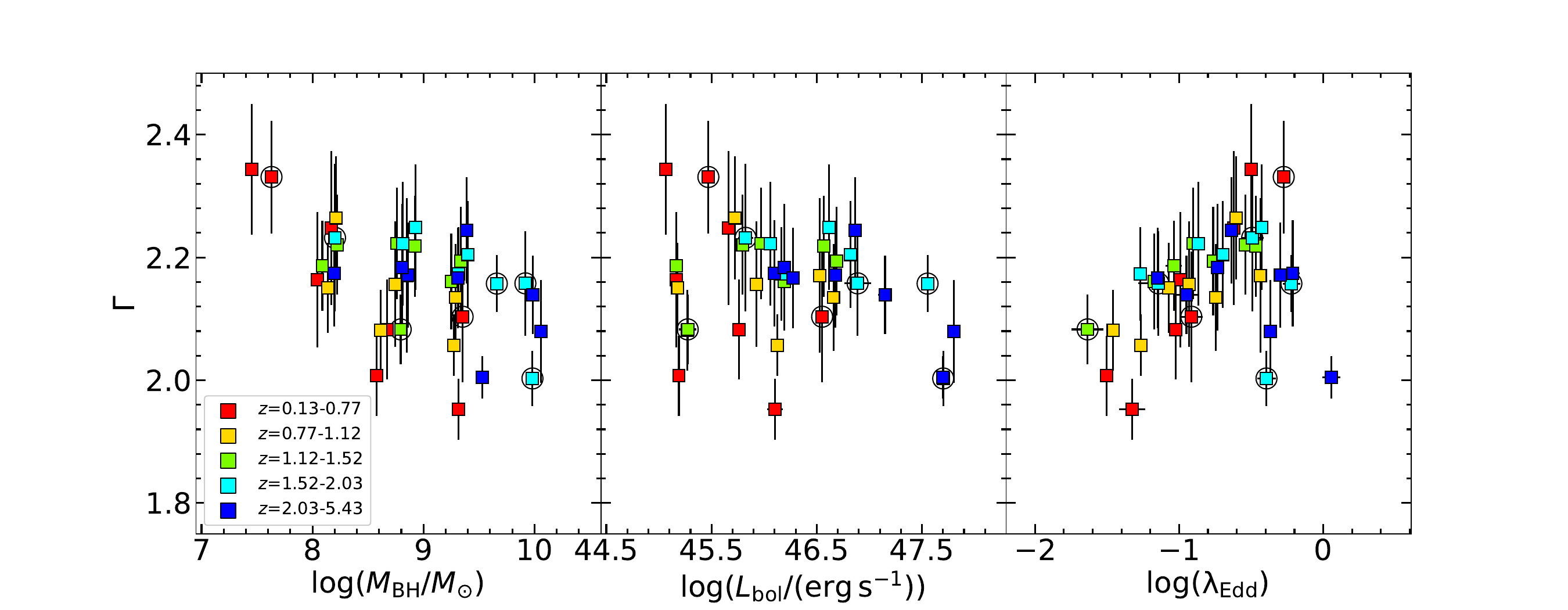}
\caption{Values of $\Gamma$ for our X-ray stacks across the parameter space. Bins not meeting the representativeness criterion are marked by a circle.}
\label{fig:x_pars}
\end{figure*}

We note, however, that the X-ray stacks are less informative than those produced in the UV--optical for at least two reasons: 1) they are not computed using the actual X-ray spectra, but rather photometric proxies, which are not sensitive to the the peculiar features of each spectrum other than the steepness of the spectrum itself; 2) the quality cuts performed in the X-ray selection (especially that concering the lower limit on the photon index, i.e. $\Gamma_{\rm ph} - \delta \Gamma_{\rm ph}$\,>\,1.7, with $\delta \Gamma_{\rm ph}$ being the uncertainty) are more conservative, given the generally lower quality of the X-ray data, and have thus a much higher impact in obliterating the spectral diversity than the optical cuts.
Nonetheless, the analysis of the slope of the stacked X-ray spectra provides valuable information about the goodness of our selection as well as the possible evolution of the X-ray properties within our parameter space. In Fig. \ref{fig:x_pars} we recover the well-known correlation between $\Gamma$\,--\,$\log\lambda_{\rm Edd}$ \citep{shemmer2008hard, risaliti2009sloan, jin2012combined, brightman2013statistical, trakhtenbrot2017bat}, with a correlation index $S$\,=\,0.38. Interestingly, the correlation index is marginally larger for the $\Gamma$\,--\,$\log M_{\rm BH}$ ($S$\,=\,$-$0.41), but this 
could be due to the way the photometric data were grouped to produce the stacks, rather than to an intrinsically tighter correlation. 

In theory, the dependence of the 2 keV fluxes on $M_{\rm BH}$ or $\lambda_{\rm Edd}$ implies that adding one of these parameters, or both, to the \lxluv space, the dispersion could be further narrowed down. For instance, the dependence on $M_{\rm BH}$, proportional to the \ion{Mg}{ii} FWHM, has been explored in \citet{signorini2023quasars}, where, however, it has been shown that the inclusion of this parameter has a negligible effect in improving the intrinsic dispersion of the relation. At the same time, the zero-covariance between the X-ray flux estimated at the pivot point (i.e., the effective-area-weighted average energy of any given band of interest) and $\Gamma$ (see Section 4 in L20) assures that even a systematic steepening of the photon index would not affect our estimates of the X-ray luminosity.

% PLOT Ha/Hb

%##########################################################################
\section{Discussion}
\label{sec:discussion}

The suitability of quasars as ``standardized candles'' relies upon the consistency of the method (i.e. the absence of selection/analysis biases, or correction thereof) and the intrinsic non-evolution of quasar spectra with redshift. Although increasingly larger samples are desirable, our group has been lately devoting a substantial effort to investigate any possible issues affecting the L20 sample with SDSS spectral coverage. In this context, this work has been chiefly focused on the analysis of the average spectral properties of 2,316 of the sources described in L20. 
The main results obtained through the various analyses performed in the previous sections are: 1) The overall shape of the composite spectra (apart from emission-line properties) does not show any clear evolution across the explored parameter space. 2) Host-galaxy contamination is not a major issue in our sources below $z$\,$\leq$\,0.8, and we therefore expect it to be completely negligible for the bulk of our sample, at higher redshfit and luminosity. 3) Our sample is compatible, on average, with a low degree of intrinsic extinction, $E(B-V)$\,$\lesssim$\,0.01. 4) The deviation of the cosmographic best fit of the quasar--SN\,Ia Hubble-Lema\^{\i}tre diagram from the $\rm{\Lambda}$CDM model reported by our group is unlikely to be caused by dust reddening, based on the spectral properties of our sample.
These findings have far-reaching consequences on the physical properties of the sources described here, as well as on the solidity of the criteria adopted to select our cosmological sample.

The similarity of quasar spectra has led through the years to the production of several composite spectra (e.g. \citealt{francis1991high}, VB01, \citealt{zheng97, brotherton2001composite, harris2016composite}), which have found large employment as templates for the typical quasar emission, in terms of line and continuum properties. In our case, building high signal-to-noise composite spectra allowed us to explore the spectral diversity within the parameter space underlying SMBH accretion.
We do not report any clear evidence for a redshift-dependent evolution of the average continua in our sample, as the agreement with either the global L20 or the independent VB01 composite spectra does not deteriorate with redshift. This appears to hold up to very high redshift, where luminous quasars have been observed to show spectral properties very akin to those seen at lower redshift (e.g., Fig. 1 in \citealt{mortlock2011luminous}). Under the paradigm that such emission is driven by SMBH accretion, these results would imply that the emission mechanism underlying the observed spectra does not vary significantly in the blue quasars spanning the last 12 Gyr.
Although we do observe the expected correlations between line equivalent width and continuum luminosity (i.e. the Baldwin effect, \citealt{baldwin1977}) and between line width and BH mass (virial relations; e.g., \citealt{peterson04}), the non-evolution of the continuum shape of our composite with any of the quantities in our parameter space is remarkable (see Fig. \ref{fig:synopt_uv} in Appendix \ref{app:synopt}).

Contamination from the host-galaxy light is a known issue when trying to isolate the AGN component (e.g., \citealt{baldwin1981classification, kauffmann03}). This effect is capable, in principle, of inducing the overestimate of the actual 2500-\AA\ fluxes, by introducing a spurious UV contribution and eventually biasing the estimate of the cosmological parameters based on the quasar \lxluv relation. For this reason, our selection criteria require steep photometric continua to single out sources where the contribution from the galactic stellar component is negligible with respect to the AGN emission (\citealt{elvis1994atlas, richards2006spectral, elvis2012spectral, krawczyk2013mean, hao2013}). As shown in Section \ref{sec:hgc}, host-galaxy contamination does not play a major role in our sample, as we generally observe steep continua, coupled with the lack of strong stellar absorption features and weak narrow emission lines from stellar photoionisation. These findings are particularly evident on the UV side of the spectrum, where our composites are dominated by higher redshift, luminous objects, in which the host contribution is expected to be negligible above $L_{\rm bol}\sim 10^{46}$ erg s$^{-1}$ (\citealt{shen2011}). A detailed disentanglement of the host-galaxy and quasar emission by means, for instance, of a principal component analysis (e.g., \citealt{yip2004spectral}) is beyond the scope of this work. Nonetheless, we can roughly estimate the average incidence of host-galaxy light by comparing the absorption features between our full composite and VB01. The EW of the \ion{Ca}{ii}\,$\lambda$3934 absorption 
in our composite ($0.7\pm 0.4$\,\AA\,) is similar to that observed in the VB01 template ($0.9\pm 0.1$\,\AA\,), yet a fair comparison is hampered by the differences in the resolution and the signal to noise between the two templates. This goes in the direction of our sample having an average host contribution comparable to that of VB01, deemed 7--15\% at such wavelengths, with an incidence decreasing bluewards.

Recently, \citet{zajavcek2023extinction} have investigated the possibility that luminosity distance measurements based on the \lxluv relation are biased by extinction on a sample of 58 quasars at $z$\,$\la$\,1.7 with available reverberation mapping data (\citealt{khadka2023quasar}). In the latter work, the authors used \ion{Mg}{ii} time-delays to infer distances by means of the the radius--luminosity relation (\ion{Mg}{ii} $R$--$L$ relation, with intrinsic dispersion $\sigma$\,$\sim$\,0.3 dex; \citealp[]{khadka2022consistency}). Assuming that the difference between the distances based on the \ion{Mg}{ii} $R$--$L$ and \lxluv relations were attributable to local reddening, and adopting different cosmological models, they derive an intrinsic colour excess $E(B-V)$\,$\simeq$\,0.002--0.003, an amount of extinction that would be intrinsic to all type-1 quasars. However, such an effect, even if present, would still not be enough to explain the reported high-redshift tension with the current cosmological model.

In this framework, the shape of the quasar continuum can be interpreted as ultimately resulting by the combination of two main parameters: the intrinsic spectral slope of the UV spectrum and the column density of gas and dust causing the extinction. The degeneracy between these two parameters could be capable, in principle, of producing similar spectra by combining increasingly larger amounts of dust in high redshift, luminous sources with intrinsically steeper spectra (e.g., \citealt{xie2015apparent}).
%In a similar way, \cite{weaver2022dust} disentangled host-galaxy and quasar disc emission taking advantage of the variability appearing from long-term photometric campaigns on 9,242 SDSS Stripe 82 quasars. By adopting a standard \citet{shakura1973black} spectrum ($L_{\nu}\sim \nu^{\alpha_{\nu}}, \, \alpha_{\nu}=1/3$) and an SMC extinction law, they found a median $E(B-V)$\,=\,0.18\,$\pm$\,0.13. A steeper spectrum with $\alpha_{\nu}$\,=\,5/7 (akin to the \citealt{mummery2020spectral} typical slope), suggested by their best-fit residuals, requires the median colour excess to increase to $E(B-V)$\,=\,0.28\,$\pm$\,0.15. However, it is worth mentioning that in all the aforementioned studies no colour cuts were performed, thus the comparison with our results is not straightforward. 

There are several reasons to believe instead that dust extinction, even if present in small amounts in our sample, is not systematically biasing our results. First, the photometric selection, besides excluding sources where the host galaxy significantly reddens the UV--optical SED, also singles out objects where the amount of dust extinction is expected to be negligible. Indeed, assuming that the photometric slopes of the average quasar SED described in \citet{richards2006spectral} are truly representative of blue quasars, by selection, the colour excess of our sources should not exceed $E(B-V)$\,$\simeq$\,0.1. We also note that the incidence of significantly reddened objects with $E(B-V)$\,$>$\,0.1, at least in early SDSS data (whose average properties are encapsulated in the VB01 template) was estimated to be modest, below $<$\,2\% according to \citet{hopkins2004dust}. In this work, we took a step further, searching a posteriori for spectral signatures of dust reddening, but we found none. Moreover, the distribution of the broad-line Balmer decrements from the H$\alpha$ and H$\beta$ reported in the WS22 catalogue for the low-redshift tail ($z$\,<\,0.57) of our sample is in excellent agreement with other samples made of allegedly unobscured sources (e.g., \citealt{zhou2006comprehensive,dong2007broad}). In Section \ref{sec:extinction} we showed that the spectral shape of our global template is consistent with the VB01 template reddened by minimal levels of local extinction, i.e. $E(B-V)$\,$\lesssim$\,0.01. Similar conclusions are also drawn in \citet{reichard2003continuum} who estimate the intrinsic reddening in non-BAL AGN from the early SDSS release to be $E(B-V)\sim 0.004$. This implies, under the assumption that the VB01 template is made of mostly blue unobscured quasars, a low incidence of reddening (dust/gas extinction) and galactic contamination.
Finally, the similarity of the spectral shape between our sample and other quasar composite spectra from several surveys designed to select luminous blue quasars (\citealt{francis1991high, brotherton2001composite}; VB01; \citealt{selsing2016x, harris2016composite}) can be interpreted as a proof of the lack of significant reddening, and a clue of the the non-evolution through cosmic time of the intrinsic emission properties of quasars.
The other possible explanations to interpret the overall agreement among composite spectra from different samples involve a certain degree of fine-tuning. Several samples of quasars on different scales of luminosity and redshift should have intrinsically steep spectral shapes and significant amounts of dust, but still producing similar spectral shapes. As a corollary to this consideration, we should also conclude that all the quasars used to build the average SEDs available in the literature (\citealt{richards2006spectral, krawczyk2013mean, gu2013evolution}), whose colours have been adopted to select our sample, are also affected by the same issue. 

If we were to attribute the tension with the $\mathrm{\Lambda}$CDM in the Hubble--Lema\^{\i}tre diagram to dust reddening, a constant extinction with redshift would not be enough. A differential extinction, increasing with redshift, would be required, entailing increasingly steeper intrinsic spectra combining in a such a way to deliver observed spectral shapes virtually independent on redshift. 

Yet, this introduces additional questions, as larger intrinsic luminosities would also imply larger SMBH masses, enlarging the discrepancy between the largest possible SMBH masses in the Eddington-limited regime and those observed, unless the accretion efficiency varies systematically at high redshift (see, for example, \citealt{inayoshi2020assembly} and \citealt{lusso2022dawn} for recent reviews on this topic). In addition, significant local extinction $E(B-V) \sim 0.1$ would imply an intrinsic UV flux larger by a factor whose exact value depends on extinction curve, but on average between 3 and 4 times at 1450\,\AA. In turn, assuming as a rough approximation a first order constant bolometric correction, the intrinsic bolometric luminosity should be accordingly larger than the observed one. Therefore, the total mass in black holes in the Universe as evaluated by \citet{soltan1982}, would need a significant increase with increasing redshift, unless the $M_{\rm BH}$ values are already underestimated by similar amounts. The same argument applies when considering greybody-like extinction curves (e.g. \citealt{gaskell2004nuclear}), which do not modify the continuum slope, but underestimate the bolometric luminosities.

The discrepancy between the observed DM and those derived in the $\Lambda CDM$ framework cannot even be solved invoking for cosmic opacity on the 2500\,\AA\, photons which would cause the observed DM to diminish. In addition, an opposite effect on DM would arise when considering cosmic opacity acting on X-ray photons. We can thus consider the decrease in the DM caused by the dimming of the UV emission as an upper limit to the detachment from the $\rm{\Lambda}$CDM model caused by free electrons extinction along the line of sight. In quantitative terms, the extrapolation of the cosmic dimming values derived for SNe Ia in $B$ band in \citealt{zhang2008dimming, hu2017investigating} lie below 0.02 at $z<3$, while our Hubble--Lema\^{\i}tre diagram (Fig.\ref{fig:hd_reddening}) shows much larger values, as $\Delta$DM approaches 1 at $z=3$.  Alternatively, some ubiquitous -- yet unknown -- material in the IGM should be responsible for a grey-body like absorption, though this option seems unlikely.

On the X-ray side, gas absorption plays an opposite role with respect to the UV counterpart, as the correction for absorption would increase the departure from the $\rm{\Lambda}$CDM model. Generally speaking, the incidence of gas absorption among our sources is minimised by the $\Gamma_{\rm ph}$\,>\,1.7 criterion. In addition, this effect is further mitigated at high redshift by the shift within the observed 0.5--2 keV band of much higher rest-frame energies, for which the effect of gas absorption becomes increasingly weaker. Indeed, in all the high-redshift objects for which a dedicated X-ray spectral analysis was performed (\citealt{vito2019, nardini2019, salvestrini2019, sacchi2022quasars}), column densities in excess of 10$^{23}$ cm$^{-2}$ would be required to modify the spectral slope and thus affect the extrapolated rest-frame 2 keV flux, but such values were generally excluded. 

As a final consideration, the results of the spectroscopic analysis described in this work ultimately prove the reliability of the photometric selection criteria adopted in our works to define a sample of quasars with typical properties. Using the photometric cuts, rather than the more time-consuming spectroscopic analysis, will allow us to considerably enlarge the cosmological sample with the forthcoming surveys (e.g. LSST, DESI), while retaining a high degree of consistency with the spectroscopic properties.

\section{Conclusions}
\label{sec:conclusions}

In this sixth paper of the series, we presented a thorough analysis of the average spectral properties of 2,316 sources with available SDSS data from the sample described in \citet{lusso2020}. Our main aims were to investigate the possible evolution of the UV--optical properties with redshift, and to study the possible systematic effects (e.g., host-galaxy contamination, dust extinction) that could undermine the validity of quasars as standard candles. To this end, we built spectral composites in the parameter space 
defined by $M_{\rm BH}$, $L_{\rm bol}$, and redshift, to investigate how the spectral properties of the cosmological sample vary as a function of the key accretion parameters and cosmic time. 
We also carried out specific tests to determine the presence of residual galactic contamination and dust reddening, based on their spectral signatures.

Our main findings are listed below:

\begin{itemize}
    \item Apart from the line emission, which varies according to well-known relations with $L_{\rm bol}$ and $M_{\rm BH}$, we report a strong similarity in terms of continuum shapes among the composite spectra across the explored parameter space. The slopes of the continuum of our composite spectra cluster around $\langle \alpha_{\lambda} \rangle = -1.50\pm0.14$, without statistically significant correlations (i.e. $|S|\leq0.3$) with the explored parameters ($M_{\rm BH}$, $L_{\rm bol}$, $z$). This points to a non-evolution of the broadband continuum emission within our sample with respect to the considered parameters.
    \item Our global composite spectrum shows a good agreement with other templates of typical blue quasars, as well as with the individual composite spectra in the various regions of the parameter space. This confirms that the L20 catalogue is, on average, made of typical objects.
    \item By comparing the composite spectrum of our low-redshift ($z$\,$\leq$\,0.8) sources with SDSS objects affected by heavy host-galaxy contamination, we were able to verify that host galaxy emission is negligible in our sample. We, estimate the host galaxy fraction at optical wavelengths to be $\lesssim$10\%, and likely even lower at UV wavelengths. 
    \item After performing several tests, as listed in Section \ref{sec:extinction}, we demonstrated that the objects constituting the L20 sample are consistent with a low degree of intrinsic extinction, with an average colour excess value $E(B-V)\lesssim 0.01$.
    \item We showed that the tension between our cosmographic best fit of the joint quasar--SN\,Ia Hubble--Lema\^{\i}tre diagram and the $\rm{\Lambda}$CDM model cannot be simply accounted for by allowing for cosmic opacity or differential dust extinction. Otherwise, a colour excess $E(B-V)\gtrsim 0.1$ increasing with $z$, accompanied by anomalously steep UV spectra, should ubiquitously affect high-$z$ sources, while delivering similar spectral shapes at all redshifts.
    \item The spectroscopic analysis carried out in this work corroborates the photometric selection criteria adopted to define a sample of quasars with typical properties. This, in turn, allows to keep employing the more efficient photometric cuts rather than a more time-consuming individual spectroscopic analysis.
    
\end{itemize}

The search for possible systematics performed in this work has broad implications for the cosmological applications of our compilation of quasars. First, the spectral confirmation that dust or galaxy contamination do not significantly affect, without invoking peculiar/anomalous explanations, our cosmological sample further corroborates the reliability of our methods and results. In this context, quasars emerge as consistent standardizable candles, and the high-redshift deviation from the current cosmological model remains outstanding. 
Secondly, as our method has proven fruitful in producing large samples of homogeneous sources, we can extend and refine its application to the forthcoming data from new facilities. As a matter of fact, the extended ROentgen Survey with an Imaging Telescope Array (eROSITA, \citealp[]{merloni2012erosita}) began populating the pivotal $z$\,$\sim$\,1 region in the X-ray domain, allowing a more robust cross-calibration between quasars and SNe\,Ia, and a more precise determination of $\rm{\Omega_{\Lambda}}$ \citep{lusso2020cosmology}. In the mid- to long-term future, the cross-match between \textit{Euclid} (\citealt{laureijs2011euclid}) and LSST (\citealt{ivezic2019lsst}) observations in the optical--UV and \textit{Athena} (and/or other proposed facilities) in the X-rays will significantly enlarge the sample, helping to fill up the region at intermediate--high redshift. New large and reliable samples will be key to test the cosmological models in synergy with other well-established probes, with the intent to precisely constrain the cosmological parameters.

% \begin{acknowledgements}

% \end{acknowledgements}

\bibliographystyle{aa} 
\bibliography{bibl}

\begin{thebibliography}{120}
\expandafter\ifx\csname natexlab\endcsname\relax\def\natexlab#1{#1}\fi

\bibitem[{Abdurro’uf {et~al.}(2022)Abdurro’uf, Accetta, Aerts,
  Silva~Aguirre, Ahumada, Ajgaonkar, Filiz~Ak, Alam, Allende~Prieto, Almeida,
  {et~al.}}]{abdurro2022seventeenth}
Abdurro’uf, N., Accetta, K., Aerts, C., {et~al.} 2022, The Astrophysical
  Journal. Supplement Series, 259

\bibitem[{Aoyama {et~al.}(2018)Aoyama, Hou, Hirashita, Nagamine, \&
  Shimizu}]{aoyama2018cosmological}
Aoyama, S., Hou, K.-C., Hirashita, H., Nagamine, K., \& Shimizu, I. 2018,
  Monthly Notices of the Royal Astronomical Society, 478, 4905

\bibitem[{{Arcodia} {et~al.}(2019){Arcodia}, {Merloni}, {Nandra}, \&
  {Ponti}}]{arcodia2019}
{Arcodia}, R., {Merloni}, A., {Nandra}, K., \& {Ponti}, G. 2019, \aap, 628,
  A135

\bibitem[{{Arnaud}(1996)}]{arnaud1996}
{Arnaud}, K.~A. 1996, Astronomical Society of the Pacific Conference Series,
  Vol. 101, {XSPEC: The First Ten Years}, ed. G.~H. {Jacoby} \& J.~{Barnes}, 17

\bibitem[{Assef {et~al.}(2010)Assef, Kochanek, Brodwin, Cool, Forman, Gonzalez,
  Hickox, Jones, Le~Floc'h, Moustakas, {et~al.}}]{assef2010low}
Assef, R., Kochanek, C., Brodwin, M., {et~al.} 2010, The Astrophysical Journal,
  713, 970

\bibitem[{{Ba{\~n}ados} {et~al.}(2018){Ba{\~n}ados}, {Venemans},
  {Mazzucchelli}, {Farina}, {Walter}, {Wang}, {Decarli}, {Stern}, {Fan},
  {Davies}, {Hennawi}, {Simcoe}, {Turner}, {Rix}, {Yang}, {Kelson}, {Rudie}, \&
  {Winters}}]{banados2018}
{Ba{\~n}ados}, E., {Venemans}, B.~P., {Mazzucchelli}, C., {et~al.} 2018, \nat,
  553, 473

\bibitem[{{Baldwin}(1977)}]{baldwin1977}
{Baldwin}, J.~A. 1977, \apj, 214, 679

\bibitem[{Baldwin {et~al.}(1981)Baldwin, Phillips, \&
  Terlevich}]{baldwin1981classification}
Baldwin, J.~A., Phillips, M.~M., \& Terlevich, R. 1981, Publications of the
  Astronomical Society of the Pacific, 93, 5

\bibitem[{Bargiacchi {et~al.}(2022)Bargiacchi, Benetti, Capozziello, Lusso,
  Risaliti, \& Signorini}]{bargiacchi2022quasar}
Bargiacchi, G., Benetti, M., Capozziello, S., {et~al.} 2022, Monthly Notices of
  the Royal Astronomical Society, 515, 1795

\bibitem[{Bargiacchi {et~al.}(2023)Bargiacchi, Dainotti, \&
  Capozziello}]{bargiacchi2023tensions}
Bargiacchi, G., Dainotti, M., \& Capozziello, S. 2023, Monthly Notices of the
  Royal Astronomical Society, 525, 3104

\bibitem[{Bargiacchi {et~al.}(2021)Bargiacchi, Risaliti, Benetti, Capozziello,
  Lusso, Saccardi, \& Signorini}]{bargiacchi2021cosmography}
Bargiacchi, G., Risaliti, G., Benetti, M., {et~al.} 2021, Astronomy \&
  Astrophysics, 649, A65

\bibitem[{Brightman {et~al.}(2013)Brightman, Silverman, Mainieri, Ueda,
  Schramm, Matsuoka, Nagao, Steinhardt, Kartaltepe, Sanders,
  {et~al.}}]{brightman2013statistical}
Brightman, M., Silverman, J., Mainieri, V., {et~al.} 2013, Monthly Notices of
  the Royal Astronomical Society, 433, 2485

\bibitem[{Brotherton {et~al.}(2001)Brotherton, Tran, Becker, Gregg,
  Laurent-Muehleisen, \& White}]{brotherton2001composite}
Brotherton, M., Tran, H.~D., Becker, R., {et~al.} 2001, The Astrophysical
  Journal, 546, 775

\bibitem[{Czerny {et~al.}(2018)Czerny, Beaton, Bejger, Cackett, Dall’Ora,
  Holanda, Jensen, Jha, Lusso, Minezaki, {et~al.}}]{czerny2018astronomical}
Czerny, B., Beaton, R., Bejger, M., {et~al.} 2018, Space Science Reviews, 214,
  1

\bibitem[{Dawson {et~al.}(2012)Dawson, Schlegel, Ahn, Anderson, Aubourg,
  Bailey, Barkhouser, Bautista, Beifiori, Berlind, {et~al.}}]{dawson2012baryon}
Dawson, K.~S., Schlegel, D.~J., Ahn, C.~P., {et~al.} 2012, The Astronomical
  Journal, 145, 10

\bibitem[{Dong {et~al.}(2007)Dong, Wang, Wang, Yuan, Zhou, Dai, \&
  Zhang}]{dong2007broad}
Dong, X., Wang, T., Wang, J., {et~al.} 2007, Monthly Notices of the Royal
  Astronomical Society, 383, 581

\bibitem[{Dong {et~al.}(2005)Dong, Zhou, Wang, Wang, Li, \&
  Zhou}]{dong2005partially}
Dong, X.-B., Zhou, H.-Y., Wang, T.-G., {et~al.} 2005, The Astrophysical
  Journal, 620, 629

\bibitem[{Duras {et~al.}(2020)Duras, Bongiorno, Ricci, Piconcelli, Shankar,
  Lusso, Bianchi, Fiore, Maiolino, Marconi, {et~al.}}]{duras2020universal}
Duras, F., Bongiorno, A., Ricci, F., {et~al.} 2020, Astronomy \& Astrophysics,
  636, A73

\bibitem[{Eisenstein {et~al.}(2011)Eisenstein, Weinberg, Agol, Aihara, Prieto,
  Anderson, Arns, Aubourg, Bailey, Balbinot, {et~al.}}]{eisenstein2011sdss}
Eisenstein, D.~J., Weinberg, D.~H., Agol, E., {et~al.} 2011, The Astronomical
  Journal, 142, 72

\bibitem[{Elvis {et~al.}(2012)Elvis, Hao, Civano, Brusa, Salvato, Bongiorno,
  Capak, Zamorani, Comastri, Jahnke, {et~al.}}]{elvis2012spectral}
Elvis, M., Hao, H., Civano, F., {et~al.} 2012, The Astrophysical Journal, 759,
  6

\bibitem[{{Elvis} \& {Karovska}(2002)}]{elviskarovska2002}
{Elvis}, M. \& {Karovska}, M. 2002, \apjl, 581, L67

\bibitem[{Elvis {et~al.}(1994)Elvis, Wilkes, McDowell, Green, Bechtold,
  Willner, Oey, Polomski, \& Cutri}]{elvis1994atlas}
Elvis, M.~S., Wilkes, B.~J., McDowell, J.~C., {et~al.} 1994, The Astrophysical
  Journal Supplement Series

\bibitem[{Evans(1989)}]{evans1989photometric}
Evans, D. 1989, Astronomy and Astrophysics Supplement Series (ISSN 0365-0138),
  vol. 78, no. 2, May 1989, p. 249-268. Research supported by SERC., 78, 249

\bibitem[{{Evans} {et~al.}(2010){Evans}, {Primini}, {Glotfelty}, {Anderson},
  {Bonaventura}, {Chen}, {Davis}, {Doe}, {Evans}, {Fabbiano}, {Galle}, {Gibbs},
  {Grier}, {Hain}, {Hall}, {Harbo}, {He}, {Houck}, {Karovska}, {Kashyap},
  {Lauer}, {McCollough}, {McDowell}, {Miller}, {Mitschang}, {Morgan},
  {Mossman}, {Nichols}, {Nowak}, {Plummer}, {Refsdal}, {Rots}, {Siemiginowska},
  {Sundheim}, {Tibbetts}, {Van Stone}, {Winkelman}, \& {Zografou}}]{evans2010}
{Evans}, I.~N., {Primini}, F.~A., {Glotfelty}, K.~J., {et~al.} 2010, \apjs,
  189, 37

\bibitem[{{Faucher-Gigu{\`e}re} {et~al.}(2008){Faucher-Gigu{\`e}re}, {Lidz},
  {Hernquist}, \& {Zaldarriaga}}]{faucher08}
{Faucher-Gigu{\`e}re}, C.-A., {Lidz}, A., {Hernquist}, L., \& {Zaldarriaga}, M.
  2008, \apj, 688, 85

\bibitem[{Fitzpatrick(1999)}]{fitzpatrick1999correcting}
Fitzpatrick, E.~L. 1999, Publications of the Astronomical Society of the
  Pacific, 111, 63

\bibitem[{Francis {et~al.}(1991)Francis, Hewett, Foltz, Chaffee, Weymann,
  Morris, {et~al.}}]{francis1991high}
Francis, P., Hewett, P.~C., Foltz, C.~B., {et~al.} 1991

\bibitem[{Gallerani {et~al.}(2010)Gallerani, Maiolino, Juarez, Nagao, Marconi,
  Bianchi, Schneider, Mannucci, Oliva, Willott,
  {et~al.}}]{gallerani2010extinction}
Gallerani, S., Maiolino, R., Juarez, Y., {et~al.} 2010, Astronomy \&
  Astrophysics, 523, A85

\bibitem[{Gaskell \& Ferland(1984)}]{gaskell1984theoretical}
Gaskell, C.~M. \& Ferland, G.~J. 1984, Publications of the Astronomical Society
  of the Pacific, 96, 393

\bibitem[{Gaskell {et~al.}(2004)Gaskell, Goosmann, Antonucci, \&
  Whysong}]{gaskell2004nuclear}
Gaskell, C.~M., Goosmann, R.~W., Antonucci, R.~R., \& Whysong, D.~H. 2004, The
  Astrophysical Journal, 616, 147

\bibitem[{Goodrich(1990)}]{goodrich1990pa}
Goodrich, R.~W. 1990, Astrophysical Journal, Part 1 (ISSN 0004-637X), vol. 355,
  May 20, 1990, p. 88-93., 355, 88

\bibitem[{Gordon {et~al.}(2016)Gordon, Fouesneau, Arab, Tchernyshyov, Weisz,
  Dalcanton, Williams, Bell, Bianchi, Boyer, {et~al.}}]{gordon2016panchromatic}
Gordon, K.~D., Fouesneau, M., Arab, H., {et~al.} 2016, The Astrophysical
  Journal, 826, 104

\bibitem[{Greene \& Ho(2005)}]{greene2005estimating}
Greene, J.~E. \& Ho, L.~C. 2005, The Astrophysical Journal, 630, 122

\bibitem[{Gu(2013)}]{gu2013evolution}
Gu, M. 2013, The Astrophysical Journal, 773, 176

\bibitem[{Haardt \& Maraschi(1991)}]{haardt1991two}
Haardt, F. \& Maraschi, L. 1991, Astrophysical Journal, Part 2-Letters (ISSN
  0004-637X), vol. 380, Oct. 20, 1991, p. L51-L54., 380, L51

\bibitem[{Haardt \& Maraschi(1993)}]{haardt1993x}
Haardt, F. \& Maraschi, L. 1993, The Astrophysical Journal, 413, 507

\bibitem[{Halpern \& Steiner(1983)}]{halpern1983low}
Halpern, J. \& Steiner, J. 1983, Astrophysical Journal, Part 2-Letters to the
  Editor (ISSN 0004-637X), vol. 269, June 15, 1983, p. L37-L41., 269, L37

\bibitem[{{Hao} {et~al.}(2013){Hao}, {Elvis}, {Bongiorno}, {Zamorani},
  {Merloni}, {Kelly}, {Civano}, {Celotti}, {Ho}, {Jahnke}, {Comastri}, {Trump},
  {Mainieri}, {Salvato}, {Brusa}, {Impey}, {Koekemoer}, {Lanzuisi}, {Vignali},
  {Silverman}, {Urry}, \& {Schawinski}}]{hao2013}
{Hao}, H., {Elvis}, M., {Bongiorno}, A., {et~al.} 2013, \mnras, 434, 3104

\bibitem[{Harris {et~al.}(2016)Harris, Jensen, Suzuki, Bautista, Dawson, Vivek,
  Brownstein, Ge, Hamann, Herbst, {et~al.}}]{harris2016composite}
Harris, D.~W., Jensen, T.~W., Suzuki, N., {et~al.} 2016, The Astronomical
  Journal, 151, 155

\bibitem[{Hopkins {et~al.}(2004)Hopkins, Strauss, Hall, Richards, Cooper,
  Schneider, Berk, Jester, Brinkmann, \& Szokoly}]{hopkins2004dust}
Hopkins, P.~F., Strauss, M.~A., Hall, P.~B., {et~al.} 2004, The Astronomical
  Journal, 128, 1112

\bibitem[{Hu {et~al.}(2017)Hu, Yu, \& Wang}]{hu2017investigating}
Hu, J., Yu, H., \& Wang, F. 2017, The Astrophysical Journal, 836, 107

\bibitem[{Inayoshi {et~al.}(2020)Inayoshi, Visbal, \&
  Haiman}]{inayoshi2020assembly}
Inayoshi, K., Visbal, E., \& Haiman, Z. 2020, Annual Review of Astronomy and
  Astrophysics, 58, 27

\bibitem[{Ivezi{\'c} {et~al.}(2019)Ivezi{\'c}, Kahn, Tyson, Abel, Acosta,
  Allsman, Alonso, AlSayyad, Anderson, Andrew, {et~al.}}]{ivezic2019lsst}
Ivezi{\'c}, {\v{Z}}., Kahn, S.~M., Tyson, J.~A., {et~al.} 2019, The
  Astrophysical Journal, 873, 111

\bibitem[{Jalan {et~al.}(2023)Jalan, Rakshit, Woo, Kotilainen, \&
  Stalin}]{jalan2023empirical}
Jalan, P., Rakshit, S., Woo, J.-J., Kotilainen, J., \& Stalin, C. 2023, Monthly
  Notices of the Royal Astronomical Society: Letters, 521, L11

\bibitem[{Jin {et~al.}(2024)Jin, Lusso, Ward, Done, \&
  Middei}]{jin2024wavelength}
Jin, C., Lusso, E., Ward, M., Done, C., \& Middei, R. 2024, Monthly Notices of
  the Royal Astronomical Society, 527, 356

\bibitem[{Jin {et~al.}(2012)Jin, Ward, \& Done}]{jin2012combined}
Jin, C., Ward, M., \& Done, C. 2012, Monthly Notices of the Royal Astronomical
  Society, 425, 907

\bibitem[{{Kauffmann} {et~al.}(2003)}]{kauffmann03}
{Kauffmann}, G. {et~al.} 2003, \mnras, 346, 1055

\bibitem[{{Kellermann} {et~al.}(1989){Kellermann}, {Sramek}, {Schmidt},
  {Shaffer}, \& {Green}}]{kellermann89}
{Kellermann}, K.~I., {Sramek}, R., {Schmidt}, M., {Shaffer}, D.~B., \& {Green},
  R. 1989, \aj, 98, 1195

\bibitem[{Khadka {et~al.}(2022)Khadka, Zaja{\v{c}}ek, Panda,
  Mart{\'\i}nez-Aldama, \& Ratra}]{khadka2022consistency}
Khadka, N., Zaja{\v{c}}ek, M., Panda, S., Mart{\'\i}nez-Aldama, M.~L., \&
  Ratra, B. 2022, Monthly Notices of the Royal Astronomical Society, 515, 3729

\bibitem[{Khadka {et~al.}(2023)Khadka, Zaja{\v{c}}ek, Prince, Panda, Czerny,
  Mart{\'\i}nez-Aldama, Jaiswal, \& Ratra}]{khadka2023quasar}
Khadka, N., Zaja{\v{c}}ek, M., Prince, R., {et~al.} 2023, Monthly Notices of
  the Royal Astronomical Society, 522, 1247

\bibitem[{Kratzer \& Richards(2015)}]{kratzer2015mean}
Kratzer, R.~M. \& Richards, G.~T. 2015, The Astronomical Journal, 149, 61

\bibitem[{Krawczyk {et~al.}(2013)Krawczyk, Richards, Mehta, Vogeley, Gallagher,
  Leighly, Ross, \& Schneider}]{krawczyk2013mean}
Krawczyk, C.~M., Richards, G.~T., Mehta, S.~S., {et~al.} 2013, The
  Astrophysical Journal Supplement Series, 206, 4

\bibitem[{{La Franca} {et~al.}(2014){La Franca}, {Bianchi}, {Ponti},
  {Branchini}, \& {Matt}}]{lafranca2014}
{La Franca}, F., {Bianchi}, S., {Ponti}, G., {Branchini}, E., \& {Matt}, G.
  2014, \apjl, 787, L12

\bibitem[{Laureijs {et~al.}(2011)Laureijs, Amiaux, Arduini, Augueres,
  Brinchmann, Cole, Cropper, Dabin, Duvet, Ealet,
  {et~al.}}]{laureijs2011euclid}
Laureijs, R., Amiaux, J., Arduini, S., {et~al.} 2011, arXiv preprint
  arXiv:1110.3193

\bibitem[{{Lenart} {et~al.}(2023){Lenart}, {Bargiacchi}, {Dainotti},
  {Nagataki}, \& {Capozziello}}]{lenart2023}
{Lenart}, A.~{\L}., {Bargiacchi}, G., {Dainotti}, M.~G., {Nagataki}, S., \&
  {Capozziello}, S. 2023, \apjs, 264, 46

\bibitem[{Lusso(2020)}]{lusso2020cosmology}
Lusso, E. 2020, Frontiers in Astronomy and Space Sciences, 7, 8

\bibitem[{Lusso \& Risaliti(2017)}]{lusso2017quasars}
Lusso, E. \& Risaliti, G. 2017, Astronomy \& Astrophysics, 602, A79

\bibitem[{{Lusso} {et~al.}(2020){Lusso}, {Risaliti}, {Nardini}, {Bargiacchi},
  {Benetti}, {Bisogni}, {Capozziello}, {Civano}, {Eggleston}, {Elvis},
  {Fabbiano}, {Gilli}, {Marconi}, {Paolillo}, {Piedipalumbo}, {Salvestrini},
  {Signorini}, \& {Vignali}}]{lusso2020}
{Lusso}, E., {Risaliti}, G., {Nardini}, E., {et~al.} 2020, \aap, 642, A150

\bibitem[{Lusso {et~al.}(2022)Lusso, Valiante, \& Vito}]{lusso2022dawn}
Lusso, E., Valiante, R., \& Vito, F. 2022, arXiv preprint arXiv:2205.15349

\bibitem[{{Lusso} {et~al.}(2015){Lusso}, {Worseck}, {Hennawi}, {Prochaska},
  {Vignali}, {Stern}, \& {O'Meara}}]{lusso2015}
{Lusso}, E., {Worseck}, G., {Hennawi}, J.~F., {et~al.} 2015, \mnras, 449, 4204

\bibitem[{{Lusso} {et~al.}(2010)}]{lusso2010}
{Lusso}, E. {et~al.} 2010, \aap, 512, A34

\bibitem[{Ma {et~al.}(2023)Ma, Li, Gu, Jiang, Hou, Qin, \&
  Bian}]{ma2023variability}
Ma, Y.-S., Li, S.-J., Gu, C.-S., {et~al.} 2023, Monthly Notices of the Royal
  Astronomical Society, 522, 5680

\bibitem[{Madau(1995)}]{madau1995radiative}
Madau, P. 1995, The Astrophysical Journal, 441, 18

\bibitem[{Magdziarz \& Zdziarski(1995)}]{magdziarz1995angle}
Magdziarz, P. \& Zdziarski, A.~A. 1995, Monthly Notices of the Royal
  Astronomical Society, 273, 837

\bibitem[{Marconi {et~al.}(2004)Marconi, Risaliti, Gilli, Hunt, Maiolino, \&
  Salvati}]{marconi2004local}
Marconi, A., Risaliti, G., Gilli, R., {et~al.} 2004, Monthly Notices of the
  Royal Astronomical Society, 351, 169

\bibitem[{{Markwardt}(2009)}]{Markwardt2009}
{Markwardt}, C.~B. 2009, in Astronomical Society of the Pacific Conference
  Series, Vol. 411, Astronomical Data Analysis Software and Systems XVIII, ed.
  D.~A. {Bohlender}, D.~{Durand}, \& P.~{Dowler}, 251

\bibitem[{{Marziani} \& {Sulentic}(2014)}]{marzianisulentic2014}
{Marziani}, P. \& {Sulentic}, J.~W. 2014, Advances in Space Research, 54, 1331

\bibitem[{Mej{\'\i}a-Restrepo {et~al.}(2022)Mej{\'\i}a-Restrepo, Trakhtenbrot,
  Koss, Oh, Den~Brok, Stern, Powell, Ricci, Caglar, Ricci,
  {et~al.}}]{mejia2022bass}
Mej{\'\i}a-Restrepo, J.~E., Trakhtenbrot, B., Koss, M.~J., {et~al.} 2022, The
  Astrophysical Journal Supplement Series, 261, 5

\bibitem[{{Menzel} {et~al.}(2016){Menzel}, {Merloni}, {Georgakakis}, {Salvato},
  {Aubourg}, {Brandt}, {Brusa}, {Buchner}, {Dwelly}, {Nandra}, {P{\^a}ris},
  {Petitjean}, \& {Schwope}}]{menzel2016}
{Menzel}, M.-L., {Merloni}, A., {Georgakakis}, A., {et~al.} 2016, \mnras, 457,
  110

\bibitem[{{Merloni}(2003)}]{merloni2003}
{Merloni}, A. 2003, \mnras, 341, 1051

\bibitem[{Merloni {et~al.}(2012)Merloni, Predehl, Becker, B{\"o}hringer,
  Boller, Brunner, Brusa, Dennerl, Freyberg, Friedrich,
  {et~al.}}]{merloni2012erosita}
Merloni, A., Predehl, P., Becker, W., {et~al.} 2012, arXiv preprint
  arXiv:1209.3114

\bibitem[{Mor{\'e}(1978)}]{more1978levenberg}
Mor{\'e}, J.~J. 1978, in Numerical analysis (Springer), 105--116

\bibitem[{Mortlock {et~al.}(2011)Mortlock, Warren, Venemans, Patel, Hewett,
  McMahon, Simpson, Theuns, Gonz{\'a}les-Solares, Adamson,
  {et~al.}}]{mortlock2011luminous}
Mortlock, D.~J., Warren, S.~J., Venemans, B.~P., {et~al.} 2011, Nature, 474,
  616

\bibitem[{{Nardini} {et~al.}(2019){Nardini}, {Lusso}, {Risaliti}, {Bisogni},
  {Civano}, {Elvis}, {Fabbiano}, {Gilli}, {Marconi}, {Salvestrini}, \&
  {Vignali}}]{nardini2019}
{Nardini}, E., {Lusso}, E., {Risaliti}, G., {et~al.} 2019, \aap, 632, A109

\bibitem[{Neugebauer {et~al.}(1979)Neugebauer, Oke, Becklin, \&
  Matthews}]{neugebauer1979absolute}
Neugebauer, G., Oke, J., Becklin, E., \& Matthews, K. 1979, Astrophysical
  Journal, Part 1, vol. 230, May 15, 1979, p. 79-94., 230, 79

\bibitem[{{Nicastro}(2000)}]{nicastro2000}
{Nicastro}, F. 2000, \apjl, 530, L65

\bibitem[{Novikov \& Thorne(1973)}]{novikov1973astrophysics}
Novikov, I.~D. \& Thorne, K.~S. 1973, Black holes (Les astres occlus), 1, 343

\bibitem[{Osterbrock(1977)}]{osterbrock1977spectrophotometry}
Osterbrock, D.~E. 1977, The Astrophysical Journal, 215, 733

\bibitem[{P{\^a}ris {et~al.}(2018)P{\^a}ris, Petitjean, Aubourg, Myers,
  Streblyanska, Lyke, Anderson, Armengaud, Bautista, Blanton,
  {et~al.}}]{paris2018sloan}
P{\^a}ris, I., Petitjean, P., Aubourg, {\'E}., {et~al.} 2018, Astronomy \&
  Astrophysics, 613, A51

\bibitem[{{Peterson} {et~al.}(2004){Peterson}, {Ferrarese}, {Gilbert}, {Kaspi},
  {Malkan}, {Maoz}, {Merritt}, {Netzer}, {Onken}, {Pogge}, {Vestergaard}, \&
  {Wandel}}]{peterson04}
{Peterson}, B.~M., {Ferrarese}, L., {Gilbert}, K.~M., {et~al.} 2004, \apj, 613,
  682

\bibitem[{{Prevot} {et~al.}(1984){Prevot}, {Lequeux}, {Prevot}, {Maurice}, \&
  {Rocca-Volmerange}}]{prevot84}
{Prevot}, M.~L., {Lequeux}, J., {Prevot}, L., {Maurice}, E., \&
  {Rocca-Volmerange}, B. 1984, \aap, 132, 389

\bibitem[{Prochaska {et~al.}(2009)Prochaska, Worseck, \&
  O’Meara}]{prochaska2009direct}
Prochaska, J.~X., Worseck, G., \& O’Meara, J.~M. 2009, The Astrophysical
  Journal, 705, L113

\bibitem[{Rakshit {et~al.}(2020)Rakshit, Stalin, \&
  Kotilainen}]{rakshit2020spectral}
Rakshit, S., Stalin, C., \& Kotilainen, J. 2020, The Astrophysical Journal
  Supplement Series, 249, 17

\bibitem[{Reichard {et~al.}(2003)Reichard, Richards, Hall, Schneider, Berk,
  Fan, York, Knapp, \& Brinkmann}]{reichard2003continuum}
Reichard, T.~A., Richards, G.~T., Hall, P.~B., {et~al.} 2003, The Astronomical
  Journal, 126, 2594

\bibitem[{Richards {et~al.}(2006)Richards, Lacy, Storrie-Lombardi, Hall,
  Gallagher, Hines, Fan, Papovich, Berk, Trammell,
  {et~al.}}]{richards2006spectral}
Richards, G.~T., Lacy, M., Storrie-Lombardi, L.~J., {et~al.} 2006, The
  Astrophysical Journal Supplement Series, 166, 470

\bibitem[{{Richards} {et~al.}(2001)}]{richards01}
{Richards}, G.~T. {et~al.} 2001, \aj, 121, 2308

\bibitem[{Risaliti \& Lusso(2015)}]{risaliti2015hubble}
Risaliti, G. \& Lusso, E. 2015, The Astrophysical Journal, 815, 33

\bibitem[{Risaliti \& Lusso(2019)}]{risaliti2019cosmological}
Risaliti, G. \& Lusso, E. 2019, Nature Astronomy, 3, 272

\bibitem[{Risaliti {et~al.}(2009)Risaliti, Young, \& Elvis}]{risaliti2009sloan}
Risaliti, G., Young, M., \& Elvis, M. 2009, The Astrophysical Journal, 700, L6

\bibitem[{Sacchi {et~al.}(2022)Sacchi, Risaliti, Signorini, Lusso, Nardini,
  Bargiacchi, Bisogni, Civano, Elvis, Fabbiano, {et~al.}}]{sacchi2022quasars}
Sacchi, A., Risaliti, G., Signorini, M., {et~al.} 2022, Astronomy \&
  Astrophysics, 663, L7

\bibitem[{{Salvestrini} {et~al.}(2019){Salvestrini}, {Risaliti}, {Bisogni},
  {Lusso}, \& {Vignali}}]{salvestrini2019}
{Salvestrini}, F., {Risaliti}, G., {Bisogni}, S., {Lusso}, E., \& {Vignali}, C.
  2019, \aap, 631, A120

\bibitem[{Schlafly \& Finkbeiner(2011)}]{schlafly2011measuring}
Schlafly, E.~F. \& Finkbeiner, D.~P. 2011, The Astrophysical Journal, 737, 103

\bibitem[{{Scolnic} {et~al.}(2018){Scolnic}, {Jones}, {Rest}, {Pan},
  {Chornock}, {Foley}, {Huber}, {Kessler}, {Narayan}, {Riess}, {Rodney},
  {Berger}, {Brout}, {Challis}, {Drout}, {Finkbeiner}, {Lunnan}, {Kirshner},
  {Sand ers}, {Schlafly}, {Smartt}, {Stubbs}, {Tonry}, {Wood-Vasey}, {Foley},
  {Hand}, {Johnson}, {Burgett}, {Chambers}, {Draper}, {Hodapp}, {Kaiser},
  {Kudritzki}, {Magnier}, {Metcalfe}, {Bresolin}, {Gall}, {Kotak}, {McCrum}, \&
  {Smith}}]{scolnic2018}
{Scolnic}, D.~M., {Jones}, D.~O., {Rest}, A., {et~al.} 2018, \apj, 859, 101

\bibitem[{Selsing {et~al.}(2016)Selsing, Fynbo, Christensen, \&
  Krogager}]{selsing2016x}
Selsing, J., Fynbo, J.~P., Christensen, L., \& Krogager, J.-K. 2016, Astronomy
  \& Astrophysics, 585, A87

\bibitem[{Shakura \& Sunyaev(1973)}]{shakura1973black}
Shakura, N.~I. \& Sunyaev, R.~A. 1973, Astronomy and Astrophysics, 24, 337

\bibitem[{Shemmer {et~al.}(2008)Shemmer, Brandt, Netzer, Maiolino, \&
  Kaspi}]{shemmer2008hard}
Shemmer, O., Brandt, W., Netzer, H., Maiolino, R., \& Kaspi, S. 2008, The
  Astrophysical Journal, 682, 81

\bibitem[{{Shen} {et~al.}(2011){Shen}, {Richards}, {Strauss}, {Hall},
  {Schneider}, {Snedden}, {Bizyaev}, {Brewington}, {Malanushenko},
  {Malanushenko}, {Oravetz}, {Pan}, \& {Simmons}}]{shen2011}
{Shen}, Y., {Richards}, G.~T., {Strauss}, M.~A., {et~al.} 2011, \apjs, 194, 45

\bibitem[{Signorini {et~al.}(2023)Signorini, Risaliti, Lusso, Nardini,
  Bargiacchi, Sacchi, \& Trefoloni}]{signorini2023quasars}
Signorini, M., Risaliti, G., Lusso, E., {et~al.} 2023, Astronomy \&
  Astrophysics, 676, A143

\bibitem[{{Soltan}(1982)}]{soltan1982}
{Soltan}, A. 1982, \mnras, 200, 115

\bibitem[{{Tananbaum} {et~al.}(1979){Tananbaum}, {Avni}, {Branduardi}, {Elvis},
  {Fabbiano}, {Feigelson}, {Giacconi}, {Henry}, {Pye}, {Soltan}, \&
  {Zamorani}}]{avnitananbaum79}
{Tananbaum}, H., {Avni}, Y., {Branduardi}, G., {et~al.} 1979, \apjl, 234, L9

\bibitem[{Trakhtenbrot {et~al.}(2017)Trakhtenbrot, Ricci, Koss, Schawinski,
  Mushotzky, Ueda, Veilleux, Lamperti, Oh, Treister,
  {et~al.}}]{trakhtenbrot2017bat}
Trakhtenbrot, B., Ricci, C., Koss, M.~J., {et~al.} 2017, Monthly Notices of the
  Royal Astronomical Society, 470, 800

\bibitem[{{Vanden Berk} {et~al.}(2001)}]{vandenberk2001}
{Vanden Berk}, D.~E. {et~al.} 2001, \aj, 122, 549

\bibitem[{{Vignali} {et~al.}(2003){Vignali}, {Brandt}, \&
  {Schneider}}]{vignali03}
{Vignali}, C., {Brandt}, W.~N., \& {Schneider}, D.~P. 2003, \aj, 125, 433

\bibitem[{{Vito} {et~al.}(2019){Vito}, {Brandt}, {Bauer}, {Calura}, {Gilli},
  {Luo}, {Shemmer}, {Vignali}, {Zamorani}, {Brusa}, {Civano}, {Comastri}, \&
  {Nanni}}]{vito2019}
{Vito}, F., {Brandt}, W.~N., {Bauer}, F.~E., {et~al.} 2019, \aap, 630, A118

\bibitem[{Wang {et~al.}(2021)Wang, Yang, Fan, Hennawi, Barth, Banados, Bian,
  Boutsia, Connor, Davies, {et~al.}}]{wang2021luminous}
Wang, F., Yang, J., Fan, X., {et~al.} 2021, The Astrophysical Journal Letters,
  907, L1

\bibitem[{Wang {et~al.}(2020)Wang, Songsheng, Li, Du, \&
  Zhang}]{wang2020parallax}
Wang, J.-M., Songsheng, Y.-Y., Li, Y.-R., Du, P., \& Zhang, Z.-X. 2020, Nature
  Astronomy, 4, 517

\bibitem[{Wang {et~al.}(2009)Wang, Zhou, Grupe, Yuan, Dong, \& Lu}]{wang2009x}
Wang, T., Zhou, H., Grupe, D., {et~al.} 2009, The Astronomical Journal, 137,
  4002

\bibitem[{{Webb} {et~al.}(2020){Webb}, {Coriat}, {Traulsen}, {Ballet}, {Motch},
  {Carrera}, {Koliopanos}, {Authier}, {de la Calle}, {Ceballos}, {Colomo},
  {Chuard}, {Freyberg}, {Garcia}, {Kolehmainen}, {Lamer}, {Lin}, {Maggi},
  {Michel}, {Page}, {Page}, {Perea-Calderon}, {Pineau}, {Rodriguez}, {Rosen},
  {Santos Lleo}, {Saxton}, {Schwope}, {Tom{\'a}s}, {Watson}, \&
  {Zakardjian}}]{webb2020}
{Webb}, N.~A., {Coriat}, M., {Traulsen}, I., {et~al.} 2020, arXiv e-prints,
  arXiv:2007.02899

\bibitem[{Wild {et~al.}(2011)Wild, Groves, Heckman, Sonnentrucker, Armus,
  Schiminovich, Johnson, Martins, \& LaMassa}]{wild2011optical}
Wild, V., Groves, B., Heckman, T., {et~al.} 2011, Monthly Notices of the Royal
  Astronomical Society, 410, 1593

\bibitem[{Wu \& Shen(2022)}]{wu2022catalog}
Wu, Q. \& Shen, Y. 2022, The Astrophysical Journal Supplement Series, 263, 42

\bibitem[{Xie {et~al.}(2015)Xie, Shen, Shao, \& Yin}]{xie2015apparent}
Xie, X., Shen, S., Shao, Z., \& Yin, J. 2015, The Astrophysical Journal
  Letters, 802, L16

\bibitem[{Yang {et~al.}(2020)Yang, Wang, Fan, Hennawi, Davies, Yue, Banados,
  Wu, Venemans, Barth, {et~al.}}]{yang2020poniua}
Yang, J., Wang, F., Fan, X., {et~al.} 2020, The Astrophysical Journal Letters,
  897, L14

\bibitem[{Yip {et~al.}(2004)Yip, Connolly, Berk, Ma, Frieman, SubbaRao, Szalay,
  Richards, Hall, Schneider, {et~al.}}]{yip2004spectral}
Yip, C., Connolly, A., Berk, D.~V., {et~al.} 2004, The Astronomical Journal,
  128, 2603

\bibitem[{Yoshii {et~al.}(2014)Yoshii, Kobayashi, Minezaki, Koshida, \&
  Peterson}]{yoshii2014new}
Yoshii, Y., Kobayashi, Y., Minezaki, T., Koshida, S., \& Peterson, B.~A. 2014,
  The Astrophysical Journal Letters, 784, L11

\bibitem[{Zafar {et~al.}(2015)Zafar, M{\o}ller, Watson, Fynbo, Krogager, Zafar,
  Saturni, Geier, \& Venemans}]{zafar2015extinction}
Zafar, T., M{\o}ller, P., Watson, D., {et~al.} 2015, Astronomy \& Astrophysics,
  584, A100

\bibitem[{Zaja{\v{c}}ek {et~al.}(2023)Zaja{\v{c}}ek, Czerny, Khadka, Prince,
  Panda, Mart{\'\i}nez-Aldama, \& Ratra}]{zajavcek2023extinction}
Zaja{\v{c}}ek, M., Czerny, B., Khadka, N., {et~al.} 2023, arXiv preprint
  arXiv:2305.08179

\bibitem[{Zappacosta {et~al.}(2023)Zappacosta, Piconcelli, Fiore, Saccheo,
  Valiante, Vignali, Vito, Volonteri, Bischetti, Comastri,
  {et~al.}}]{zappacosta2023hyperluminous}
Zappacosta, L., Piconcelli, E., Fiore, F., {et~al.} 2023, arXiv preprint
  arXiv:2305.02347

\bibitem[{Zhang(2008)}]{zhang2008dimming}
Zhang, P. 2008, The Astrophysical Journal, 682, 721

\bibitem[{{Zheng} {et~al.}(1997){Zheng}, {Kriss}, {Telfer}, {Grimes}, \&
  {Davidsen}}]{zheng97}
{Zheng}, W., {Kriss}, G.~A., {Telfer}, R.~C., {Grimes}, J.~P., \& {Davidsen},
  A.~F. 1997, \apj, 475, 469

\bibitem[{Zhou {et~al.}(2006)Zhou, Wang, Yuan, Lu, Dong, Wang, \&
  Lu}]{zhou2006comprehensive}
Zhou, H., Wang, T., Yuan, W., {et~al.} 2006, The Astrophysical Journal
  Supplement Series, 166, 128

\end{thebibliography}

\begin{appendix}

\twocolumn
\section{Stacking methods}
\label{app:stack_opt}
To build the stacks, we considered different techniques to resample the flux values, average them, and estimate the uncertainty in each spectral channel. The abbreviations in italics are used to denote these options in Table B.1.

%\textbf{Stacking algorithm}

%\begin{itemize}
\subsection{Stacking algorithm}

    %\item 
--  Direct stacking ($avg$): %$\longrightarrow$ 
    each spectrum was rebinned on a fixed wavelength grid designed to contain all the possible wavelengths in the same redshift bin. Then, each spectrum was normalised (see below). 
    In each spectral channel the final flux was computed as the mean, median, or geometric mean of the normalised non-zero fluxes. The uncertainty on the flux on a spectral channel was computed as the standard deviation or the inter-percentile deviation divided by the number of objects contributing to the spectral channel.\\
    %\item 
--  MC resampling ($mc$): %$\longrightarrow$ 
    a target number was decided before the stacking (generally $N_{\rm target}$\,=\,10,000). Then, we randomly extracted $N_{\rm target}$ spectra allowing for replacement and created a mock spectrum by randomizing the flux according to the uncertainty vector, in the same way as described in Section \ref{sec:fits} to generate the mock spectra. The mock spectra are subject to the same procedure as in the case of direct stacking.\\
    %\item 
--  Bootstrap resampling ($bs$): %$\longrightarrow$ 
    a target number was decided before the stacking (generally $N_{\rm target}$\,=\,200). Then we randomly extracted $N_{\rm draw}$ spectra allowing for replacement and stacked them together to create the \textit{i}-th composite spectrum. $N_{\rm draw}$ was chosen to be equal to the total number of good spectra in the bin, according to the quality criteria defined in Sec. \ref{sec:stacking}. This procedure was performed $N_{\rm target}$ times, building the distribution of the mean fluxes in each spectral channel. The final spectrum in each spectral channel was obtained as the mean of the mean fluxes and the uncertainty as the standard deviation.

%\end{itemize}

%\textbf{normalisation}
\subsection{Normalisation}

%\begin{itemize}
    %\item 
--  Monochromatic ($mcr$): %$\longrightarrow$ 
    the spectrum is normalised by its value at a certain wavelength $\Bar{\lambda}$, generally chosen in a relatively emission-free region. Actually, the normalisation is not performed using the exact monochromatic interpolated flux, but rather the mean flux values in a wavelength range (10--100\,\AA) around $\Bar{\lambda}$, in order to avoid anomalously high/low fluxes due to bad pixels.\\
    %\item 
--  Integral ($itg$): %$\longrightarrow$ 
    the spectrum is normalised by its integral over the whole wavelength range.
%\end{itemize}

%\textbf{Average estimator}
\subsection{Average estimator}

%\begin{itemize}
    %\item 
--  Mean: %$\longrightarrow$ 
    the direct mean is the most straightforward way to average the flux in each spectral channel. However, especially close to the wings of the spectra where the SNR lowers, composites are more prone to anomalously high fluxes, likely due to bad pixels.\\
%    \item 
--  Median: %$\longrightarrow$ 
    using the median value in each spectral channel is safer when there are strong outliers, such as bad pixels, which can significantly alter the average flux. Moreover, the median spectrum preserves the relative fluxes of the emission features (see Sec. 3 in VB01).\\
%    \item 
--  Geometric mean ($gmean$): %$\longrightarrow$ 
    the geometric mean spectrum preserves instead the global continuum shape. Indeed, it can be shown (Sec. 3 in VB01) that the geometric mean of a distribution of spectra describable as power laws is equivalent to a power law with a spectral index given by the arithmetic mean of the spectral indices $\langle f_{\lambda}\rangle_{gmean}$\,$\propto$\,$\lambda^{-\langle\alpha_{\lambda}\rangle}$.
%\end{itemize}

%\textbf{Uncertainty estimator}
\subsection{{Uncertainty estimator}}

%\begin{itemize}
    %\item 
--  Standard deviation ($std$): %$\longrightarrow$ 
    the uncertainty is estimated as the standard deviation of the non-zero normalised fluxes divided by the square root of the objects contributing to a spectral channel.\\
    %\item 
--  Semi--interpercentile range ($pct$): %$\longrightarrow$ 
    the uncertainty is estimated as half the difference between the $99^{\rm th}$ and $1^{\rm st}$ percentile values divided by the square root of the objects contributing to a spectral channel.
%\end{itemize}

\section{The test bin}
\label{app:test_bin}
To test how the different stacking options could affect the final result we performed all the 36 possible combinations of the stacking options using the averagely populated ($N_{\rm obj}=77$) bin `cB4'. For the MC (bootstrap) stacks 10,000 (200) re-samples were performed.

\begin{figure}[h!]
\centering
\includegraphics[scale=0.27]{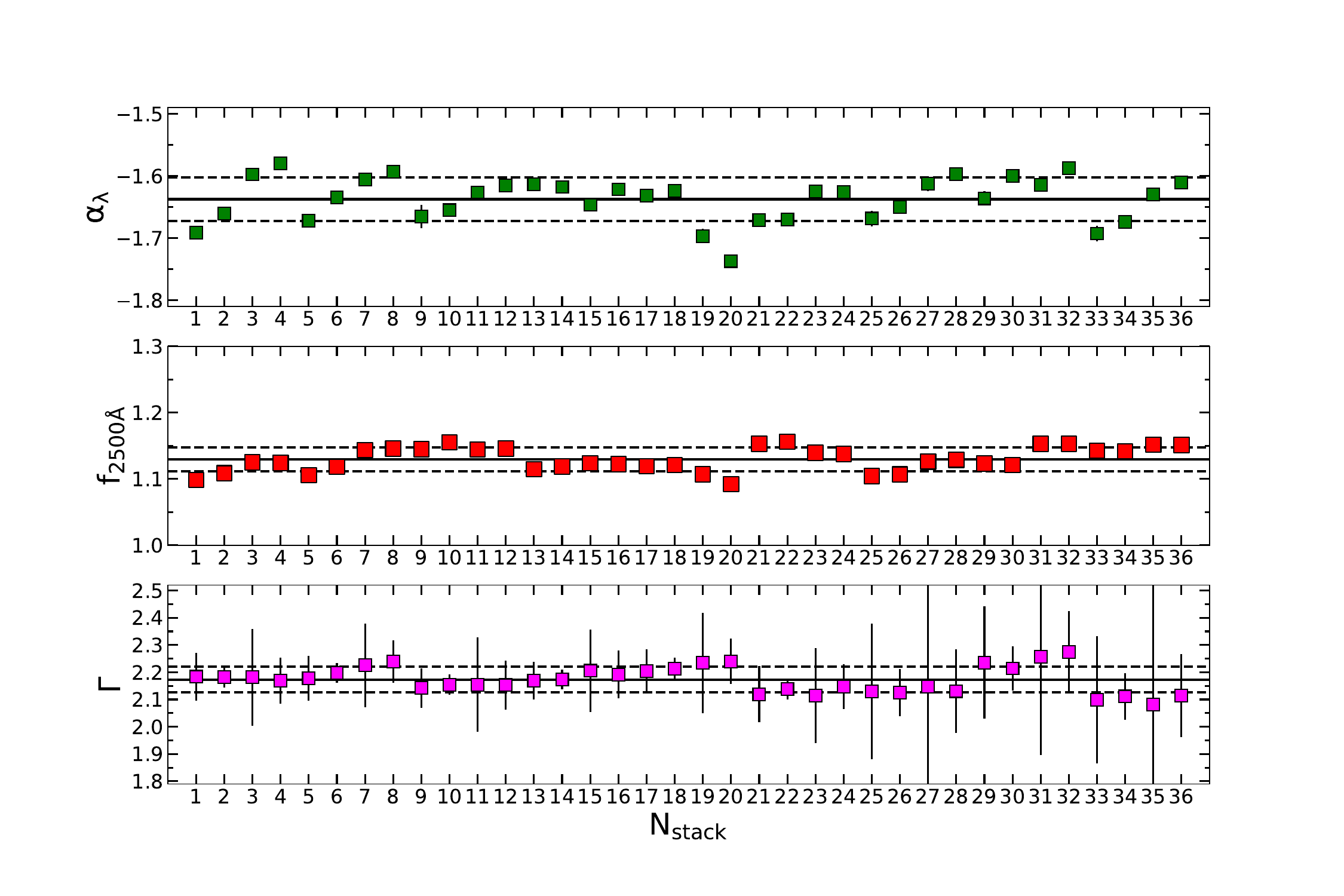}
\caption{Stack-to-stack variations of $\alpha_{\lambda}$ (green), 2500-\AA\ monochromatic flux (red), and X-ray $\Gamma$ (violet). The solid black line represents the mean value, dashed lines enclose values within a standard deviation. Stacks are numbered according to Table B.1.}
\label{fig:testbin_prop}
\end{figure}

We performed the spectral fits of all the test-bin stacks, in order to assess whether the slope of the continuum and the monochromatic fluxes were affected by the choices adopted for building the stack. In Fig. \ref{fig:testbin_prop} we show that $\alpha_{\lambda}$, $f_{\lambda}$, and $\Gamma$ are barely sensitive to the different stacking assumptions, as their relative variations are of the order of 2\%.

\begin{table*}[h!]
\begin{tabular}{ |p{1cm}|p{3cm}||p{1cm}|p{3cm}||p{1cm}|p{3cm}|  }

 \hline
 $N_{\rm stack}$ & stack features & $N_{\rm stack}$ & stack features & $N_{\rm stack}$ & stack features\\
 \hline
 1 & avg gmean int pct  & 13 & bs gmean int pct & 25 & mc gmean int pct\\
 2 & avg gmean int std  & 14 & bs gmean int std & 26 & mc gmean int std\\
 3 & avg gmean mcr pct  & 15 & bs mean itg  std   & 27 & mc gmean mcr pct\\
 4 & avg gmean mcr std  & 16 & bs gmean mcr std & 28 & mc gmean mcr std\\
 5 & avg mean int pct     & 17 & bs mean int pct    & 29 & mc mean int pct   \\
 6 & avg mean int std     & 18 & bs mean int std    & 30 & mc mean int std   \\
 7 & avg mean mcr pct     & 19 & bs mean mcr pct    & 31 & mc mean mcr pct   \\
 8 & avg mean mcr std     & 20 & bs mean mcr std    & 32 & mc mean mcr std   \\
 9 & avg median int pct   & 21 & bs median int pct  & 33 & mc median int pct \\
 10 & avg median int std  & 22 & bs median int std  & 34 & mc median int std \\
 11 & avg median mcr pct  & 23 & bs median mcr pct  & 35 & mc median mcr pct \\
 12 & avg median mcr std  & 24 & bs median mcr std  & 36 & mc median mcr std \\
 
\hline
\end{tabular}
%\label{tbl:test_bin_table}

\caption{Labelling of the test-bin stacks. Abbreviations correspond to those defined in the list of stacking options in Section \ref{app:stack_opt}. }

\end{table*}

\section{Impact of the extinction curve}
\label{app:extinction_curves}
\begin{figure}[h!]
\centering
\includegraphics[width=\linewidth,clip]{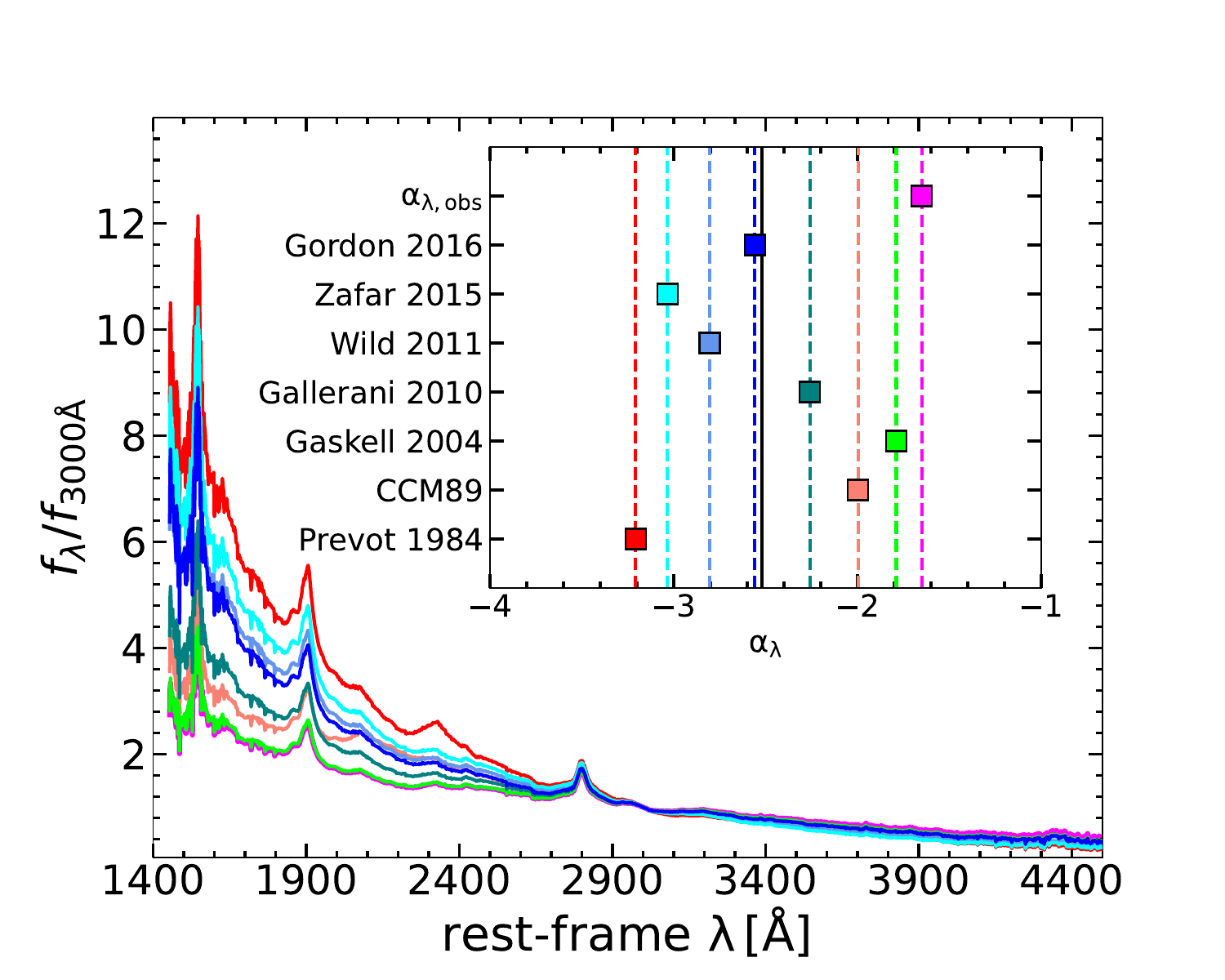}
\caption{The test-bin composite spectrum de-reddened according to different extinction curves as reported in the inset and assuming $E(B-V)=0.1$ and $R_V=3.1$. The slopes produced from the spectral fits according to the different extinction curves are represented in the inset. The spectrum and the slope of the continuum before the de-reddening ($\alpha_{\lambda,\ obs}$) are reported in magenta as references. The slope derived assuming the extinction curve from \cite{gordon2016panchromatic} is the closest to the average (black line).}
\label{fig:ext_test}
\end{figure}

With the end to explore the effect of the choice of the extinction curve adopted to redden or de-redden the composites throughout this work, we tested different extinction curves on the composite spectrum in the `cB4' bin. We included SMC-like curves (\citealt{prevot84, gordon2016panchromatic}), as well as AGN-derived extinction curves (\citealt{gaskell2004nuclear, gallerani2010extinction}, excluding BALs; \citealt{wild2011optical,zafar2015extinction}). After de-reddening the composite we performed the spectral fit to infer the slope of the continuum $\alpha_{\lambda}$ and compare it to that of the actual composite. We show the results of this procedure in Fig.\ref{fig:ext_test}. The slope of the de-reddened spectrum assuming the \citet{gordon2016panchromatic} SMC-like curve (adopted in this work) is close to the average of the distribution. 
The strongest outlier in the extinction distribution is the \citet{gaskell2004nuclear} extinction curve, which is akin to a grey-body for wavelengths below $\lesssim 3300$\,\AA, and therefore does not alter the spectral shape at short wavelengths. However, their sample is based on 72 radio-selected quasars with the addition of mostly radio-quiet objects from the LBQS survey, and it is possible, as already suggested by \citet{hopkins2004dust}, that the flattening of their extinction curve is driven by the spectral properties of the radio-loud quasars in their sample. Our choice of a SMC-like curve as representative of the intrinsic quasar extinction curve seems also justified by the findings of the latter work. There, the authors found a red tail in the distribution of SDSS photometric colours according to different redshifts. This trend proved to be consistent with SMC-like reddening, but not with Large Magellanic Cloud- or Milky Way-like.

\section{Table of results}
In Table \ref{tbl:results} we report all the average accretion parameters of the bins in which the parameter space is divided, as well as the spectral properties of the composites.

\onecolumn

\begin{table*}[h!]
%\begin{tabular}{|l|l|l|l|l|l|l|l|l|l|l|l|l|}
\centering
\begin{tabular}{|c|c|c|c|c|c|c|c|c|c|c|c|c|}

\hline\hline
bin &  $N_{\rm obj}$ & $\log({M_{\rm BH}/M_{\odot}})$ &  $\log({L_{\rm bol}/\rm{erg\,s^{-1}}})$ &  $\log(\lambda_{\rm Edd})$ & $\alpha_{\lambda}$ & $f_{2500\,\AA}$ & $\Gamma$ & $\Delta \alpha$ \\
\hline \noalign{\smallskip}
aC1 & 6   & \pc 7.63  $\pm$ 0.02 & 45.47 $\pm$ 0.02 & $-$0.27 $\pm$ 0.02 & $-$1.432 $\pm$ 0.005 & 1.298 $\pm$ 0.001 & 2.33 $\pm$ 0.09 &    0.014$\pm$ 0.002 \\
aD1 & 40  & \pc 7.45  $\pm$ 0.03 & 45.07 $\pm$ 0.03 & $-$0.50 $\pm$ 0.03 & $-$1.541 $\pm$ 0.073 & 1.324 $\pm$ 0.018 & 2.34 $\pm$ 0.11 & $-$0.060$\pm$ 0.010 \\
aC2 & 109 & \pc 8.17  $\pm$ 0.02 & 45.66 $\pm$ 0.02 & $-$0.62 $\pm$ 0.02 & $-$1.484 $\pm$ 0.007 & 1.311 $\pm$ 0.002 & 2.25 $\pm$ 0.13 & $-$0.022$\pm$ 0.003 \\
aD2 & 126 & \pc 8.04  $\pm$ 0.02 & 45.16 $\pm$ 0.02 & $-$0.99 $\pm$ 0.02 & $-$1.437 $\pm$ 0.025 & 1.299 $\pm$ 0.006 & 2.16 $\pm$ 0.11 &    0.011$\pm$ 0.002 \\
aC3 & 141 & \pc 8.67  $\pm$ 0.01 & 45.76 $\pm$ 0.02 & $-$1.02 $\pm$ 0.02 & $-$1.584 $\pm$ 0.007 & 1.335 $\pm$ 0.002 & 2.08 $\pm$ 0.08 & $-$0.090$\pm$ 0.014 \\
aD3 & 39  & \pc 8.58  $\pm$ 0.02 & 45.19 $\pm$ 0.02 & $-$1.50 $\pm$ 0.03 & $-$1.623 $\pm$ 0.021 & 1.344 $\pm$ 0.005 & 2.01 $\pm$ 0.07 & $-$0.117$\pm$ 0.018 \\
aB4 & 6   & \pc 9.35  $\pm$ 0.05 & 46.55 $\pm$ 0.05 & $-$0.92 $\pm$ 0.08 & $-$2.025 $\pm$ 0.028 & 1.446 $\pm$ 0.007 & 2.10 $\pm$ 0.11 & $-$0.394$\pm$ 0.060 \\
aC4 & 13  & \pc 9.32  $\pm$ 0.05 & 46.10 $\pm$ 0.08 & $-$1.33 $\pm$ 0.09 & $-$1.919 $\pm$ 0.013 & 1.419 $\pm$ 0.003 & 1.95 $\pm$ 0.05 & $-$0.321$\pm$ 0.049 \\
bC2 & 60  & \pc 8.21  $\pm$ 0.02 & 45.72 $\pm$ 0.03 & $-$0.61 $\pm$ 0.03 & $-$1.342 $\pm$ 0.026 & 1.277 $\pm$ 0.006 & 2.26 $\pm$ 0.10 &    0.077$\pm$ 0.012 \\
bD2 & 45  & \pc 8.14  $\pm$ 0.03 & 45.18 $\pm$ 0.02 & $-$1.07 $\pm$ 0.04 & $-$1.374 $\pm$ 0.151 & 1.285 $\pm$ 0.035 & 2.15 $\pm$ 0.07 &    0.054$\pm$ 0.010 \\
bB3 & 20  & \pc 8.85  $\pm$ 0.04 & 46.53 $\pm$ 0.03 & $-$0.44 $\pm$ 0.04 & $-$1.751 $\pm$ 0.003 & 1.376 $\pm$ 0.001 & 2.17 $\pm$ 0.13 & $-$0.205$\pm$ 0.031 \\
bC3 & 253 & \pc 8.75  $\pm$ 0.01 & 45.93 $\pm$ 0.02 & $-$0.93 $\pm$ 0.02 & $-$1.487 $\pm$ 0.024 & 1.311 $\pm$ 0.006 & 2.16 $\pm$ 0.10 & $-$0.023$\pm$ 0.004 \\
bD3 & 43  & \pc 8.61  $\pm$ 0.02 & 45.27 $\pm$ 0.02 & $-$1.46 $\pm$ 0.03 & $-$1.409 $\pm$ 0.023 & 1.293 $\pm$ 0.005 & 2.08 $\pm$ 0.07 &    0.030$\pm$ 0.005 \\
bB4 & 23  & \pc 9.29  $\pm$ 0.03 & 46.66 $\pm$ 0.05 & $-$0.75 $\pm$ 0.04 & $-$1.664 $\pm$ 0.004 & 1.354 $\pm$ 0.001 & 2.13 $\pm$ 0.09 & $-$0.145$\pm$ 0.022 \\
bC4 & 32  & \pc 9.27  $\pm$ 0.03 & 46.12 $\pm$ 0.04 & $-$1.26 $\pm$ 0.04 & $-$1.521 $\pm$ 0.007 & 1.320 $\pm$ 0.002 & 2.06 $\pm$ 0.05 & $-$0.047$\pm$ 0.007 \\
cC2 & 35  & \pc 8.22  $\pm$ 0.02 & 45.80 $\pm$ 0.04 & $-$0.54 $\pm$ 0.04 & $-$1.445 $\pm$ 0.011 & 1.301 $\pm$ 0.003 & 2.22 $\pm$ 0.08 &    0.036$\pm$ 0.000 \\
cD2 & 9   & \pc 8.09  $\pm$ 0.05 & 45.17 $\pm$ 0.05 & $-$1.04 $\pm$ 0.06 & $-$1.419 $\pm$ 0.020 & 1.295 $\pm$ 0.005 & 2.19 $\pm$ 0.07 &    0.053$\pm$ 0.001 \\
cB3 & 49  & \pc 8.92  $\pm$ 0.02 & 46.57 $\pm$ 0.02 & $-$0.47 $\pm$ 0.03 & $-$1.565 $\pm$ 0.003 & 1.330 $\pm$ 0.001 & 2.22 $\pm$ 0.08 & $-$0.045$\pm$ 0.000 \\
cC3 & 231 & \pc 8.76  $\pm$ 0.01 & 45.97 $\pm$ 0.02 & $-$0.90 $\pm$ 0.11 & $-$1.451 $\pm$ 0.008 & 1.303 $\pm$ 0.002 & 2.22 $\pm$ 0.09 &    0.032$\pm$ 0.000 \\
cD3 & 6   & \pc 8.79  $\pm$ 0.05 & 45.27 $\pm$ 0.08 & $-$1.63 $\pm$ 0.02 & $-$1.333 $\pm$ 0.011 & 1.275 $\pm$ 0.003 & 2.08 $\pm$ 0.06 &    0.110$\pm$ 0.001 \\
cB4 & 77  & \pc 9.34  $\pm$ 0.02 & 46.69 $\pm$ 0.02 & $-$0.76 $\pm$ 0.02 & $-$1.568 $\pm$ 0.004 & 1.331 $\pm$ 0.001 & 2.19 $\pm$ 0.09 & $-$0.046$\pm$ 0.000 \\
cC4 & 61  & \pc 9.25  $\pm$ 0.01 & 46.19 $\pm$ 0.02 & $-$1.17 $\pm$ 0.02 & $-$1.495 $\pm$ 0.006 & 1.313 $\pm$ 0.002 & 2.16 $\pm$ 0.08 &    0.002$\pm$ 0.000 \\
dC2 & 9   & \pc 8.20  $\pm$ 0.05 & 45.82 $\pm$ 0.07 & $-$0.49 $\pm$ 0.08 & $-$1.137 $\pm$ 0.192 & 1.230 $\pm$ 0.043 & 2.23 $\pm$ 0.12 &    0.201$\pm$ 0.034 \\
dB3 & 66  & \pc 8.93  $\pm$ 0.02 & 46.62 $\pm$ 0.02 & $-$0.43 $\pm$ 0.02 & $-$1.453 $\pm$ 0.005 & 1.303 $\pm$ 0.001 & 2.25 $\pm$ 0.10 & $-$0.021$\pm$ 0.000 \\
dC3 & 141 & \pc 8.81  $\pm$ 0.02 & 46.06 $\pm$ 0.02 & $-$0.87 $\pm$ 0.02 & $-$1.577 $\pm$ 0.024 & 1.333 $\pm$ 0.006 & 2.22 $\pm$ 0.10 & $-$0.108$\pm$ 0.002 \\
dA4 & 7   & \pc 9.66  $\pm$ 0.05 & 47.55 $\pm$ 0.04 & $-$0.22 $\pm$ 0.06 & $-$1.557 $\pm$ 0.004 & 1.328 $\pm$ 0.001 & 2.16 $\pm$ 0.05 & $-$0.094$\pm$ 0.001 \\
dB4 & 173 & \pc 9.40  $\pm$ 0.01 & 46.82 $\pm$ 0.02 & $-$0.70 $\pm$ 0.02 & $-$1.479 $\pm$ 0.011 & 1.309 $\pm$ 0.003 & 2.20 $\pm$ 0.09 & $-$0.039$\pm$ 0.000 \\
dC4 & 69  & \pc 9.32  $\pm$ 0.02 & 46.16 $\pm$ 0.03 & $-$1.27 $\pm$ 0.03 & $-$1.374 $\pm$ 0.051 & 1.285 $\pm$ 0.012 & 2.17 $\pm$ 0.08 &    0.035$\pm$ 0.001 \\
dA5 & 7   & \pc 9.98  $\pm$ 0.03 & 47.70 $\pm$ 0.06 & $-$0.39 $\pm$ 0.06 & $-$1.560 $\pm$ 0.123 & 1.329 $\pm$ 0.030 & 2.00 $\pm$ 0.04 & $-$0.096$\pm$ 0.008 \\
dB5 & 6   & \pc 9.92  $\pm$ 0.02 & 46.89 $\pm$ 0.13 & $-$1.14 $\pm$ 0.14 & $-$1.336 $\pm$ 0.020 & 1.276 $\pm$ 0.005 & 2.16 $\pm$ 0.08 &    0.061$\pm$ 0.001 \\
eC2 & 22  & \pc 8.20  $\pm$ 0.05 & 46.10 $\pm$ 0.04 & $-$0.21 $\pm$ 0.05 & $-$1.308 $\pm$ 0.019 & 1.269 $\pm$ 0.004 & 2.17 $\pm$ 0.09 &    0.169$\pm$ 0.004 \\
eB3 & 96  & \pc 8.86  $\pm$ 0.02 & 46.68 $\pm$ 0.02 & $-$0.30 $\pm$ 0.03 & $-$1.491 $\pm$ 0.016 & 1.312 $\pm$ 0.004 & 2.17 $\pm$ 0.09 &    0.052$\pm$ 0.001 \\
eC3 & 63  & \pc 8.81  $\pm$ 0.03 & 46.19 $\pm$ 0.02 & $-$0.73 $\pm$ 0.03 & $-$1.558 $\pm$ 0.018 & 1.329 $\pm$ 0.004 & 2.18 $\pm$ 0.10 &    0.010$\pm$ 0.000 \\
eA4 & 26  & \pc 9.53  $\pm$ 0.04 & 47.70 $\pm$ 0.04 & $-$0.06 $\pm$ 0.06 & $-$1.623 $\pm$ 0.154 & 1.344 $\pm$ 0.038 & 2.00 $\pm$ 0.03 & $-$0.032$\pm$ 0.003 \\
eB4 & 174 & \pc 9.39  $\pm$ 0.02 & 46.87 $\pm$ 0.02 & $-$0.64 $\pm$ 0.02 & $-$1.447 $\pm$ 0.017 & 1.302 $\pm$ 0.004 & 2.24 $\pm$ 0.09 &    0.080$\pm$ 0.002 \\
eC4 & 23  & \pc 9.31  $\pm$ 0.04 & 46.27 $\pm$ 0.02 & $-$1.15 $\pm$ 0.03 & $-$1.152 $\pm$ 0.012 & 1.234 $\pm$ 0.003 & 2.17 $\pm$ 0.08 &    0.268$\pm$ 0.006 \\
eA5 & 44  & \pc 10.06 $\pm$ 0.03 & 47.80 $\pm$ 0.04 & $-$0.37 $\pm$ 0.04 & $-$1.452 $\pm$ 0.004 & 1.303 $\pm$ 0.001 & 2.08 $\pm$ 0.08 &    0.077$\pm$ 0.002 \\
eB5 & 12  & \pc 9.98  $\pm$ 0.06 & 47.15 $\pm$ 0.06 & $-$0.95 $\pm$ 0.09 & $-$1.732 $\pm$ 0.030 & 1.371 $\pm$ 0.008 & 2.14 $\pm$ 0.06 & $-$0.101$\pm$ 0.003 \\

\hline

\end{tabular}
\caption{Relevant quantities for each bin of the parameter space.}
\label{tbl:results}

\end{table*}

\twocolumn
\section{Correlation index tables}
\label{app:correlations}

In Table \ref{tbl:tbl_sl_pars} we report the correlation index $S$ between the slope $\alpha_{\lambda}$ and the accretion parameters, including/excluding the bins ruled out by the representativeness criterion. 

% #############################################################

\begin{table*}[h!]
\centering
\begin{tabular}{||c | c | c | c||} 
 \hline
 subsample & $\alpha_{\lambda}$\,--\,$\log M_{\rm BH}$ & $\alpha_{\lambda}$\,--\,$\log L_{\rm bol}$ & $\alpha_{\lambda}$\,--\,$\log\lambda_{\rm Edd}$ \\ [0.5ex] 
 \hline\hline
 all & $(-0.300,0.067)$ & $(-0.263,0.110)$ & $(-0.006,0.972)$ \\ 
 \hline
 selected & $(-0.281,0.125)$ & $(-0.229,0.214)$ & $(-0.017,0.928)$ \\ %[1ex] 
 \hline

\end{tabular}
\caption{Correlation index and associated $P-value$ $(S,P)$ pairs for the UV and optical slopes against the accretion parameters. In the top row all the bins are included in the computation, in the bottom only those meeting the representativeness criterion.}
\label{tbl:tbl_sl_pars}
\end{table*}
% #############################################################

\section{Reddening expression}
\label{app:dm_reddening}

Here we explicitly derive Eq. \ref{eq:dm_ext}. We begin from Eq. 7 in L20, in order to express the distance modulus DM starting from the luminosity distance $d_L$ computed from the \lxluv relation:
\begin{equation*}
\begin{split}
    {\rm DM} & = 5 \left[\frac{\log f_{\rm X} - \beta -\gamma\,(\log f_{2500\,\AA}+27.5)}{2\gamma -1} - \frac{1}{2} \log (4\pi) +28.5 \right] \\
    & - 5 \, \log (10 \, {\rm pc}).
\end{split}
\end{equation*}
Keeping only the dependence on $f_{2500\AA}$, DM can be written as:
\begin{equation*}
    {\rm DM} = 5 \, \frac{\gamma}{2(\gamma - 1)} \log f_{2500\,\AA} + \theta,
\end{equation*}
where %the term 
$\theta$ encapsulates all the terms where $f_{2500\,\AA}$ does not appear. Let us now assume, by hypothesis, that the observed UV fluxes are significantly extincted, to the point that the discrepancy from the $\rm{\Lambda}$CDM model that we report is driven by such extinction. Then, we can denote the observed DM as:
\begin{equation*}
    {\rm DM_{ext}} = - 5\, \frac{\gamma}{2(\gamma - 1)} \log f_{2500\,\AA, \rm ext} + \theta,
\end{equation*}
and the intrinsic (unextincted) DM as:
\begin{equation*}
    {\rm DM_{int}} = - 5\, \frac{\gamma}{2(\gamma - 1)} \log f_{2500\,\AA, \rm int} + \theta.
\end{equation*}
The intrinsic UV flux is related to the extincted one by:
\begin{equation*}
    f_{2500\,\AA, \rm int } = f_{2500\,\AA, \rm ext} \, 10^{0.4 \, e_{\lambda} \, R_{V} \, E(B-V)}
\end{equation*}
where $e_{\lambda}$ is the assumed extinction curve, $R_{V}$ is the total to selective extinction ratio, and $E(B-V)$ is the color excess. We can then write DM$_{\rm int}$ in terms of the extincted UV flux: 
\begin{equation*}
%    {\rm DM_{int}} = - 5\, \frac{\gamma}{2(\gamma - 1)} \log f_{2500\,\AA, \rm ext} - 5\, \frac{\gamma}{2(\gamma - 1)} \,  0.4 e_{\lambda} \, R_{V} \, E(B-V)  + \theta
{\rm DM_{int}} = - 5\, \frac{\gamma}{2(\gamma - 1)} \, [ \log  f_{2500\,\AA, \rm ext} +  0.4 e_{\lambda} \, R_{V} \, E(B-V) ]  + \theta.
\end{equation*}
The difference between the intrinsic and the observed distance moduli ($\Delta$DM) thus becomes:
\begin{equation*}
    {\rm \Delta DM} = {\rm DM_{int}} - {\rm DM_{ext}} = - 5 \,\frac{\gamma}{2(\gamma - 1)}  \,  0.4 e_{\lambda} \, R_{V} \, E(B-V).
\end{equation*}
Supposing that the DM$_{\rm int}$ is the DM corresponding to the $\rm{\Lambda}$CDM model, $\Delta$DM can be evaluated by imposing  DM$_{\rm int}$ = DM$_{\rm \Lambda CDM}$. By isolating the $E(B-V)$ term, Eq. \ref{eq:dm_ext} is found:
\begin{equation*}
    E(B-V) = \frac{\Delta \rm DM}{-\frac{5}{2}\frac{\gamma}{\gamma - 1} \, 0.4  \, R_V \, \frac{A_{2500\,\AA}}{A_V}}.
\end{equation*}

\section{Synopsis of the UV stacks}
\label{app:synopt}

\begin{figure*}[h!]
\centering
\includegraphics[scale=0.50]{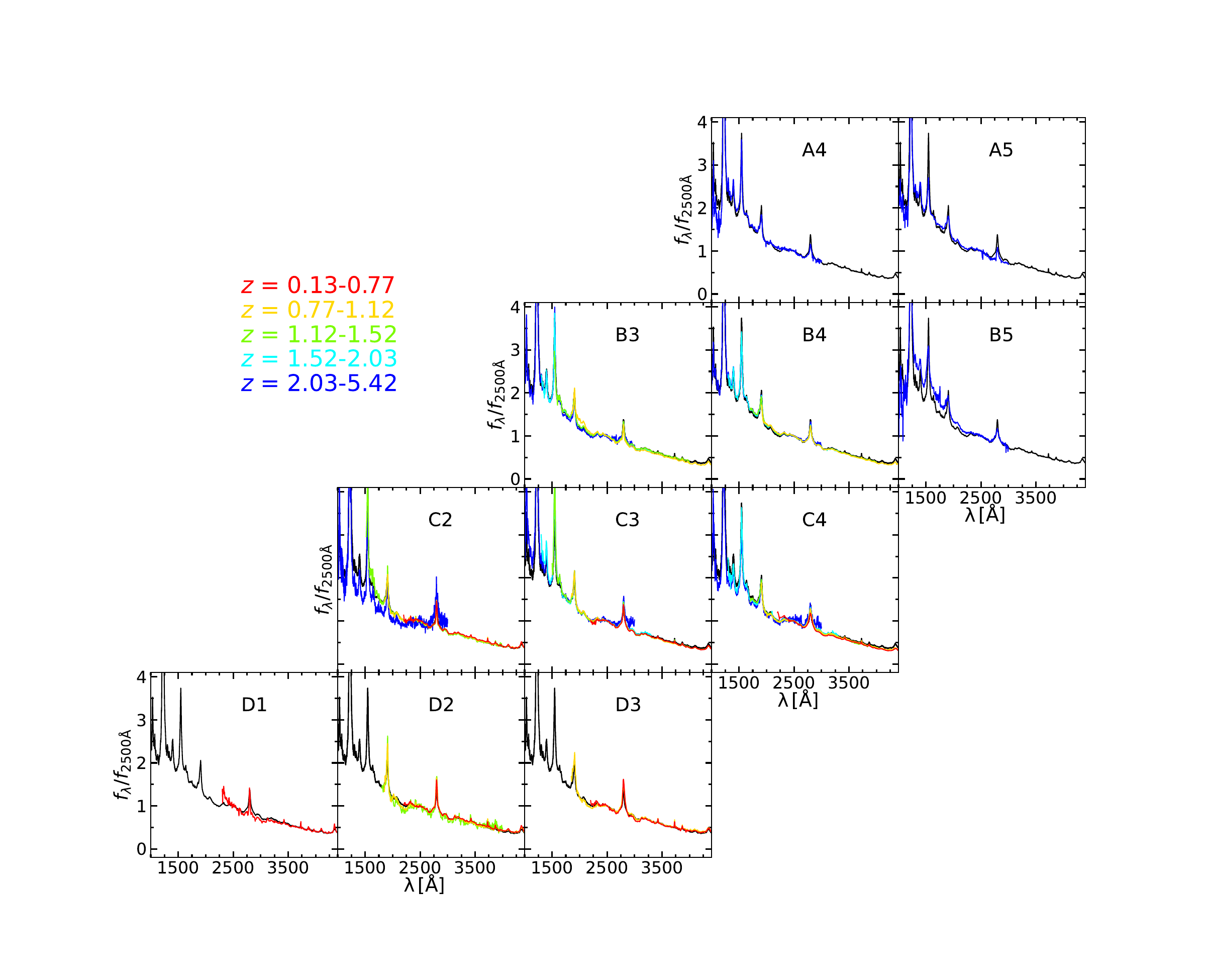}
\caption{Similarity of the stacked spectra across the parameter space. All the spectra are normalised by their average emission between 2490--2510 \AA. The black spectrum represents the average spectrum of our full sample. Only spectra of the bins meeting the representativeness criterion are included in the plot.}
\label{fig:synopt_uv}
\end{figure*}

In Fig. \ref{fig:synopt_uv}, we show all the UV spectral stacks across the parameter space in different colours according to their redshift bin, along with the average spectrum of our whole sample in black as a reference. Regions without any spectrum are those discarded in Sec. \ref{sec:paramspace}. Only the UV side for the spectra is shown, as the optical one is only available in the lower redshift bins.

\end{appendix}

\end{document}